\documentclass[12pt]{iopart}

\usepackage{iopams} 
\usepackage{graphicx}
\usepackage{longtable}
\usepackage{cite}
\usepackage{hyperref}

\newcommand{\kviadd}{Kernfysisch Versneller Instituut (KVI), University of
  Groningen, 
Zernikelaan 25, 9747 AA Groningen, The Netherlands}
\newcommand{\bochum}{Institute f\"ur Theoretische Physik II, Ruhr-Universit\"at Bochum,
D-44870 Bochum, Germany}

\newcommand{\juelichias}{Institute for Advanced Simulation, Institut f\"ur Kernphysik, and 
J\"ulich Center for Hadron Physics, Forschungszentrum J\"ulich, D-52425 J\"ulich, Germany}

\newcommand{\beq}{\begin{equation}}
\newcommand{\eeq}{\end{equation}}
\newcommand{\beqa}{\begin{eqnarray}}
\newcommand{\eeqa}{\end{eqnarray}}
\newcommand{\fet}[1]{\mbox{\boldmath $#1$}}

\begin{document}

\title{Signatures of three-nucleon interactions in few-nucleon systems}

\author{N.~Kalantar-Nayestanaki}
\address{\kviadd}
\ead{nasser@kvi.nl}
\author{E.~Epelbaum}
\address{\bochum}
\ead{evgeny.epelbaum@rub.de}
\author{J.G. Messchendorp}
\address{\kviadd}
\ead{messchendorp@kvi.nl}
\author{A.~Nogga}
\address{\juelichias}
\ead{a.nogga@fz-juelich.de}
\begin{abstract}
Recent experimental results in three-body systems have unambiguously shown
that calculations based only on nucleon-nucleon forces fail to accurately 
describe many experimental observables and one needs to include effects 
which are beyond the realm of the two-body potentials. This conclusion owes 
its significance to the fact that experiments and calculations can both be 
performed with a high accuracy. In this review, both theoretical and experimental 
achievements of the past decade will be underlined. Selected results will be presented. 
The discussion on the effects of the three-nucleon forces is, however, limited to the 
hadronic sector. It will be shown
that despite the major successes in describing these seemingly simple systems, there
are still clear discrepancies between data and the state-of-the-art calculations. 
\end{abstract}

\maketitle

\section{Introduction}
\label{sec:introduction}
\begin{figure}
\begin{center}
\label{BEs}
\itshape
\includegraphics[width=0.99\linewidth]{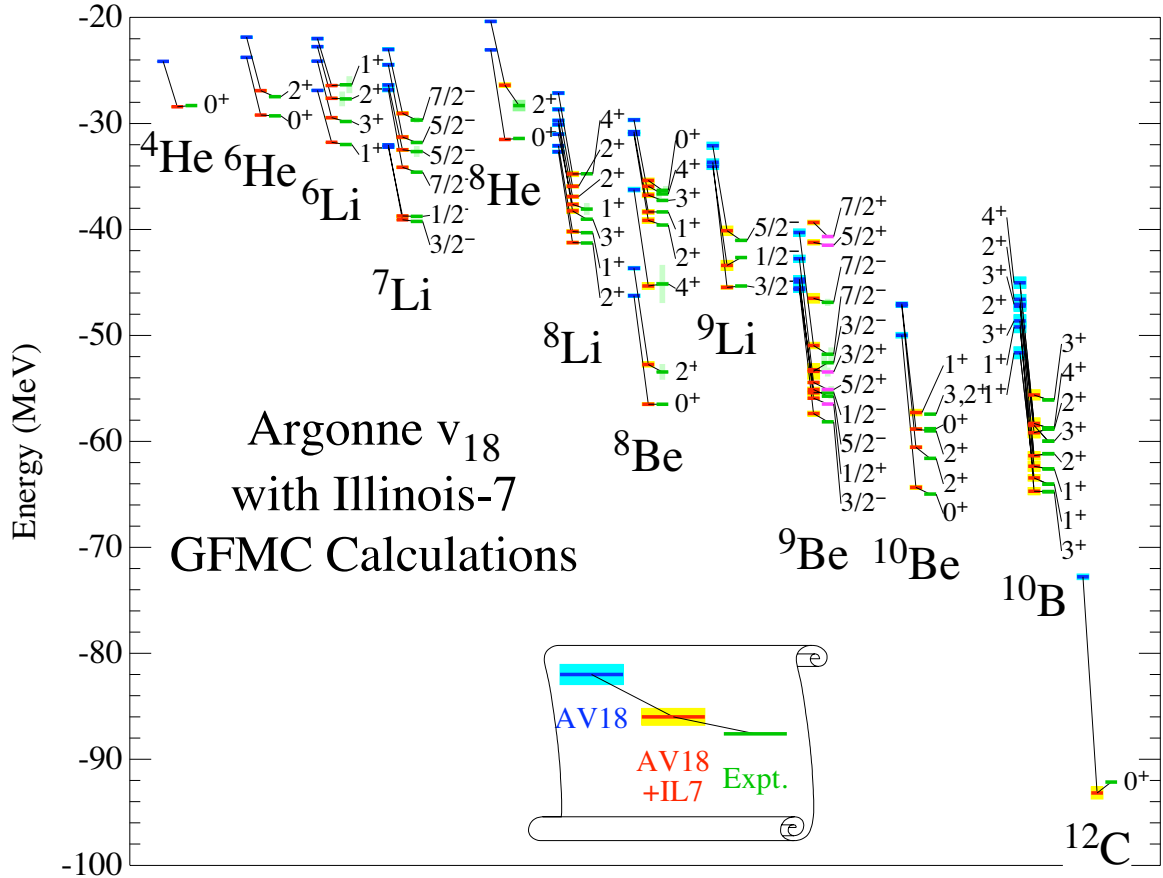}
\caption{Binding energies of light nuclei. The experimental values are compared with 
Green's Function Monte Carlo calculations performed with only a two-nucleon potential (AV18, blue/dark grey) 
and with the addition of a three-nucleon potential (IL7, yellow/light grey) \cite{Pieper:priv} (color online).
}
\end{center}
\end{figure}

The ultimate goal of nuclear physics is to understand the properties of nuclei
and reactions involving them. Given the smallness of the typical energy scales in
nuclear physics, such as e.g.~the nuclear binding energies (BEs), it appears
appropriate to formulate the nuclear N-body problem  
in terms of the non-relativistic Schr\"odinger equation. 
In a first approximation, the two-nucleon potential is sufficient to
describe the bulk of the few-nucleon observables at low and intermediate energies. 
At present, a number of semi-phenomenological two-nucleon models 
are available which provide an accurate description of the nucleon-nucleon 
(NN) scattering data below the pion production threshold with a $\chi^2$ per 
degree of freedom of the order $\sim$1. Recent advances in the development 
of few-body methods coupled with a significant increase in computational 
resources allow one to perform accurate microscopic calculations of 
three- and even four-nucleon scattering observables and of the spectra of 
light nuclei. This opens the door for precise tests of the underlying
dynamics and, in particular, of the role and structure of the three-nucleon
force (3NF). One of the simplest and most extensively studied three-nucleon
observables is the BE of the triton. It is well known to be significantly 
underestimated by the existing two-nucleon potentials \cite{Nogga:2000uu,Nogga:2002qp}
\footnote{Note,
however, that phase-equivalent (at low energy) non-local NN potentials can be
constructed which reproduce the $^3$H BE \cite{Doleschall:2003ha}.}. 
A similar underbinding occurs for other light nuclei as well 
\cite{Pieper:2001mp}. This is shown in Fig.~\ref{BEs} where the experimental 
binding energies of light nuclei are compared with exact calculations including two and
three-body forces. The need to go beyond the two-nucleon force (2NF) is rather evident. 

\begin{figure}
\begin{center}
\itshape
\includegraphics[width=0.8\linewidth]{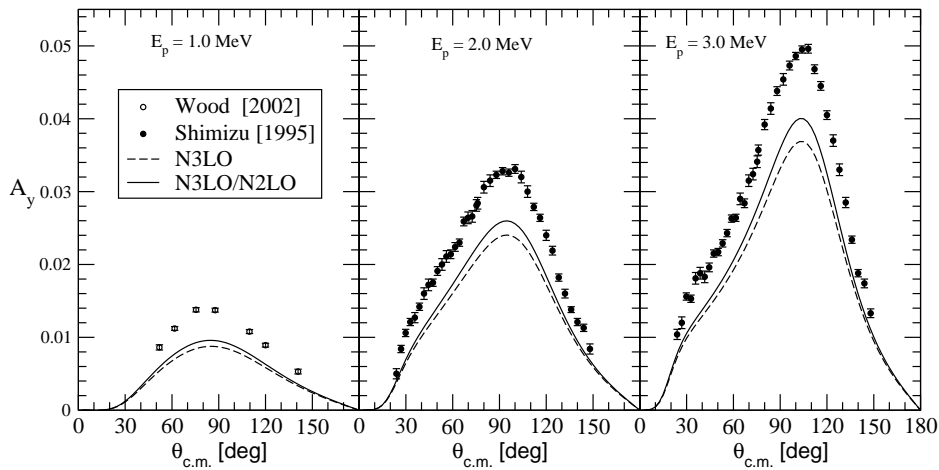}
\caption{Vector analyzing power in elastic proton-deuteron scattering at three incident proton energies below the break-up threshold.
The dashed line represents the result based on the two-nucleon potential derived 
at next-to-next-to-next-to-leading order chiral EFT while the solid 
line also includes the effect of the chiral 3NF at next-to-next-to-leading order 
\cite{Marcucci:2009xf}.
The data are from 
Refs.~\cite{Wood:2001gb,Shimizu:1995zz}.}
\label{aypuzzle}
\end{center}
\end{figure}

\begin{figure}
\begin{center}
\itshape
\includegraphics[width=0.6\linewidth]{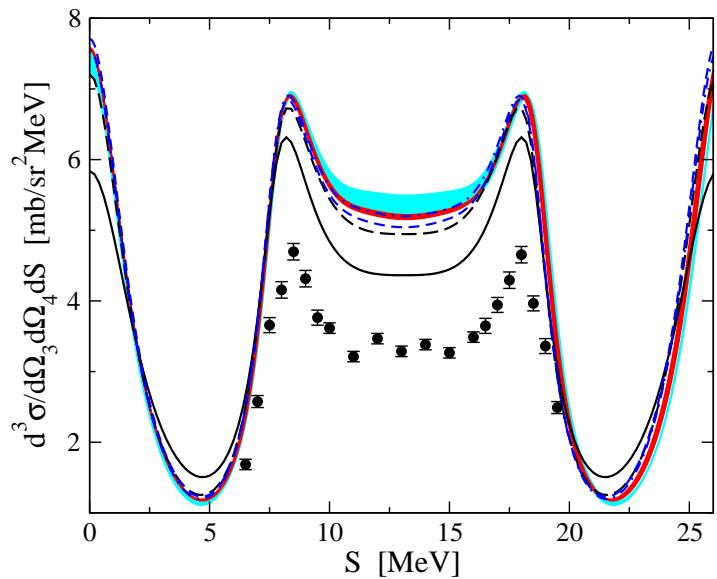}
\caption{Results for the deuteron break-up cross sections at 
  $E_d = 19$ MeV as a function of 
  the kinematical variable $S$, with $\alpha
  = 56^\circ$ \cite{Ley:2006hu} and taken at the symmetric constant 
  relative energy (SCRE) configuration. Light/cyan (dark/red) shaded bands depict 
  predictions from chiral EFT at NLO
  (N$^2$LO). Short-dashed and solid (long-dashed) lines show the
  results based on the combination of the CD-Bonn 2000 2NF
  and TM99 3NF and the coupled-channel calculations including the explicit
  $\Delta$ with (without) Coulomb interaction, respectively. The results 
  based on the
  CD-Bonn 2000 2NF overlap with the chiral EFT band at N$^2$LO (color online).
}
\label{spacestar}
\end{center}
\end{figure}

Three-nucleon continuum observables have also been explored by several
groups. While the differential cross section of 
elastic nucleon-deuteron (N$d$) scattering at  incident beam energies below~30 MeV
is rather well described using solely NN potentials, 
a large relative discrepancy with the data, known as the $A_y$-puzzle, is observed
for the analyzing power; see Fig.~\ref{aypuzzle}. The data are shown for energies which are below the break-up threshold. However, the puzzle remains and only diminishes at energies of around 30~MeV \cite{Gloeckle:1995jg}. One should, however, note that the actual values
of the analyzing powers are rather small  so that very small corrections 
of the nuclear Hamiltonian might be able to resolve this puzzle. 
We are, therefore, not convinced that this specific observable is a good
indicator for significant failures of the 3N Hamiltonian.
In section~\ref{sec:fourbody}, we will further discuss this issue also
for the 4N system.
The theoretical results 
shown in the figure are obtained within the framework of an effective field theory 
(EFT), see the next section for more details, but are very similar to the 
predictions of various potential models 
\cite{Marcucci:2009xf}. Tensor-analyzing powers and spin-transfer 
coefficients are generally rather well described at low energies using solely 
2NFs but the results of these calculations start to deviate from 
the data as the energy increases. This was clearly demonstrated for cross sections by Sagara et al. \cite{Sagara:1994zz,Sagara:2010fbs}. In addition to the elastic channel, the 
break-up reaction offers a rich kinematics and as such provides a good 
testing ground for the structure of the nuclear force. In general, the calculations
at low energies agree rather well with the experimental results. However, there
are a couple of observables which show major discrepancies with the theoretical 
predictions. One example is the well-known space-star kinematics shown in Fig.~\ref{spacestar}. It should be mentioned, however, that the preliminary results from a recent measurement at Tsukuba tandem laboratory show much a better agreement with the results of the calculations than the data shown in Fig.~\ref{spacestar} \cite{Sagara:priv}.

The observed discrepancies between data and calculations based solely on 2NFs 
are usually viewed as an indication of the existence of a 3NF. Indeed, 3NFs 
which cannot be reduced to pair-wise NN interactions arise naturally in the
context of a meson-exchange theory and at the more fundamental level of QCD.
At present, several phenomenological 3NF models exist which are typically based
on the two-pion exchange contribution and will be discussed in section
\ref{sec:theory}. Despite some remarkable successes of the phenomenological 
approach, many problems still remain open; see section \ref{sec:results3nf} for
explicit examples. In addition, there are obvious conceptual deficiencies such
as e.g.~the lack of a consistent treatment of 2NFs and 3NFs in the same framework. On the 
other hand, significant progress has been achieved recently in understanding the properties of few-nucleon systems
within the framework of chiral EFT.  This approach is linked to QCD via its symmetries and
allows one to analyze the low-energy properties of hadronic systems in a
systematic and controlled way. In addition, it offers a natural explanation
for the observed hierarchy of nuclear forces: 
$\langle V_{2N} \rangle \gg \langle V_{3N} \rangle  \gg  \langle V_{4N} \rangle$.

In this review, some of the experimental observables in 
proton-deuteron ({\it pd}) scattering will be discussed along with the theoretical 
developments which are taking place. The experimental investigations have 
been performed at various laboratories for a large part of the phase space. 
Here, we restrict ourselves mainly to elastic and break-up observables in the 
medium energy region of 50 to 250 MeV incident nucleon energy.

A number of recent review articles on the nuclear forces and their 
applications to few- and many-body systems are available. 
While Refs.~\cite{Bedaque:2002mn,Beane:2000fx} 
focus mainly on applications of EFT techniques to few- and many-nucleon 
systems, a comprehensive review of the structure of nuclear forces in 
chiral EFT and selected 
applications to nucleon-deuteron scattering and light nuclei can 
be found in Ref.~\cite{Epelbaum:2005pn}. Recent developments in these 
and related topics 
are also considered in a more broadly focused review article \cite{Epelbaum:2008ga}.  
For a broader overview on current topics related to few-nucleon
systems, we refer to \cite{Arenhovel:2004nr}. 
Further, a compilation of experimental and theoretical information 
on the mass 3 systems is presented in Ref.~\cite{Purcell:2010npa}.  For a recent
review article describing experimental investigations of discrepancies in three-nucleon
reactions at low energies see Ref.~\cite{Sagara:2010fbs}. Going beyond nuclear systems, 
for an overview on pionless EFT and universality in few-body systems with 
large scattering length, the reader is referred to \cite{Hammer:2010kp}. 
The present review article is focussed more strongly on 3NFs and confronting their 
effects with experimental data in nucleon-deuteron scattering and, 
to a lesser extent, four-nucleon scattering and the spectra of light nuclei.  

Our paper is organized as follows.
In section \ref{sec:nnforce}, the nucleon-nucleon force is introduced 
and briefly discussed. This section is followed by a section on 
3NFs and the $A=3$ systems. 
Latest developments on the theoretical side will be discussed in some detail. In 
addition, a survey of the experimental data will be made providing the grounds
for a systematic comparison with the results of various model calculations. 
Section~\ref{sec:fourbody} is devoted to the discussion of exact 
calculations of four-body systems and the investigation of the three-body 
force effects in them. This will
be done based on the small existing database for these systems. 
The manifestation of 3NFs in systems where $A>4$ will be discussed in section \ref{sec:manybody}. Here, 
recent theoretical developments will be outlined. Throughout this review, we will
limit ourselves to hadronic systems due to lack of space. For a treatment of various
aspects of the electromagnetic probe of few-body systems where the role of the 3NF is
also important, we refer the reader to Refs.~\cite{Golak:2005iy,Carlson:1997qn,Bacca:2008tb,Quaglioni:2007eg}. 
Notice further that nucleon-nucleon bremsstrahlung and its implications for our knowledge about 
3NFs have also been extensively studied \cite{Michaelian:1990vw,Przewoski:1992zz,Jetter:1994zz,Eden:1995rf,Martinus:1998zz,Nakayama:2009yz,Yasuda:1999wa,Messchendorp:1999prl,Messchendorp:1999zz,Huisman:1999zz,Huisman:2000plb,Hoefman:2000jn,Messchendorp:2000plb,Messchendorp:2000sm,Huisman:2001yc,Volkerts:2003bk,Volkerts:2004prl,Mahjour-Shafiei:2009epj,MahjourShafiei:2004vg,Johansson:2008gy,MehmandoostKhajehDad:2005sk,Messchendorp:2008as}. Finally, in section \ref{sec:conclusions}, 
conclusions are drawn and an outlook is given for the future.

\section{Nucleon-nucleon interaction in two-body systems}
\label{sec:nnforce}

\subsection{Theoretical framework}
\label{theory2nf}

The nuclear force problem is one of the oldest but still current problems in
nuclear physics. It is of a crucial importance for understanding the
properties of atomic nuclei and, more generally, strongly interacting hadronic matter.  
The strong interaction between the nucleons emerges due to the residual color force between
quarks and gluons -- the elementary constituents of the colorless nucleons. 
The
conventional way to describe the nuclear force utilizes the meson-exchange picture, which goes back
to the seminal work by Yukawa \cite{Yukawa:1935xg}. His idea, followed by the
experimental discovery of  $\pi$- and heavier
mesons ($\rho$, $\omega$, $\ldots$), stimulated the development of
boson-exchange models that have 
laid the foundations for the construction of high-precision phenomenological 
nucleon-nucleon (NN) potentials. Dispersion relations were also employed to construct the
two-pion exchange contribution, see e.g.~\cite{Cottingham:1973wt,Jackson:1975be}.

The most general structure of a non-relativistic two-nucleon potential can be
expressed in terms of just a few operators. In momentum space, the operator
basis can, for example, be chosen as 
$ \fet 1_{\rm spin}$, $\vec \sigma_1 \cdot \vec \sigma_2$,  
$(\vec \sigma_1 \cdot \vec q \, )( \vec \sigma_2 \cdot \vec q \, )$, 
$(\vec \sigma_1 \cdot \vec k )( \vec \sigma_2 \cdot \vec k)$,  $i (\vec  \sigma_1 + \vec \sigma_2 )
\cdot \vec q \times
\vec k $ and $(\vec \sigma_1 \cdot \vec q \times \vec k \,)( \vec \sigma_2 \cdot
\vec q \times \vec k$).  
Here, $\vec \sigma_i$ are the Pauli spin matrices of the nucleon $i$, $\vec q
\equiv \vec p \, ' - \vec p$, $\vec k \equiv \vec p \, ' + \vec p$
and $\vec{p}$
($\vec{p}~'$) refers to initial (final) nucleon momenta in the center-of-mass system.
The isospin structure of the two-nucleon force falls into the four
different classes \cite{Henley:1979ig}: isospin-invariant (class I), 
isospin-breaking but charge-symmetry conserving
(class-II) and  charge-symmetry breaking contributions (classes III and
IV). Class II interactions break isospin invariance but respect charge symmetry. 
They are usually referred to as charge-independence
breaking and are e.g.~responsible for the difference between isovector \emph{np} and 
the average of the \emph{nn} and \emph{pp} phase shifts. 
Class IV (III) forces do (not) cause isospin mixing in the two-nucleon
system. For a recent review on charge-symmetry breaking see
Ref.~\cite{Miller:2006tv}.

The general strategy in constructing phenomenological potential models 
such as e.g.~the CD-Bonn 2000 \cite{Machleidt:2000ge}, Argonne $V_{18}$
(AV18) \cite{Wiringa:1994wb} and 
Nijmegen I, II potentials \cite{Stoks:1994wp} involves incorporating 
the proper long-range tail of the nuclear force due to the electromagnetic
interaction (Coulomb interaction \cite{Austen:1983te}, vacuum polarization
\cite{Uehling:1935aa,Durand:1957zz} and magnetic moment interaction
\cite{Stoks:1990us}) and the 
one-pion exchange potential and parameterizing the medium- and short-range 
contributions~\footnote{In the following, the term phenomenological potential 
models will be used to refer to the models which reproduce the two-nucleon scattering
data with $\chi^2/{\rm datum} \sim 1$, such as the ones specified in the text.}.
With about 40 to 50  adjustable parameters, these potentials
provide an excellent description of a few thousand of low-energy proton-proton and 
neutron-proton scattering data 
with   $\chi^2/{\rm datum} \sim 1$. 
We refer the reader to Ref.~\cite{Machleidt:2001rw}
for a review article on high-precision potentials.  For a recent
high-precision potential model within the covariant spectator theory see
Ref.~\cite{Gross:2008ps}. Notice that this particular model is designed for 
using in the manifestly covariant Spectator equations rather than the nonrelativistic
Schr\"odinger equations. Effects due to the inclusion of the $\Delta$(1232) isobar as an 
explicit degree of freedom have also been investigated. Coupled-channel potential models involving NN$\to$N$\Delta$ 
transitions are presented in Refs.~\cite{Haidenbauer:1993pw,Deltuva:2003wm}. 
In particular, the model of Deltuva et al.~\cite{Deltuva:2003wm} 
based on the CD-Bonn potential leads to a description of the NN data 
which is comparable to the high-precision phenomenological potentials. 
It should be noted that the 
$\Delta \Delta$  channels are not included in this calculation. 
Note further that the two models of \cite{Haidenbauer:1993pw,Deltuva:2003wm} 
lead to considerably different results for NN$\to$N$\Delta$ transition amplitudes. 
These differences are, however, not visible in the two-nucleon observables.

While the conventional approach to the nuclear force problem outlined above
enjoys many successes and is frequently used in nuclear
structure and reaction calculations, it suffers from having only  
a very loose connection to QCD. 
In addition, it does not  provide a way for assigning
theoretical uncertainties and leaves the question of constructing consistent
many-body forces and exchange currents open, see \cite{Epelbaum:2008ga} for an extensive discussion. 

The obvious drawbacks of the phenomenological framework can be overcome by
employing chiral EFT, a systematic and
model-independent approach to study the low-energy hadron dynamics in harmony
with the symmetry pattern of QCD. This method exploits an approximate,
spontaneously broken chiral symmetry of QCD with two flavors of the $u$- and $d$-quarks and, to a
 lesser extent, with three flavors of  the $u$-, $d$- and $s$-quarks. 
These symmetry/symmetry-breaking patterns manifest themselves in the
hadron spectrum and provide a natural explanation of the very
small (compared to other hadrons) masses of pions which are 
identified with the corresponding 
Goldstone bosons. Moreover, the nature of the Goldstone boson implies
that pions only interact weakly at low energy. These features are at the heart 
of chiral perturbation theory. In this framework, low-energy dynamics
of pions and nucleons is 
described in terms of the  most general effective Lagrangian 
formulated in terms of hadronic degrees of freedom and featuring the same
chiral symmetry (breaking) pattern as QCD. It contains infinitely many 
local interactions with increasing number of derivatives and/or quark mass
insertions due to the explicit breaking of chiral symmetry. 
Each term in the effective Lagrangian is multiplied by an a-priori
unknown constant, the so-called low-energy constant (LEC). The values of the
LECs are, in principle, calculable from QCD but can also be determined from
the data, see e.g.~\cite{Ecker:2007dj}.  The resulting effective Lagrangian can be applied to compute low-energy
pion and single-nucleon observables in a systematic way by making a perturbative
expansion in powers of $Q/\Lambda_\chi$
\cite{Weinberg:1978kz,Gasser:1983yg}.  Here, $Q$ and
$\Lambda_\chi$ refer to the soft scale associated with 
low external momenta or $M_\pi$ and the hard chiral-symmetry breaking scale
of the order of 1 GeV, respectively. 
A recent review on the methodology and applications of chiral
perturbation theory in the Goldstone-Boson and single-nucleon sectors 
can be found in \cite{Bernard:2007zu}. 

Chiral EFT can, however, not be directly applied to low-energy few-nucleon
scattering. The strong nature of the nuclear force that 
manifests itself in the appearance of self-bound atomic nuclei invalidates
a naive application of perturbation theory. Weinberg pointed out that the
breakdown of perturbation theory can be traced back to the infrared
enhancement of reducible diagrams that involve few-nucleon cuts 
 \cite{Weinberg:1990rz,Weinberg:1991um}. The irreducible part of the amplitude
that gives rise to the nuclear force is, however, not affected by the infrared
enhancement and is thus accessible within chiral EFT. These important
observations have triggered an intense research activity towards the
systematic derivation of the nuclear forces in chiral EFT and their
applications to the nuclear few- and many-body problem. 

It should also be emphasized that an EFT can be formulated that is only valid at typical momenta well below the 
pion mass. This framework allows to take into account the unnaturally large NN scattering lengths 
but looses the connection with the chiral symmetry of QCD.  
It has been used with large success not only for nuclear systems 
\cite{Braaten:2004rn,Bedaque:2002mn}. This so-called pion-less EFT requires a 3NF in leading order. 
The corresponding hard scale is given by the pion mass and, therefore, 
intermediate-energy observables cannot be 
predicted within this framework. In the following, we restrict ourselves to chiral EFT.

\begin{figure}[t]
\includegraphics[width=0.99\textwidth]{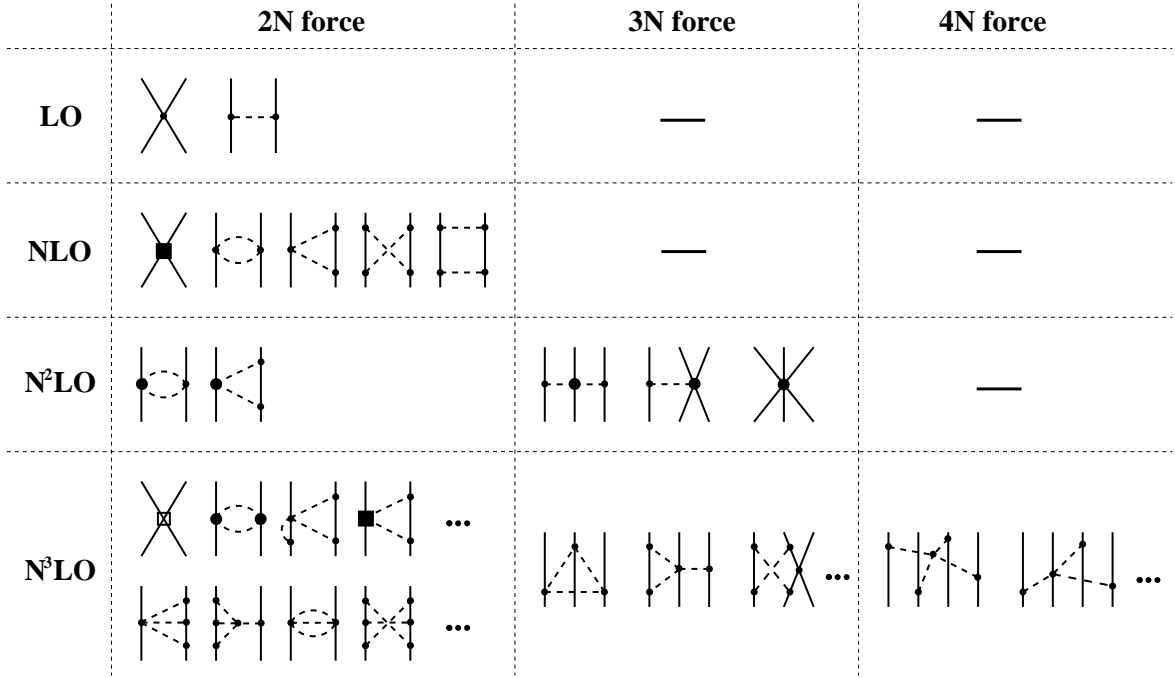}
\caption{Hierarchy of nuclear forces in chiral EFT based on Weinberg's power
  counting. Solid and dashed lines denote nucleons
  and pions, respectively. Solid dots, filled circles and filled squares
  refer, respectively, to the leading, subleading and sub-subleading vertices in the
  effective Lagrangian. The crossed square denotes 2N contact interactions with 4
  derivatives.} \label{fig1}
\end{figure}

\begin{figure}[t]
\includegraphics[width=1.0\textwidth]{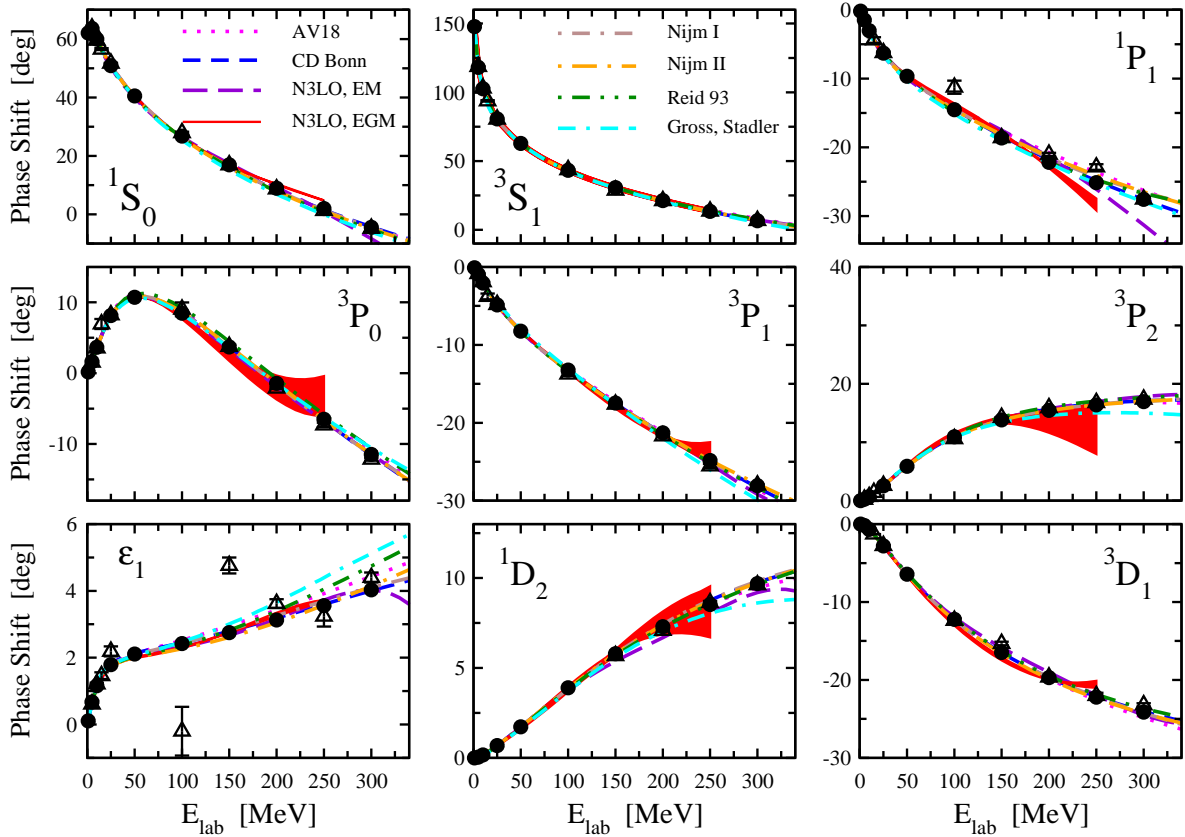}
\caption{Selected neutron-proton phase shifts for various potentials in 
  comparison with the Nijmegen \cite{Stoks:1993tb} (filled circles) and Virginia Tech
  \cite{Arndt:2007qn} (open triangles) PWA  (color online). The band
for EGM N$^3$LO corresponds to a variation of the cutoffs in the natural 
range providing a (rough) estimate of the theoretical uncertainty
at this order, see Ref.~\cite{Epelbaum:2004fk} for more details.} \label{fig:phases}
\end{figure}

The EFT expansion of the nuclear force based on the standard chiral power
counting (i.e.~assuming that all operators in the effective Lagrangian scale
according to a naive dimensional analysis) is visualized in Fig.~\ref{fig1}. 
It provides a natural qualitative explanation of
the observed hierarchy of two-, three- and more-nucleon forces 
with  $\langle V_{\rm 2N} \rangle \gg \langle V_{\rm 3N} \rangle  \gg \langle V_{\rm 4N} \rangle \ldots $.  
The expansion of the two-nucleon force (2NF)
has the form 
\begin{equation}
V_{\rm 2N} = V_{\rm 2N}^{(0)} + V_{\rm 2N}^{(2)} + V_{\rm 2N}^{(3)} + V_{\rm
  2N}^{(4)} + \ldots \,, 
\end{equation}
with the superscripts referring to the power of the expansion parameter
$Q/\Lambda_\chi$. The long-range part of the nuclear force is 
dominated by $1\pi$-exchange with the $2\pi$-exchange contributions
starting at next-to-leading order (NLO). The expressions for the potential up
to next-to-next-to-leading order (N$^2$LO) in the heavy-baryon formulation
\cite{Jenkins:1990jv,Bernard:1992qa} are rather compact and have been
independently derived by several authors using a variety of different methods 
\cite{Ordonez:1995rz,Friar:1994zz,Kaiser:1997mw,Epelbaum:1998ka}.  
$2\pi$- and $3\pi$-exchange contributions at next-to-next-to-next-to-leading
order (N$^3$LO) have been worked out by Kaiser
\cite{Kaiser:2001pc,Kaiser:2001at,Kaiser:1999ff,Kaiser:1999jg} and are
considerably more involved, see also
Ref.~\cite{Entem:2002sf}.  While the $2\pi$-exchange at NLO and, especially,
at N$^2$LO generates a rather strong potential at distances of the order of
the inverse pion mass \cite{Kaiser:1997mw}, the leading $3\pi$-exchange contributions turn out to be negligible. The N$^3$LO contributions to the
$2\pi$-exchange potential were also derived  in the covariant formulation of
chiral EFT \cite{Becher:1999he} by Higa et al.~\cite{Higa:2003jk,Higa:2003sz}.
The low-energy constants entering the pion-exchange contributions up to
N$^3$LO are known from pion-nucleon scattering and related processes
\cite{Fettes:1998ud,Fettes:1999wp,Buettiker:1999ap}. We note 
that some of them, especially the ones from $\mathcal{L}_{\pi N}^{(3)}$, are
presently not very accurately determined. 
In the isospin limit, the short-range part of the potential involves 2, 9, 24
independent terms at LO, NLO/N$^2$LO, N$^3$LO, respectively. The corresponding
LECs were fixed from the two-nucleon data leading to the accurate N$^3$LO 
potentials of Entem and Machleidt (EM) \cite{Entem:2003ft} and Epelbaum, Gl\"ockle,
Mei{\ss}ner (EGM) \cite{Epelbaum:2004fk}. 
These two potentials differ in the treatment of relativistic corrections (including the 
form of the employed dynamical equation),  isospin-breaking terms and the form 
of the regulator functions. There are also differences in the adopted values of certain pion-nucleon low-energy constants. Finally, EGM provide an estimation of the theoretical uncertainty by means of the variation of the cutoffs in some natural ranges. 
This important issue is not addressed in \cite{Entem:2003ft}, where a single excellent fit 
to neutron-proton (proton-proton) scattering data with $\chi^2/\mbox{datum} =1.10$ ($\chi^2/\mbox{datum} =1.50$) in the energy range from $0$ to $290$ MeV is given. 
We refer the reader to Refs.~\cite{Entem:2003ft,Epelbaum:2004fk} for more details. 
In Fig.~\ref{fig:phases}, we compare NN phase shifts obtained with these models with 
predictions from phenomenological potentials from Nijmegen \cite{Stoks:1994wp}, 
AV18 \cite{Wiringa:1994wb}, CD-Bonn \cite{Machleidt:2000ge} and from 
Gross and Stadler \cite{Gross:2008ps}. We emphasize that the chiral expansion 
of the two-nucleon scattering observables within the formulations of Refs.~\cite{Entem:2003ft,Epelbaum:2004fk} is expected to converge at energies well below the pion production threshold. To extend to higher energies, it is necessary to explicitly take into account the momentum scale $Q \sim \sqrt{m_N M_\pi}$ associated with the real pion production. This has not been pursued so far for two-nucleon scattering.  For a detailed discussion of the convergence of the chiral expansion for nucleon-nucleon scattering the reader is referred to Ref.~\cite{Epelbaum:2004fk}.
 
The consistency of the Weinberg power counting for short-range operators 
has been questioned by several
authors, see e.g.~Refs.~\cite{Kaplan:1998tg,Nogga:2005hy}. The meaning of the
non-perturbative renormalization of the Schr\"odinger equation in the context
of chiral EFT and the implications on the power counting are currently under
discussion, see Refs.~\cite{Lepage:1997cs,Gegelia:2004pz,Birse:2005um,PavonValderrama:2005wv,Epelbaum:2006pt,
Epelbaum:2009sd} and references therein for a sample of different views on this issue. 
Note that while several different schemes have been proposed in the
literature, a real alternative  to 
the Weinberg approach
for practical calculations is not yet available. 

Isospin-breaking (IB) contributions to the nuclear force have been extensively
studied in the framework of chiral EFT. Within the Standard Model,
isospin violation has its origin in the different quark masses and the
electromagnetic interactions. Chiral EFT is well suited to
explore the implications of these two effects which lead to a string of 
IB terms in the effective hadronic Lagrangian which are 
proportional either to the quark mass difference or the fine structure
constant. IB contributions to the nuclear forces can then be worked out
straightforwardly leading to a similar hierarchy for many-body forces
as in the isospin-symmetric case. We emphasize, however, that different
counting rules are used in the literature 
to relate the additional expansion parameters for IB contributions
(i.e.~the quark mass difference and the fine structure
constant) with $Q/\Lambda_\chi$. In the two-nucleon force, the dominant
IB contribution is due to the different pion masses, $M_{\pi^0} \neq
M_{\pi^\pm}$, in the one-pion exchange. The resulting potential is
charge-symmetry conserving, i.e.~of class II. Charge-symmetry breaking
forces of classes III and IV are considerably weaker and are, to a large
extent, driven by the
proton-to-neutron mass 
differences in the one- and two-pion exchange contributions and the short-range terms 
\cite{vanKolck:1996rm,Friar:2004ca,Epelbaum:2005fd,Epelbaum:2004xf}. 
The power counting suggests the 
hierarchy of the two-nucleon forces with $\langle V_{2N}^I \rangle > \langle 
V_{2N}^{II} \rangle > \langle V_{2N}^{III} \rangle >  \langle V_{2N}^{IV} \rangle$  
\cite{VanKolck:1993ee} 
which is consistent with observations. It should also be emphasized that 
the purely electromagnetic contributions are strongly enhanced 
under certain kinematical conditions (low energies and/or
forward angles) due to their long-range nature.
For a more comprehensive review on various contributions to the nuclear force
the reader is referred to the recent review 
articles~\cite{Bedaque:2002mn,Epelbaum:2005pn,Epelbaum:2008ga}. 

All the developments described above are based on the EFT with pions and 
nucleons as the only degrees of freedom. On the other hand, 
the $\Delta$(1232) isobar is
known to play an important role in nuclear physics due to its low excitation
energy and strong coupling to the $\pi N$ system. The explicit inclusion of
the $\Delta$ in the EFT by treating the $\Delta$-N mass splitting as a soft
scale~\cite{Hemmert:1997ye} allows one to resum a certain
class of important contributions leading to an improved convergence. Since the
calculations involving the $\Delta$ are considerably more involved, its
contribution to the nuclear force are, at present, only worked out up to N$^2$LO
\cite{Ordonez:1995rz,Kaiser:1998wa,Krebs:2007rh,Epelbaum:2007sq}. These
studies confirm an improved convergence of the EFT expansion
compared to the $\Delta$-less theory.

\subsection{Observables and comparison with experimental data}

\begin{figure}[t]
\begin{center}
\vskip 0.6 true cm
\includegraphics[width=0.9\textwidth]{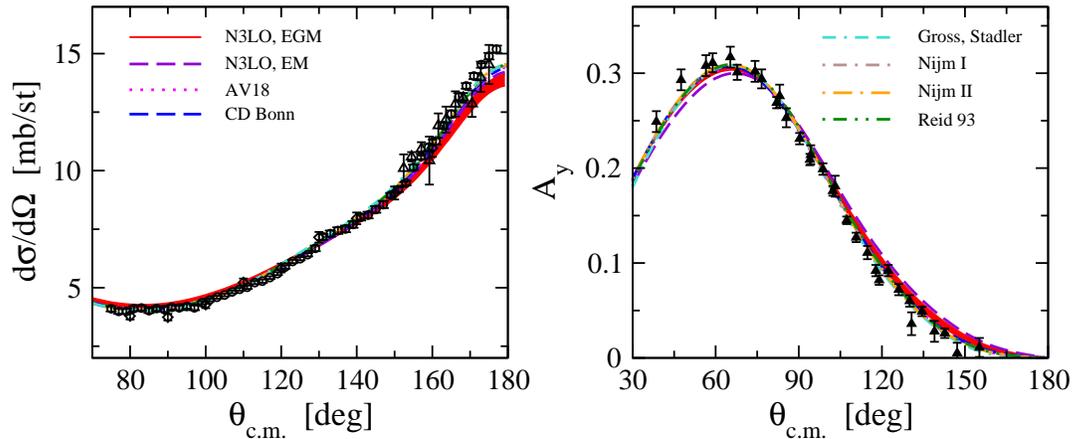}
\caption{Neutron-proton differential cross section at $E_{\rm lab} = 96$ MeV
  (left panel) and analyzing power at $E_{\rm lab} = 67.5$ MeV (right panel)
  calculated using various potentials. The data for the cross
  section/analyzing power  are from
  Refs.~\cite{Klug:2002nim,Blideanu:2004wi,Rahm:2001my}/\cite{BrogliGysin:1992npa} (color online).}  
\label{fig:NNobs} 
\end{center}
\end{figure}

Both the phenomenological potentials and the ones resulting from chiral
EFT at N$^3$LO allow for an accurate description of the low-energy
nucleon-nucleon scattering data and the deuteron properties. As representative
examples, we show in Fig.~\ref{fig:NNobs} the neutron-proton differential cross
section at $E_{\rm lab} = 96$ MeV and vector analyzing power at  $E_{\rm lab}
= 67.5$ MeV, for which fairly recent data are available.

It is worth saying a few words about the fitting procedure. In most cases,
the unknown parameters contributing to the isospin $T=1$ channels are
determined solely from the more accurate proton-proton scattering data. The
neutron-proton potentials in the $T=1$ channels are then reconstructed from the
corresponding proton-proton ones by employing corrections for the Coulomb interaction and
certain isospin-breaking effects while the $T=0$ part is
fixed to the neutron-proton data. The precise form of isospin-breaking
corrections varies between the different models. Note that the same 
procedure is also adopted in
the Nijmegen 1993 partial wave analysis (PWA). In this context we would like to emphasize, that isospin-breaking effects in the two-nucleon force lead to sizable effects in the neutron-deuteron $A_y$, see e.g.~\cite{Tornow:1998zz}. Clearly, the neutron-neutron
potentials also have to be reconstructed from the corresponding proton-proton
and neutron-proton ones since no neutron-neutron scattering data are available. The 
phenomenological potentials are typically fitted to the data below $350$~MeV, while
the EM N$^3$LO potential uses the data below $290$~MeV. The LECs entering the EGM N$^3$LO
potential are determined from a fit to the Nijmegen PWA at energies below 
$100 \ldots 200$ MeV (depending on the partial wave). 

As already pointed out in the previous section, the 
available phenomenological nucleon-nucleon
potentials provide excellent fits of the scattering data below the pion
production threshold with $\chi^2/{\rm datum} \sim 1$ or slightly
above. We further emphasize that one usually allows for some rescaling of the
data in order to minimize the resulting $\chi^2$.  Notice further that 
different groups adopt somewhat different criteria to reject
inconsistent data. For example, the data set used in the 
recent model of Gross and Stadler \cite{Gross:2008ps} includes 3788 neutron-proton data, 
3336/3010 of which are prior to 2000/1993 while rejecting 1180 data points
which are found to be statistically inconsistent. To compare, the Nijmegen PWA
was fit to 2514, AV18 to 2526 and CD-Bonn 2000 to 3058 {\it np} data. A set of
53 neutron-proton data for the differential cross section at 96 MeV and
$75^\circ \leq \theta_{\rm CM} \leq 179^\circ$ from
Ref.~\cite{Rahm:2001my} shown in the left panel of Fig.~\ref{fig:NNobs} is an 
example of data that have been rejected from the PWA.  It yields 
$\chi^2/{\rm datum} \approx 2.8$ for the current solution of the Virginia Tech
PWA~\cite{Arndt:2007qn}.
We also emphasize that a very precise recent measurement of neutron-proton $A_y$ 
at $12$ MeV at TUNL~\cite{Braun:2008eh} raised some debate about a 
possible $A_y$ problem in low-energy
neutron-proton scattering, see Ref.~\cite{Gross:2008pd} and references therein. 
The last issue is possibly related to the $A_y$ 
problem in low-energy N$d$ scattering mentioned earlier~\cite{Tornow:1998zz}. 
For our later discussions on 3NF effects at intermediate energies, the differences 
in the fitting procedures, the strategies for rejecting data and minor 
inconsistencies of different PWAs are not significant. 
Last but not least, we would also like to emphasize the large difference between the 
Nijmegen and Virginia Tech PWA regarding the values of the mixing angle $\epsilon_1$, 
see Fig.~\ref{fig:phases}, indicating that the available neutron-proton scattering 
data are not very sensitive to this particular observable.  

\section{Three-nucleon forces and $A=3$ systems}
\label{sec:3nforce}

\subsection{3NF models}
\label{sec:theory}

As discussed in the introduction, it is by now clear that one needs additional 
interactions beyond pair forces to describe nuclei and nuclear reactions 
accurately. Such interactions are many-body forces and it is believed that 
the most important many-body interactions are 3NFs. 
In this section, we briefly summarize the current status of 3NF 
models with a special emphasis on models which were used to study few-nucleon 
systems. For a more complete overview, the reader is referred to \cite{Robilotta:2008ew}. 

\begin{figure}[t]
\begin{center}
\includegraphics[width=\textwidth]{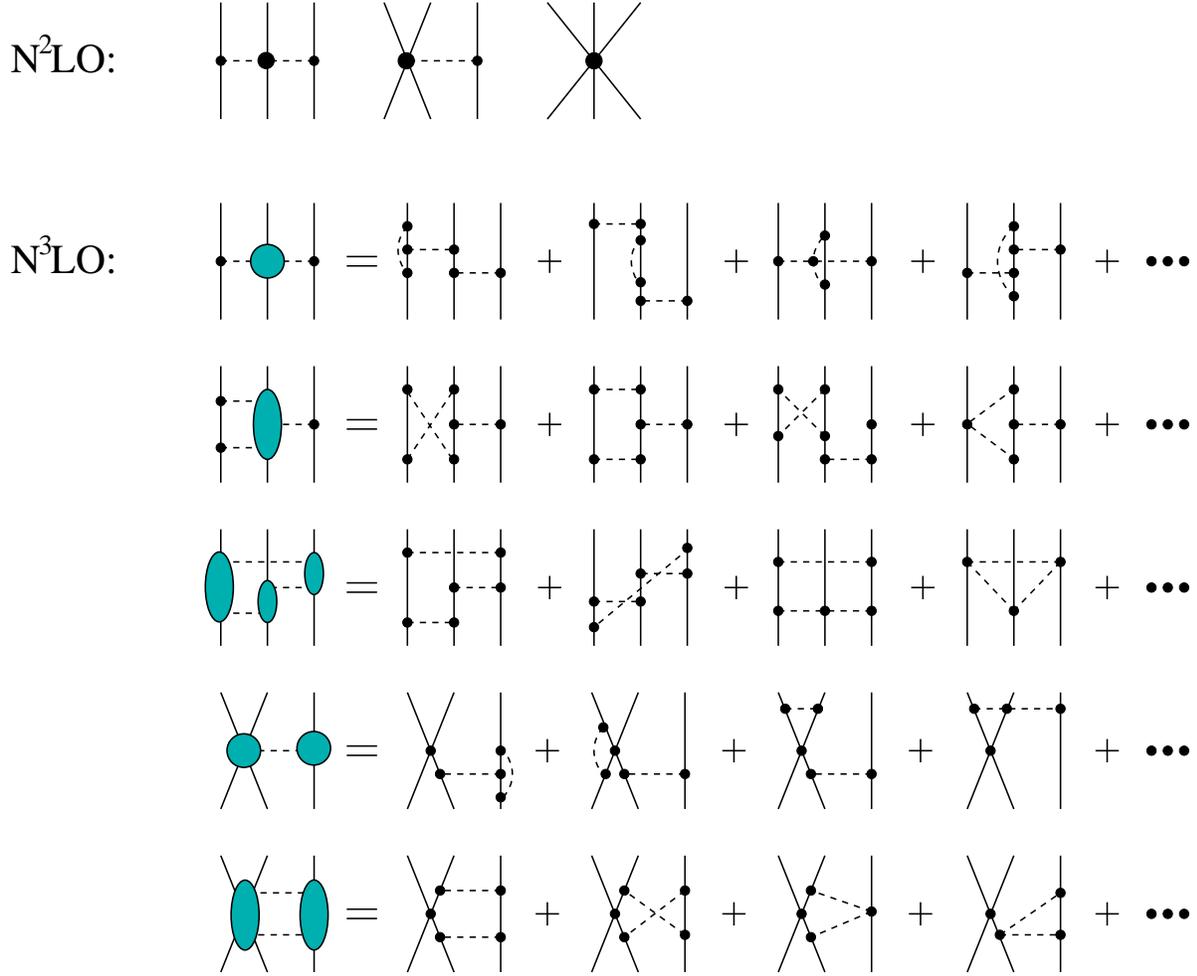}
\end{center}
\caption{\label{fig:3nffeyn}Different topologies contributing to the leading and subleading 
3NF in chiral EFT.}
\end{figure}

Qualitatively, the importance of many-body forces for nuclei has already been realized
in the early days of nuclear physics \cite{Primakoff:1939zz}. In the 1950s, the pion field theory was 
extensively used to derive nuclear forces. In this era, there were attempts to derive 
3NFs on the same footing as NN interactions \cite{Klein:1953zza}. However, it turned out to be 
impossible 
to obtain a quantitative description of nuclear interactions in this framework due to the lack 
of a systematic expansion parameter, so that 
the community finally turned to a combination of theoretical insights and phenomenology to develop 
nuclear interactions culminating in today's phenomenological and accurate 
NN interaction models. With this development, NN and 3N forces were not derived anymore 
on the same grounds. Instead it was tried to make use of dispersion relations to link 
the interactions to $\pi$N scattering. The most prominent early 3NF obtained in this 
way is the famous Fujita-Miyazawa force \cite{Fujita:1957zz} that is still at the heart 
of phenomenological 3NFs. It was assumed that the most important contribution 
to the 3N interaction is given by the left-most topology in Fig.~\ref{fig:3nffeyn}.  The blob encodes 
the $\pi$N scattering amplitude. For the 3NF, this amplitude is required below threshold so that
dispersion integrals were needed to determine the 3NF quantitatively.  
In this way, it is found that by far the most important contribution emerges from $p$-waves and 
can be linked to $\Delta$ isobars in intermediate states. 

The development of today's most widely used 3NF models began in the 1970s 
and early 1980s based on the work by Fujita and Miyazawa \cite{Fujita:1957zz} or 
the later work using current algebra by Brown and Green \cite{Brown:1969npa} 
or McKellar and Rajaraman \cite{McKellar:1968zza}. The Tucson-Melbourne 
collaboration studied the latter approach in more detail \cite{Coon:1974vc,Coon:1978gr} using 
current algebra relations to constrain the $\pi$N scattering amplitude. This lead to the 
widely used Tucson-Melbourne (TM) 3NF. Originally, the interaction contained one more 
term than the Fujita-Miyazawa force (see the discussion below).

Around the same 
time, the Urbana group found that for a decent description of saturation of nuclear matter, 
a short ranged repulsive interaction is required. They simplified the $2\pi$ exchange 
ansatz of \cite{Fujita:1957zz} to include only the $\Delta$ part but added a purely phenomenological 
short-range term. The force was then adjusted to reproduce the triton binding energy 
and the nuclear matter saturation density in conjunction with a specific NN interaction 
\cite{Carlson:1983kq}. This lead to a series of 3NFs called Urbana. The most up-to-date version 
adjusted in conjunction with AV18 is called Urbana-IX \cite{Pudliner:1995wk}. 

In parallel, Robilotta and others developed a 3NF model based on a $\pi$N amplitude from 
chiral Lagrangians resulting in 3NFs very similar to the one of the Tucson-Melbourne 
group \cite{Coelho:1984hk}  although a specific term of the orginal 
TM force was missing. Contrary to their expectations, they found that the 2$\pi$-exchange part 
of the interaction depends strongly
on the cutoff parameter employed 
in the models \cite{Coelho:1984hk,Robilotta:1986nv} which resulted in a very strong dependence of, e.g.~the 
triton binding energy  on the chosen cutoff parameter. Their analysis showed that this feature is related 
to contact interaction terms which were part of the 2$\pi$-exchange ansatz, and they argued 
that these parts should be removed \cite{Robilotta:1986nv}  which resulted in a further difference to TM. 

In Ref.~\cite{Friar:1998zt}, this issue was re-examined in the framework of chiral EFT. In this 
framework, there are terms beyond the 2$\pi$ exchange (see below). But it turns out that 
the 2$\pi$ exchange contribution has a similar structure as the 3NF of Ref.~\cite{Coelho:1984hk}. 
Specifically, it misses the additional term in the TM interaction. Ref.~\cite{Friar:1998zt} recommends 
to remove the additional term from the TM force. This finally lead to a new version \cite{Coon:2001pv} of the 
Tucson-Melbourne force commonly called TM'. We will present predictions of this model later on in this review.
Note that TM' (as the chiral 3NF) keeps the short range part of the 2$\pi$-exchange and that, for TM',  
the dependence on the cutoff is used to adjust the 3NF in conjunction with different NN models to the triton binding energy \cite{Nogga:1997mr}.

In the framework of EFT, the cutoff dependence is removed by the additional topologies of 
Fig.~\ref{fig:3nffeyn}, which enable one to absorb the dependence on the short range part of the 3NF 
in additional contact terms. 

As a further improvement of the TM model, shorter-ranged contributions were considered 
in Refs.~\cite{Ellis:1984jh,Coon:1995xz}. Their effect on the $^3$H binding energy 
was studied in \cite{Coon:1984zz,Stadler:1995wu,Adam:2003jk} showing that they 
reduce the binding energy. Unfortunately, their effect on intermediate energy $nd$ scattering 
has not been systematically studied yet. Recently, however, there is new interest in such 
interactions \cite{Skibinski:2011db} and their effect on intermediate energy 
observables will be studied in the near future. 

The most important contribution to the 3NFs discussed so far can be linked to intermediate excitations 
of the nucleons to $\Delta$ isobars. This motivated the Hanover-Lisbon group to study few-nucleon 
systems in a coupled-channel framework that allows for explicit $\Delta$ excitations 
(see section~\ref{sec:nnforce}). 
Then, few-nucleon calculations automatically contain terms related to intermediate $\Delta$ 
excitations. These terms are, however, not completely equivalent to corresponding 3NFs 
since the $\Delta$s are not treated in a static approximation and also mesons other than 
pions are taken into account. The short-range part generated by the heavy-meson exchanges 
is probably, to some extent, taken into account by the adjustment of cutoff parameters of the 3NFs 
discussed previously. But at least the dispersive corrections due to non-static $\Delta$s 
do not have a counter part in standard 3NFs models, and
they are known to be non-negligible for the triton binding energy 
\cite{Deltuva:2003fn} and N$d$ scattering \cite{Deltuva:2003zp}. The coupled-channel 
approach has the obvious advantage that 
all parameters are fixed by two-body observables and the properties of few-nucleon systems are predicted. 
On the other hand, there is no systematic way to incorporate  other effects  
into the 3NFs. Therefore, once the model fails to reproduce some data, 
systematic improvements are difficult. 
This is already the case for the triton binding energy which is slightly underpredicted in this approach.

\begin{figure}[t]
\centering 
\includegraphics[angle=0, width=.9\textwidth] {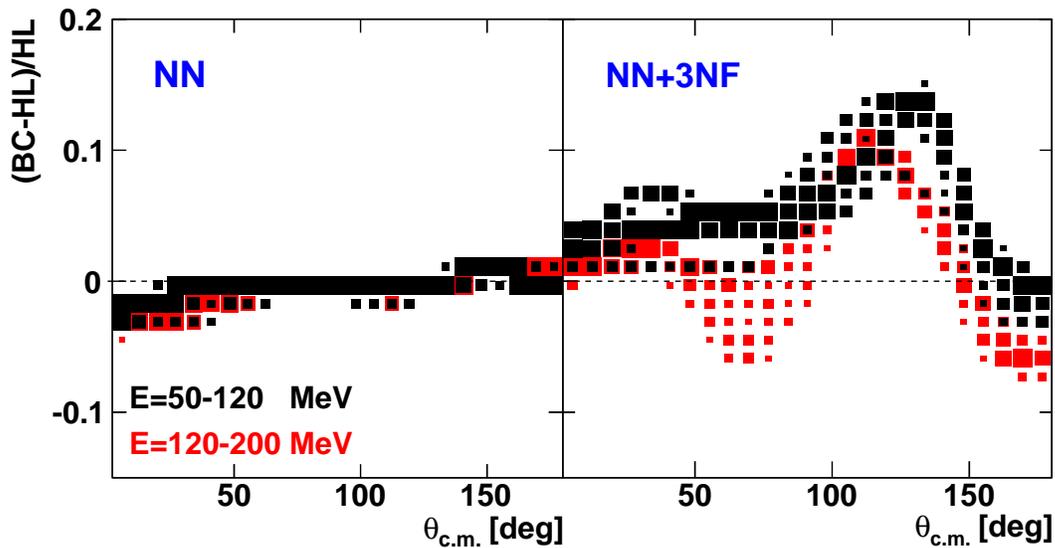} 
\caption{The relative difference between the predictions of the Bochum-Cracow (BC) and Hanover-Lisbon (HL) groups for proton-deuteron elastic-scattering cross sections as a function of $\theta_{\rm c.m.}$ for a number of incident nucleon energies between 50 to 200~MeV. In the left panel, the differences between the models are shown for the case when only the two-nucleon CD-Bonn potential has been used. For the right panel, the three-nucleon force effects have also been included in both calculations (TM' for BC and the $\Delta$ in the case of HL). The results are grouped in two bands, one for 50-120 MeV with black squares and the other for 120-200 MeV with grey/red squares (color online). In all calculations, Coulomb effects have been discarded. 
 }
\label{3n-comp} 
\end{figure}

The calculations carried out by the Hanover-Lisbon group 
also include the Coulomb interaction making a direct 
comparison to $pd$ scattering data possible \cite{Deltuva:2005cc,Deltuva:2005wx}. Recent studies 
showed that the Coulomb 
interaction becomes important specifically in elastic forward scattering and in break-up configurations 
where the two protons 
have a small relative momentum~\cite{Kievsky:2001fq,Deltuva:2005xa,Deltuva:2009qi}. 
Full calculations based on a purely nucleonic NN potential and standard 3NF 
models only recently became available~\cite{Witala:2009ws,Witala:2009zz,Deltuva:2009qi}.
The effect of the Coulomb interaction using the AV18 NN potential and the 
Urbana IX 3NF was shown~\cite{Deltuva:2009qi} to give similar results compared to that of
the calculations using the intermediate $\Delta$ approach. 
For a large part of the available experimental data, the Coulomb effect is sizeable with respect to the 3NF 
effects. From this perspective, it is favourable to use the 
calculations
by the Hanover-Lisbon group
as a benchmark for the global analysis of the world database.
On top of this, 3NF effects can be isolated using the interaction model with or without 
an explicit $\Delta$ resonance~\cite{Deltuva:2003wm}. 
Hence, 
a systematic comparison between data and predictions by 
the Hanover-Lisbon group 
provides 
us with a way to systematically study the role of 3NFs without the ambiguity of the role
of the Coulomb effect.
It turns out that in many cases the 3NF effect due to explicit 
$\Delta$s is very similar to the effects of the standard 3NF models. 
To quantify this statement, a comparison has been made for proton-deuteron elastic-scattering cross sections 
calculated for an energy range between 50 to 250 MeV. This comparison is presented in Fig.~\ref{3n-comp} where 
the difference is shown between the results of the calculations by Bochum-Cracow group using 
the CD-Bonn NN force 
and the TM' 3NF and those by the Hanover-Lisbon group 
using the CD-Bonn NN potential but now with the explicit inclusion of the $\Delta$ 
(both without taking the Coulomb effects into account). In the left panel, the relative difference 
in cross-section predictions of both approaches using only the two-nucleon CD-Bonn potential are shown where 
one observes very small differences as one would naively expect. In the right panel, the relative differences 
are again plotted including 3NF effects in both models. One can see that, depending on the 
scattering angle, differences of up to 15\% emerge between the two model calculations. 
The above mentioned model dependence should, therefore, be taken into account when making definitive 
conclusions. On the other hand, the standard calculations can be 
performed for many different combinations of NN and 3N force. This will allow to 
quantify the model dependence and to pin down observables that are sensitive to the 
spin-isospin structure of 3NFs. Therefore, the combination of both theoretical approaches will be required to analyze the existing data with high precision. 

As mentioned above, with the invention of phenomenological potential models of the 
NN interaction, the development of 3NFs became separated from the NN interactions.
While one part of the community was discussing the identification of observables 
that allow one 
to determine the ``off-shell'' part of NN interactions to develop a ``proper'' NN force \cite{Jetter:1994zz,Eden:1995rf,Martinus:1998zz,Nakayama:2009yz}
the other part was studying the properties of the different 3NF models realizing that the 
short-range 
part of the 2$\pi$-exchange interactions is strongly dependent on details of the 
3NF, like cutoffs \cite{Robilotta:1986nv}.  It then became clear that off-shell effects are 
not measurable \cite{Fearing:1999fw} and the NN and 3N interactions cannot 
be studied independently of each other \cite{Polyzou:1990isi,Amghar:1995av}.
Different choices for the NN force can be made equivalent by adding proper 3NFs. 
Conversley, there is no unique 3NF as (parts of the) 3NF can be traded for off-shell 
NN interactions, see Refs.~\cite{Stadler:1996ut,Gross:2008ps} for an explicit example. 
Within the model approach, there were attempts to derive NN and 3N interactions 
on the same grounds \cite{Eden:1996ey} but, with the invent of 
phenomenological potentials, it became a common practice 
to adjust the 3N interaction models (using e.g. a cutoff parameter) in conjunction with one of 
the phenomenological NN models to reproduce at least the $^3$H binding energy (like it always has been done
for the Urbana series of models) \cite{Nogga:1997mr}.
This approach had a phenomenological success. It turns out that many low-energy 
observables scale with the triton binding energy in the sense that all models adjusted 
in this way will predict the observables equally well\footnote{\label{efimovfoot}It turns out that such correlations are a universal consequence of the large nucleon-nucleon scattering length and also show up in atomic and molecular systems. This can be understood based on pion-less EFT \cite{Hammer:2010kp,Braaten:2004rn,Efimov:1985plb,Platter:2004zs}.} .  

This adjustment of the strength of the 3NF in conjuction with 2N force has 
been a major step 
towards a quantitative description of 
few-nucleon systems. The adjusted 3NFs and the different NN interactions gave a series of nuclear Hamiltonians 
that describe the low energy 3N scattering observables very well. 
The few exceptions became known as puzzles (see the discussion of $A_{y}$ and the space 
star anomaly below). However, the description of the binding energies of $p$-shell nuclei and 
of intermediate-energy scattering observables is worse and still dependent on the 
chosen model clearly showing that the 3N Hamiltonian is not fully understood. 
Based on the models, it became possible to identify observables that are probably 
most useful to study the structure, spin and isospin dependence of 3N forces \cite{Witala:2001by}
which triggered a series of experiments that increased the database of three-nucleon 
scattering observables tremendously (see discussion in section~\ref{sec:survey_data_3n}). We will discuss 
the data in comparison to model calculations below. 

Obviously, there is also a major interest to understand the binding and excitation energies of light nuclei based on 
NN and 3N interactions. Since these are very sensitive to small contributions to the 
Hamiltonian, the calculations also showed failures of the models \cite{Pieper:2001mp,Navratil:2003ef}. For 
the AV18 and Urbana-IX potentials, the failures were traced back to the isospin dependence of the force. Therefore, 
the Argonne-Los Alamos-Urbana collaboration added a new structure motivated by 3$\pi$-exchanges 
with intermediate $\Delta$ isobars to their model culminating in the series of Illinois interactions 
\cite{Pieper:2001ap}.  For light nuclei, the model improved the description of the data, 
but e.g.~going to neutron matter, today's realizations lead to predictions that are not in line 
with masses and radii of observed neutron stars \cite{Gandolfi:2010za}.

Besides all the previously described developments on 3NFs, 
relativistic corrections to the generally used non-relativistic framework might also contribute 
to the discrepancy between the theory and data. 
In the 3N systems, relativistic effects have thoroughly been studied for the 
bound state \cite{Kamada:2010efb,Kamada:2002qt}, 
elastic \cite{Witala:2004pv} and break-up \cite{Witala:2005nw} $nd$ scattering. 
The effects on the triton binding energy are somewhat dependent on the scheme 
with which the non-relativisitic NN interaction was matched to the relativistic one. Generally, the effect 
is repulsive such that even larger 3NF effects can be expected. Also the dependence on the 
NN model is not reduced. Therefore, clearly, 3NFs will be required to resolve the underbinding 
problem for nuclear bound states. For scattering,  relativistic effects were usually found to be small
even at intermediate energies. Only for some specific break-up configurations, the effects were sizable.
We note that $A_y$ at low energy is somewhat affected by 
relativistic effects  \cite{Miller:2007ad}. A complete calculation  however 
showed that they are far too small to resolve the well-known puzzle and even  
worsen the agreement with the data \cite{Witala:2008va}. 
Based on these results, we conclude that the inclusion of the relativistic corrections 
along these lines does not allow  to remove 
discrepancies between predictions and data.  
On the other hand, the calculations by Gross and Stadler carried out within the covariant 
spectator theory \cite{Stadler:1996ut,Gross:2008ps} demonstrate that the correct triton 
binding energy can be obtained 
from a relativistic kernel consisting only from one-boson exchange terms, provided one makes 
a specific choice for the off-shell behavior of the boson-nucleon couplings. 
The results of these studies 
provide yet another explicit evidence of the non-uniqueness of the nuclear Hamiltonian  and 
its separation into two- and three-nucleon operators. It is, however, not clear to what extent the 
findings of \cite{Stadler:1996ut,Gross:2008ps} emerge due to the relativistic treatment of the 
nuclear dynamics. 

In summary, we have today a set of nuclear interaction models based on NN and 3N forces 
that describe nuclear systems fairly well, but still show failures. It is conceivable that these 
failures are due to missing structures in the 3N forces. Improvement of the models 
is much more difficult than in the case of NN interactions since the number of possible 
momentum, spin and isospin structures is much higher. Whereas it was possible to use 
the most general operator structure of the NN interaction allowed by the symmetries as the starting point 
for the models, this is not feasible for the 3N interactions. We, therefore, need more insight 
to pin down the most important structure of 3N interactions. The models 
discussed so far are not systematically improvable and are, therefore, limited in providing
detailed insight in 3NFs. 

The most promising approach to improve nuclear interactions is presently chiral EFT
as discussed in section~\ref{sec:nnforce}. At this point, mostly the version without 
explicit $\Delta$ degrees of freedom has been used to formulate consistent 
NN and 3N (and even many-body) interactions. 
In the  formulation based on the pions and nucleons as the only explicit degrees of freedom, 
the first non-vanishing 3NF emerges at N$^2$LO in the chiral expansion (as long as one 
uses energy-independent formulations) and consists of only 
three different topologies \cite{vanKolck:1994yi,Epelbaum:2002vt}, see Fig.~\ref{fig:3nffeyn}. 
One of them is identical 
to the 2$\pi$ exchange which is the base of all phenomenological 3NF models. Its strength, however, 
is related to the strength of the corresponding terms in the NN force \cite{Epelbaum:2005pn}. 
This underlines the need of consistency in NN and 3N interactions. Two additional topologies 
are of a shorter range. The strength of the 1$\pi$-exchange contribution (Fig.~\ref{fig:3nffeyn}, middle diagram)
is related to various other processes such as e.g.~pion production in two-nucleon collisions 
\cite{Hanhart:2000gp} or $^3$H beta decay \cite{Gardestig:2006hj}.  The weak decay has already been used in \cite{Gazit:2008ma} to constrain the 3NF. 
In the available few-nucleon studies, the two low-energy constants related to the shorter-range
topologies (middle and right diagrams in Fig.~\ref{fig:3nffeyn}) are usually determined 
from few-nucleon observables. The  $^3$H binding energy is generally used as one constraint on 
these parameters. As a second constraint, often the $^4$He radius or binding energy or the $nd$ doublet 
scattering length are used. Although these observables are correlated with the $^3$H binding energy, 
it turns out that an accurate description with phenomenological interactions is only possible once 
the short-ranged 3NF terms are included in the calculations (see e.g. \cite{Kievsky:2010zr}).  

From the two-nucleon system, we know already that N$^2$LO only allows for a quantitative description of 
low-energy data. Only 
at N$^3$LO, the two-nucleon system is described  equally well by chiral interactions as by 
the phenomenological potential models up to intermediate energies. This carries over to three-nucleon observables. The predictions for these observables become rather strongly dependent on the cutoff at N$^2$LO making them useful 
for a quantitative comparison with the data only at energies below 100~MeV.  
Therefore, it is of utmost importance to extend now also the 3NF to N$^3$LO. 
In part, this has been achieved in Refs.~\cite{Ishikawa:2007zz,Bernard:2007sp} for the long-range 
part of the interaction. The remaining 
parts have also been worked out and are to be published \cite{Bernard:2011tmp}. 
At this order, there are several new structures, and it will
be interesting to study their impact on few-nucleon observables in future. Here, the power of chiral EFT 
becomes apparent. Not only that the power counting helped to identify the more important 
structures of the 3NF, but also the strength of the individual terms is related to other parts of the 
Hamiltonian. This implies that the number of parameters of the 3NF that need to be fit to 
few-nucleon data is the same in order N$^2$LO and N$^3$LO. For a recent work on the short-range 
terms beyond N$^3$LO see Ref.~\cite{Girlanda:2011fh}.

\subsection{Experimental techniques for measuring the observables}

In the past decades, detection systems suitable to 
study certain aspects of the dynamics of the three-body systems have been developed at various laboratories. 
The exact form and characteristics of a detection system clearly depend
on the reaction and the observable one is studying. For example, in the study 
of the cross section of elastic {\it pd} scattering, one needs an unpolarized source of protons or 
deuterons and a small detector which is capable of identifying one of the outgoing particles with a high energy/angle 
resolution. For this reaction, the best detector would then be a magnetic spectrometer with a small solid angle. 
The measurement of the analyzing power of the reaction would require the use of polarized sources 
of particles. To measure more complicated observables such as spin-correlation coefficients one would 
need a polarized target as well, and the investigation of spin-transfer coefficients would require, instead, the 
measurement of the polarization of the outgoing particles. Each one of these reactions will have its complications 
although most techniques are now well under control. In this section, some experimental aspects relevant for this field will be discussed. 

The use of polarized beams has been very common in the past two decades. There are three types of polarized-ion
sources: optically-pumped source, Lamb-shift source, and the atomic-beam source~\cite{Clegg:1967nim,Bechtold:1978nim,Wise:1993nim,Clegg:1995nimb,Friedrich:1995bk,Hatanaka:1997nim}. At the facilities where few-nucleon
systems are studied at intermediate energies, most sources are based on the latter type. 
In all these sources, polarizations of 60-80\% of the maximum theoretical value have been achieved. 
For a review of these sources, see~\cite{Clegg:2001as}. 
The Lamb-shift principle has also been used in the measurement of the degree of polarization of protons and deuterons \cite{Kremers:2004nim,Beijers:2006aip,Engels:2003rsi}. 

The (polarized) beams of protons and deuterons are accelerated by different techniques depending on the energy required. 
The low-energy measurements (below 25~MeV) were generally performed using a (Tandem) Van de Graaff accelerator. 
At higher energies, one uses cyclotrons to accelerate particles up to a few hundred MeV. 
At IUCF, Uppsala and J{\"u}lich, particles are also accelerated in rings \cite{Meyer:2007zzb}. Since the beam has
gone through a secondary stage of acceleration to achieve higher energies, one should 
perform polarimetry of the beam in order to 
determine the degree of the polarization of the beam. This is done using a polarimeter which is based on measuring the azimuthal asymmetries of scattered particles and knowing the analyzing powers of a well-known reaction such as proton-proton, proton-deuteron, and proton-Carbon elastic scattering. At lower energies, elastic $p$-$^3$He as well as $^3$He$(d,p)^4$He have also  been used. For a description of the principle of operation of these polarimeters, the reader is referred to \cite{Bieber:2001nim,Wells:1993nim} for intermediate energy polarimeters (between 100 to 200 MeV outgoing particle energies) and \cite{Sydow:1993nim} for the low-energy ones (below 30~MeV). 

In addition to protons and deuterons, neutron beams have also extensively been used  at TUNL, Bonn, Erlangen, PSI, Uppsala, LANSCE and RCNP \cite{Tornow:1996zz,Tornow:1998nf,Setze:2005jk,vonWitsch:1976nim,Koble:1989nim,Hohlweck:1989nim,Schoberl:1988npa,Haffter:1992npa,Klug:2002nim,Gericke:2005ef,Maeda:2007zza}. 
The obvious disadvantage of the neutron beam is that it cannot be manipulated in the beam lines 
so that experiments are more difficult. The beam intensities are also generally lower than proton or deuteron beams since neutrons are produced as a secondary beam of particles. In contrast to
charged-particle detectors where efficiencies close to 100\% can be reached, the efficiency of the neutron detectors is generally lower and requires difficult calibration \cite{GonzlezTrotter:2009nim}.
All this is not particularly a problem in these reactions as most of them involve only hadronic vertices producing high counting rates and the challenges have been tackled since Coulomb distortions are absent when one uses neutron beams.

The (polarized) beams of particles impinge on different targets of interest. For the three-body studies, proton and deuteron beams are combined with deuteron and proton targets, respectively. In the past, one used a thin foil of solid CH$_2$ or CD$_2$. These targets are easy to manufacture with a high accuracy in the target thickness. The main disadvantage of these targets is, however, the background originating from the carbon content of the target. Measurements had to be done on pure carbon targets to measure independently the contribution of, in general, unwanted background. With the advent of very thin (down to a few $\mu$m) and strong synthetic foils, liquid targets became more popular and in particular at intermediate energies where the straggling through target posed less of a problem \cite{KalantarNayestanaki:1998nim,AbdelSamad:2006xm}. These targets have the advantage of less background from the target window but their thickness has to be determined every time the target is warmed up as the bulging of the foils, and therefore the target thickness, varies by a few percent. Finally, for the ring experiments, one has to use very thin targets in the form of gas or microparticles \cite{Sperisen:1989nim,Berdoz:1989nim}. The very low target density is then compensated by the fact that the beam revolution frequency in the ring is of the order of 1~MHz increasing the effective luminosity by a factor of $10^6$. 

\begin{figure}[t]
\centering 
\includegraphics[angle=0, width=.8\textwidth] {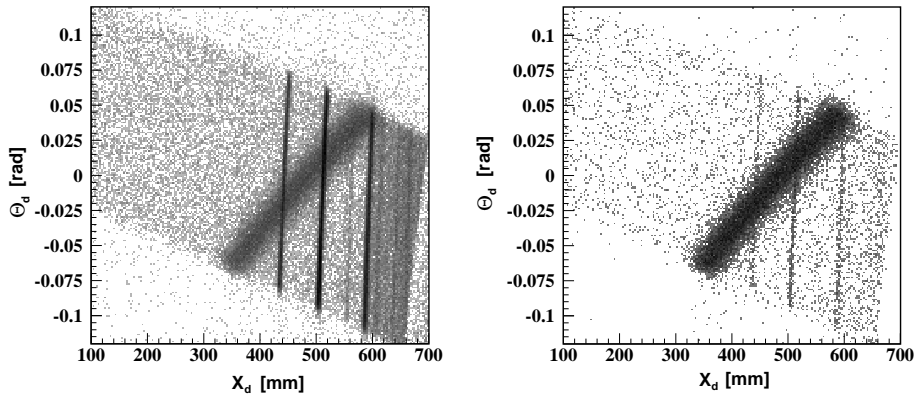} 
\caption{A spectrum of deuterons emerging at a laboratory angle of 41$^\circ$ from proton-deuteron scattering on a solid CD$_2$ target with a proton beam of 150 MeV. One can clearly see the peak of interest on a smooth background and a few excited states from the carbon content. The reason for the skewedness of the deuteron peak is the light mass of the particle entering the magnetic spectrometer (kinematical broadening). On the left, the spectrum is measured with deuterons entering the magnetic spectrometer and on the right, a coincidence has been made with a small scintillator detector placed at the appropiate angle for elastic proton-deuteron scattering. }
\label{spectrometer} 
\end{figure}

As mentioned above, the design of the detection system depends on the reaction, the configuration of interest and the energy of the particles to be detected. Magnetic spectrometers with small solid angles but a high energy and angle resolution have been used in the study of the elastic scattering processes. As an example, a spectrum is shown in Fig.~\ref{spectrometer} for 
proton-deuteron scattering from a solid CD$_2$ target. Despite the fact that a coincidence measurement has reduced the background drastically, the picture shows clearly the need to have a separate measurement on carbon to understand and subtract the background to obtain high accuracies. Most results discussed in this review article for the elastic-scattering channel have been obtained with the use of spectrometers for the detection of protons and deuterons. 

To measure more complicated spin observables such as spin-transfer coefficients or spin-correlation coefficients, one needs to measure the 
polarization of the outgoing particle in the former case or have a polarized target in combination with a polarized incident beam for the latter case. 
Successful attempts have been made in the past to measure both observables. At low energies, this has been done at Cologne and Bonn with 
the help of a polarimeter \cite{Sydow:1993nim,Hempen:1998zz}. At intermediate energies, several polarimeters have been designed and 
operated \cite{Woertche:2000qa,AmirAhmadi:2006nim,Noji:2007nim,Sakai:1996nim,Cameron:1991nim}. Since these measurements require a secondary scattering, 
both the statistical and systematic uncertainties will be larger than those of analyzing powers for instance. Due to kinematical constraints 
(low energy of the outgoing particles), the angular ranges of the measurements are also generally more limited than in the case of other observables. 
The measurements of spin-correlations coefficients have been done only at IUCF with the help of the polarized target 
used in the ring \cite{Meyer:2007zzb}. An overview of the IUCF-PINTEX setup can be found in Ref.~\cite{Rinckel:2000wd} describing
extensively the polarized beams and targets for all their double spin experiments. Polarized gas targets have also been developed elsewhere for various purposes (see Refs.~\cite{Steffens:2003rpp,Rathmann:2005nim} for a general discussion on this). Solid polarized targets from Ammonia have also been produced successfully in the past \cite{Meyer:2004nim} but have never been used in the realm of few-body physics. 

\begin{figure}[t]
\centering 
\includegraphics[angle=0, width=.8\textwidth] {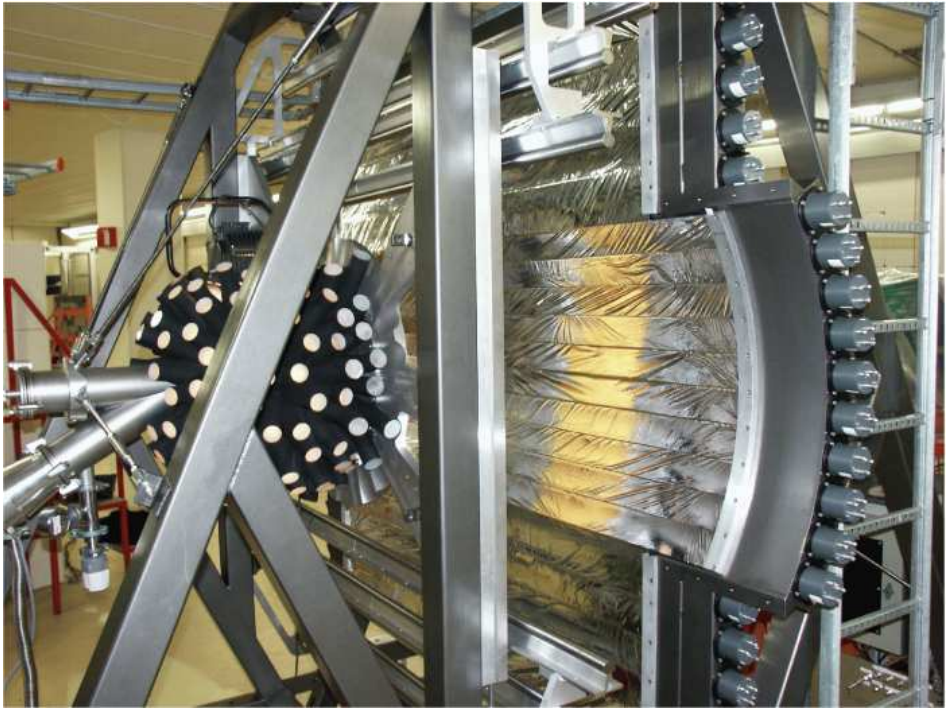} 
\caption{The $4\pi$ detection system, BINA (Big Instrument for Nuclear-polarization Analysis) at KVI. For presentation purposes, the phototubes in the backward ball and all the cables have been removed. Also the wire chamber placed to measure the particles moving to forward angles of less than $45^\circ$ is not shown (color online).}
\label{BINA} 
\end{figure}

\begin{figure}[t]
\centering 
\includegraphics[angle=0, width=.5\textwidth] {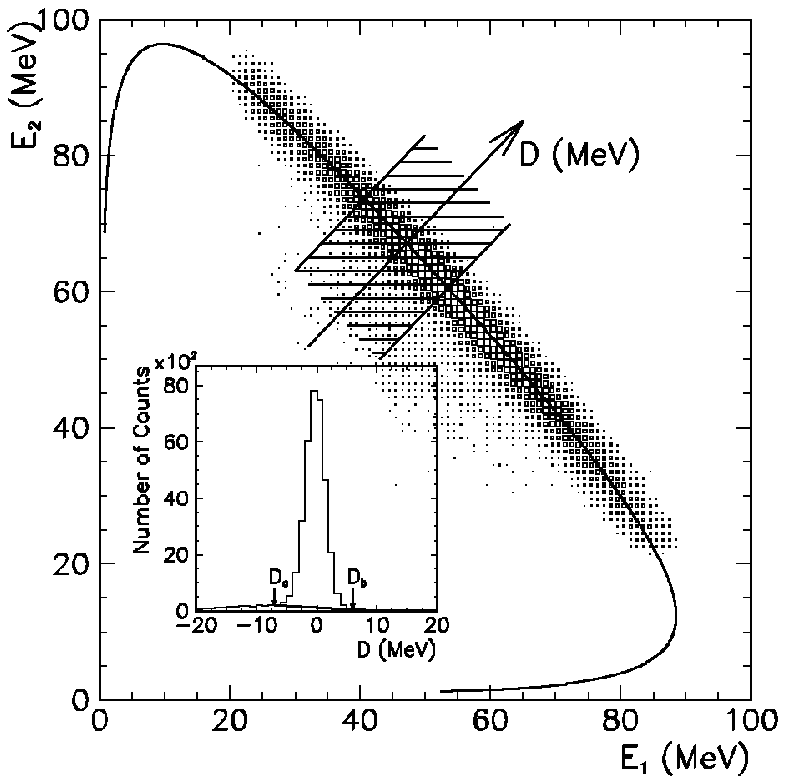} 
\caption{The measured energy correlation between the two outgoing protons in the break-up 
reaction from proton-deuteron scattering for a certain combination of proton angles 
at a beam energy of 65~MeV/nucleon. The kinematical locus, the so-called $S$-curve is 
drawn for this combination and the data follow the expected pattern with little background. 
A typical $S$-bin is indicated together with the way the data are projected on the axis 
perpendicular to this axis, D-axis. In the inset, the projected data are shown with little background. 
Reprinted with permission from~\cite{Kistryn:2003xs}. Copyright (2003) by the American Physical Society. }
\label{scurve} 
\end{figure}

In the nucleon-deuteron break-up studies, there will be three particles in the final state and, hence,
the detection system becomes more complicated. In this case, one could
either choose a very specific geometry in the measurement or employ a 
detector with a very large acceptance to accommodate for the detection 
of more than one particle in a large part of the reaction phase space. All the low-energy
measurements performed at Cologne \cite{Schieck:2001fbs}, Bonn \cite{Huhn:2001yk}, Durham (TUNL) 
\cite{Tornow:1996zz,Tornow:1998nf,Setze:2005jk} and Kyushu~\cite{Sagara:1994zz} and also the first studies at
an energy of 65~MeV/nucleon~\cite{Allet:1994zz,Allet:1996zz,Zejma:1997zz,Bodek:2001fb} have opted for
the first choice and looked at the break-up reaction for selected kinematics such as FSI, QSF, SST, SCRE etc.. Based on the
experience gained from these measurements and earlier work at KVI \cite{KalantarNayestanaki:2000nim,Kistryn:2003xs}, 
a $4\pi$ detection system was designed and built recently. This detection system, which is in operation since 2005, is shown in Fig.~\ref{BINA} and is 
capable of detecting light ions (protons, deuterons etc.), and partly neutrons, down to very low energies 
(a few MeVs in the backward part of the detector). It also acts as a polarimeter for the beam as it has a 
$\phi$ coverage of $360^\circ$. The break-up data discussed in this review have been obtained mostly 
using  detectors similar to the one discussed in Refs.~\cite{KalantarNayestanaki:2000nim,Rinckel:2000nim} or shown in 
Fig.~\ref{BINA}. Figure~\ref{scurve} shows a typical 
measurement performed in the break-up experiments in which the energy of the first outgoing particle
(proton) is plotted against that of the second one (proton). Due to energy and momentum conservation,
the detection of two particles in the final state already overdetermines the kinematics, thereby giving an 
extra handle to reduce the already low background of these measurements. The cross sections are
generally studied as a function of the arc-length $S$ for a given bin in $S$ as shown by the 
two lines in Fig.~\ref{scurve} and for a fixed angular combination of the two outgoing protons.  
Alternative ways of presenting the data also exist in the literature \cite{KurosZolnierczuk:2004xt}. 

\subsection{Survey of the experimental database for $A=3$ systems}
\label{sec:survey_data_3n}

In the study of three-nucleon force effects, many laboratories have produced data in the past three 
decades. Most of the data at low energies (up to 30~MeV), have been produced at TUNL with 
(polarized) neutrons, protons and deuterons \cite{Tornow:1996zz,Tornow:1998nf,Setze:2005jk}, 
Cologne \cite{Grossmann:1996npa,Rauprich:1991npa,Karus:1985zz,Patberg:1996prc,Przyborowski:1999sp,Duweke:2004xv,Ley:2006hu,Sydow:1998fbs,Sydow:1994npa,Klein:1973npa}, Bonn \cite{Hempen:1998zz}, Madison \cite{Kocher:1969npa}, and Kyushu with protons and deuterons \cite{Sagara:1994zz}. Measurements have also taken 
place with the injector of the cyclotron at RCNP at a deuteron energy of 26~MeV. The observables include cross sections, analyzing powers and spin-transfer coefficients. The focus of this paper is on the results at intermediate energies, from 50 MeV/nucleon to energies around the pion-production threshold. 
Various observables in elastic and break-up reactions, such as differential cross sections, 
analyzing powers, spin correlations, and spin-transfer coefficients, have been measured at 
these energies exploiting (polarized) beams of protons and deuterons, partly in combination with 
polarized targets.
The beams of (polarized) protons and deuterons have been delivered by cyclotrons, such as those at 
KVI (65-190~MeV/nucleon) \cite{Bieber:2000zz,Ermisch:2001zz,Ermisch:2003zq,Kistryn:2003xs,Ermisch:2005kf,Kistryn:2006ww,AmirAhmadi:2007tu,Stephan:2007zza,Mardanpour:2007epj,RamazaniMoghaddamArani:2008ww,Stephan:2009isi,Eslamikalantari:2009mp,Messchendorp:2009mpl,Mardanpour:2010zz,Stephan:2010zzb}, RIKEN (70-140~MeV/nucleon) \cite{Sakai:2000mm,Sekiguchi:2002sf,Sekiguchi:2004yb,Sekiguchi:2005vq}, RCNP (250~MeV/nucleon) \cite{Maeda:2007zza}, PSI (formerly known as SIN) (65~MeV/nucleon) \cite{Kistryn:1992npa,Balewski:1995npa,Allet:1994zz,Allet:1996zz,Zejma:1997zz,Bodek:2001fb}, Harvard (146~MeV/nucleon) \cite{Postma:1961pr}, and IUCF (70-200~MeV/nucleon) \cite{Stephenson:1999ex}, synchrocyclotrons such as those at Berkeley (145-220~MeV/nucleon \cite{Igo:1972npa}, Orsay (155~MeV/nucleon) \cite{Kuroda:1966np}, and Rochester (198~MeV/nucleon) \cite{Adelberger:1972pu}, synchrotrons such as that at SATURNE (95-200~MeV/nucleon for intermediate energies) \cite{Garcon:1986npa}, or accelerated in a ring such as the one at IUCF (135 and 200~MeV/nucleon) \cite{vonPrzewoski:2003ig,Meyer:2004prl}. Experiments at much higher energies of up to 1.3 GeV were also performed at SATURNE. 
In addition, neutron beams have been used at PSI (up to 65~MeV) \cite{Haffter:1992npa}, Uppsala (at 95~MeV) \cite{Mermod:2005dn} and at LANSCe (140-240~MeV) \cite{Chtangeev:2005mth} to study the three-body system with the advantage of having no Coulomb effect which, otherwise has to be accounted for in the calculations. The obvious disadvantage of the neutron beams is the quality and the intensity of the beams. Elastic scattering, the break-up reaction and radiative capture have been studied with these beams. 

\begin{figure}[t]
\centering 
\includegraphics[angle=0, width=\textwidth] {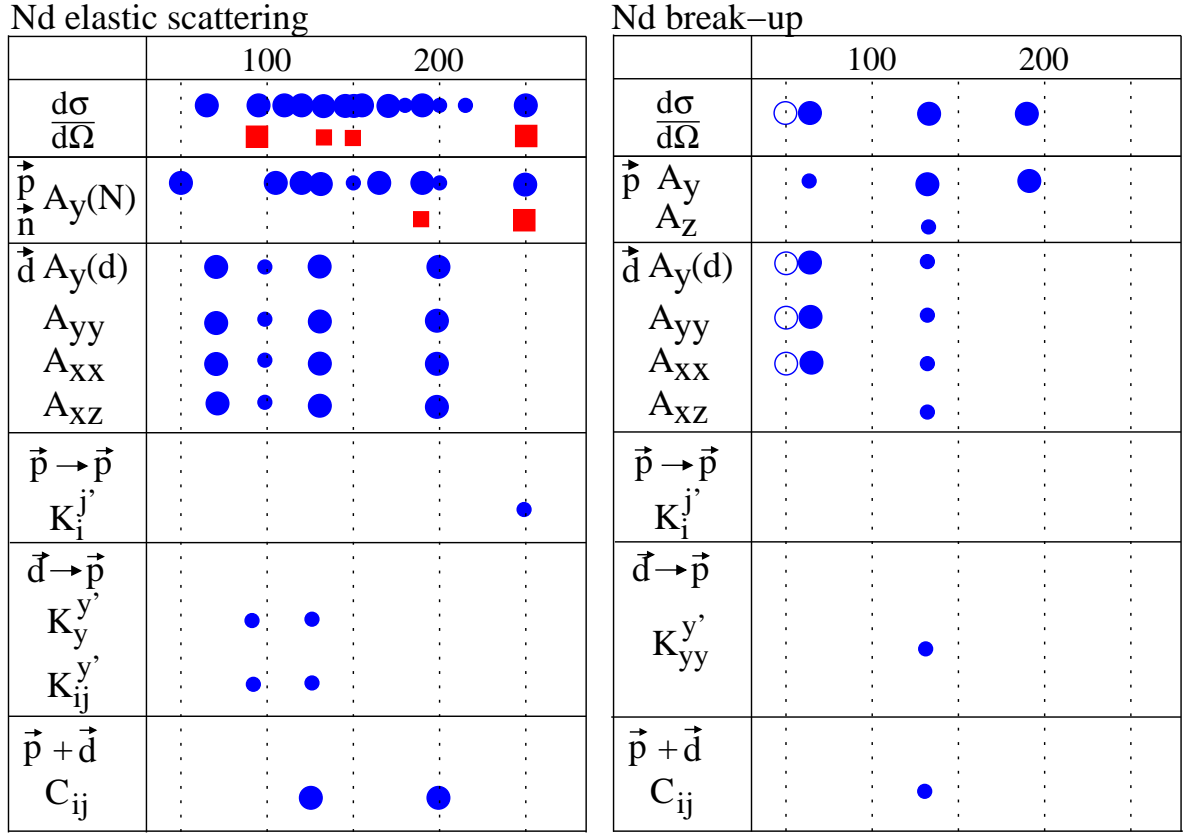} 
\caption{Overview of the observables measured at various laboratories with beams of neutrons (grey/red squares), protons and deuterons (grey/blue circles) with different energies (in units of MeV per nucleon). The size of each circle or square roughly represents the angular coverage for a particular observable at a given energy. A large circle or square refers to a (nearly) complete angular coverage, whereas for a small circle or square only a limited angular range was measured. Open circles refer to data that are presently being analyzed and not yet published (color online).}
\label{worlddatabase} 
\end{figure}

An overview of what has been measured up until now can be seen in Fig.~\ref{worlddatabase}. 
This figure is inspired by a graph presented by K.~Sekiguchi at various conferences. 
As can be seen from the figure, the density of the points is reduced as one attempts to 
measure more complicated observables. For instance, measurements at only a few energies have been 
performed for the spin-correlation coefficients which requires a polarized beam and a polarized target. 
This has been achieved at the IUCF ring with very thin polarized gas-jet targets.
Note, however, that the spin-correlation coefficients, denoted by $C_{ij}$ in Fig.~\ref{worlddatabase},
for the elastic N$d$ channel correspond to 10 observables measured over a wide angular range.
The density of the points is also higher for energies below the pion-production threshold. 
This has to do with the fact that the opening of this inelasticity complicates the calculations to the extent to which it becomes difficult to draw any conclusions about the nature of the three-nucleon forces. Also the number of points measured with the neutron beams is clearly less than that measured with proton or deuteron beams due to the difficulty in producing high-quality neutron beams. 

\subsection{A comparison of experimental data with various models}
\label{sec:results3nf}

The wealth of the data in the three-body systems can 
be presented in this short review neither in a tabular form nor in a graphical form. To familiarize the reader with the observables, several observables are first shown as a function of the kinematical variable as they were originally published but for a couple of energies. 
Out of all measured observables presented in this paper, some of them clearly demonstrate that ab-initio Faddeev calculations based on phenomenological NN potentials, even after including the Coulomb force, are not capable of describing the data in the case only two-nucleon forces are included. In addition, it will be shown that for the same observables, the addition of a phenomenological three-body forces is giving contradicting results. These conclusions will then been globally presented in figures which we refer to as a ``global analysis''. An attempt has been made to select all the available data and compare them with the calculations on a very global scale. In this way, the details of individual experiments and possible problems with them will be obscured. However, (new) features can be observed which would then point to places where more attention has to be paid to. 

\begin{figure}[t]
\centering 
\includegraphics[angle=0, width=0.5\textwidth] {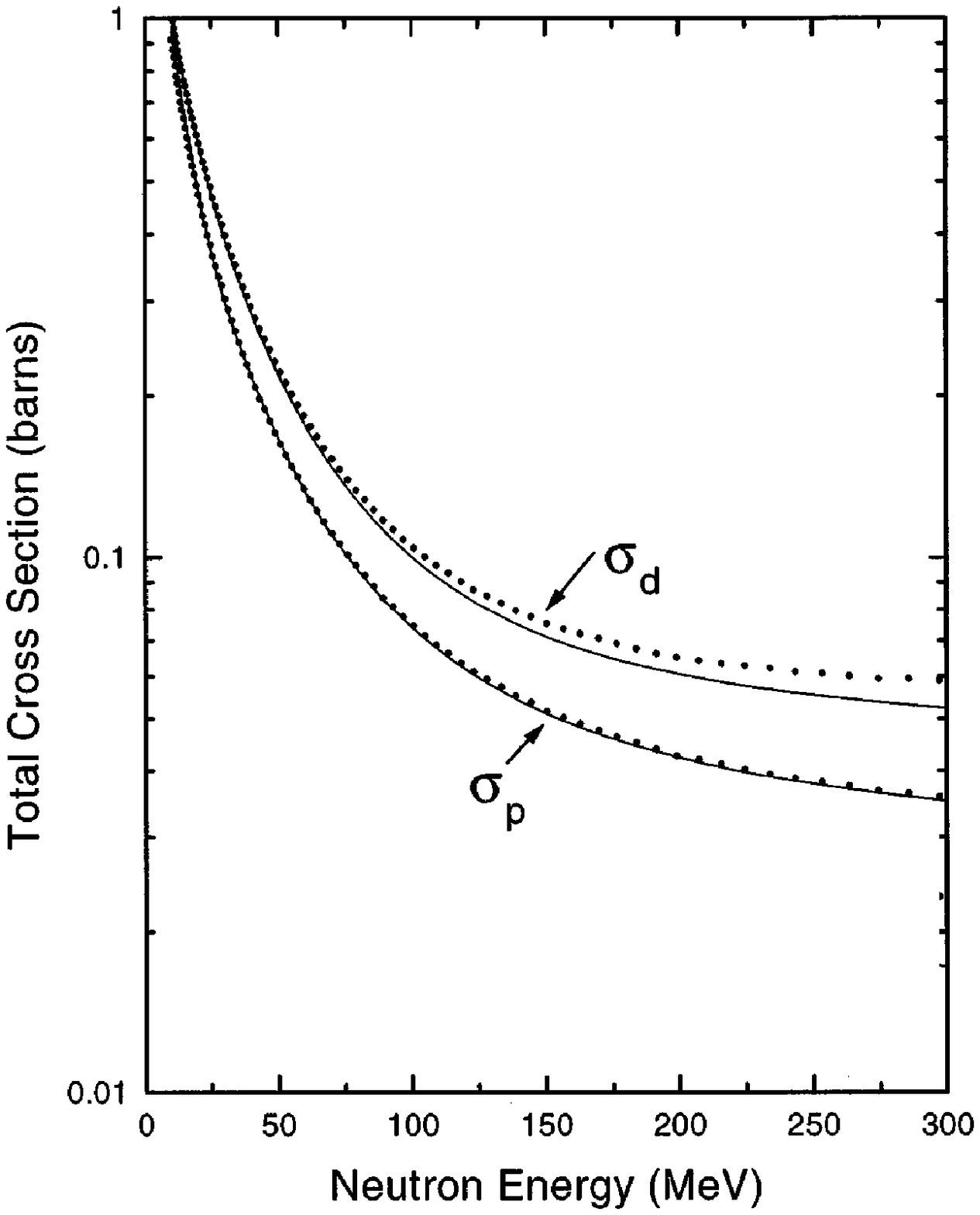} 
\caption{Total neutron-proton and neutron-deuteron cross sections as a function of incident neutron energy \cite{Abfalterer:1998zz}. The solid curves are the results of the calculations for the two- and three-body systems exploiting only 2NF. The dotted lines are data.
Reprinted with permission from~\cite{Abfalterer:1998zz}. Copyright (1998) by the American Physical Society.}
\label{totalndcross} 
\end{figure}

As illustrated in section \ref{theory2nf}, nuclear forces obey a certain hierarchy implying that 3NF effects are much smaller, on the average, than 2NFs. This can be very nicely demonstrated by the inclusive total $np$ and $nd$ scattering data measured at Los Alamos \cite{Abfalterer:1998zz} as shown in Fig.~\ref{totalndcross}. 
First calculations including the 3NF~\cite{Witala:1999sg} have shown that
current models of the 3NF can explain approximately 1/2 of deviation
of the calculations from the data. Since a recent study of relativistic
effects~\cite{Witala:2010fbs} indicates that these cannot resolve the problem,
it can be expected that an improved 3NF will remove the remaining
discrepancies. Whether shorter range 3NFs can help to resolve this
issue needs to be seen in future.
These observations demonstrate the need for {\it exclusive} measurements 
which can provide a significantly larger sensitivity to 3NF effects for 
specific regions in phase space or for other observables than total 
cross sections.

\begin{figure}[t]
\centering 
\includegraphics[angle=0, width=0.9\textwidth] {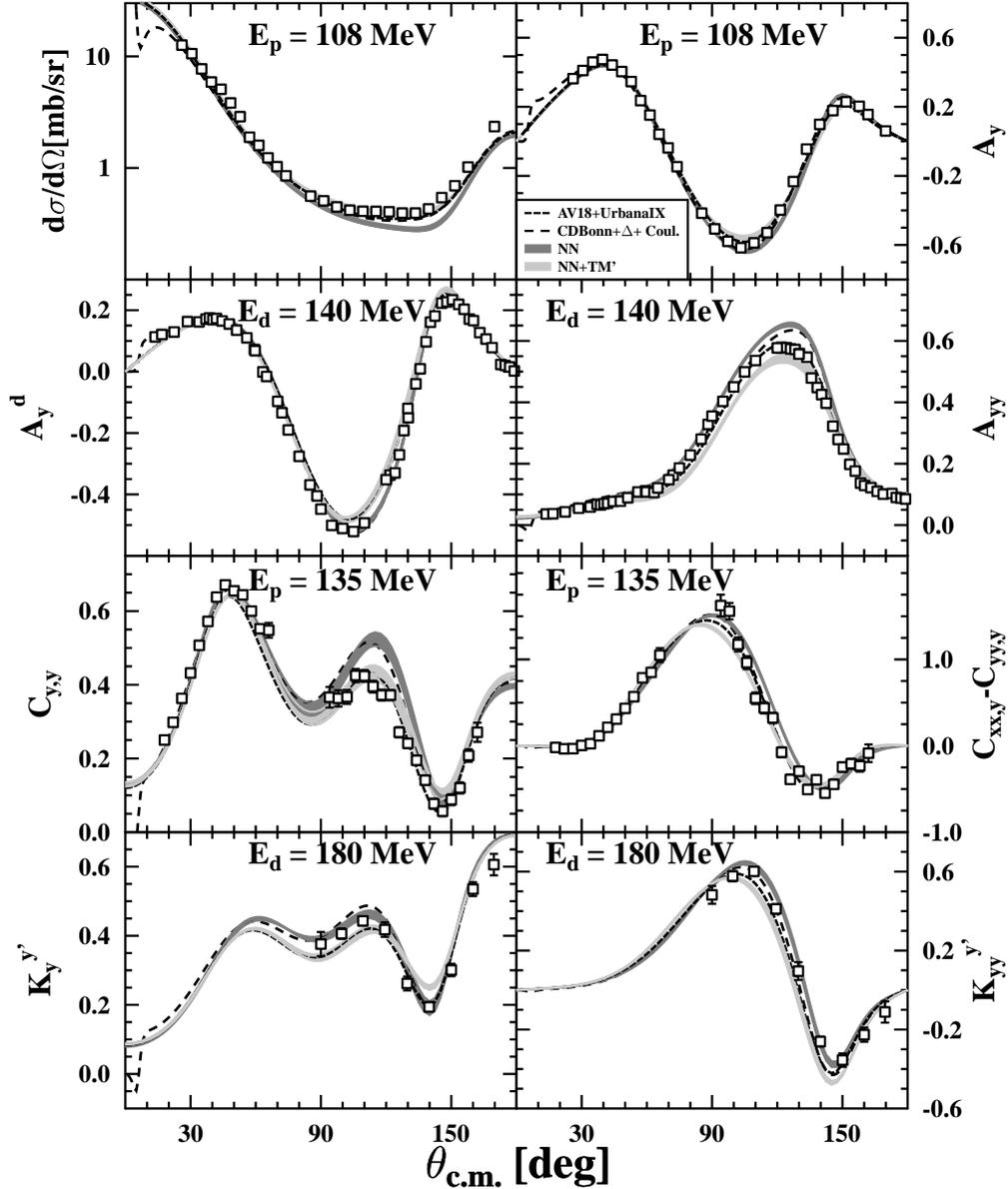} 
\vspace{-1cm}
\caption{Differential cross sections and proton analyzing powers (top row), deuteron vector and tensor analyzing powers (second row), selected spin-correlation coefficients (third row) and selected spin-transfer coefficients (fourth row) for elastic scattering at an incident energy around
100 MeV/nucleon. Errors are statistical only. The results of the calculations performed with several models including only two-nucleon (three-nucleon TM') forces are shown with dark (light) grey bands while those with AV18 and UrbanaIX is presented by the short dashed line. The long dashed line represent the results of the calculations done by the Hanover-Lisbon group including the Coulomb-force effects as well. Data are from Refs.~\cite{Ermisch:2005kf,Sekiguchi:2004yb,vonPrzewoski:2003ig,AmirAhmadi:2007tu}.
}
\label{selected-elastic1} 
\end{figure}

\begin{figure}[t]
\centering 
\includegraphics[angle=0, width=0.9\textwidth] {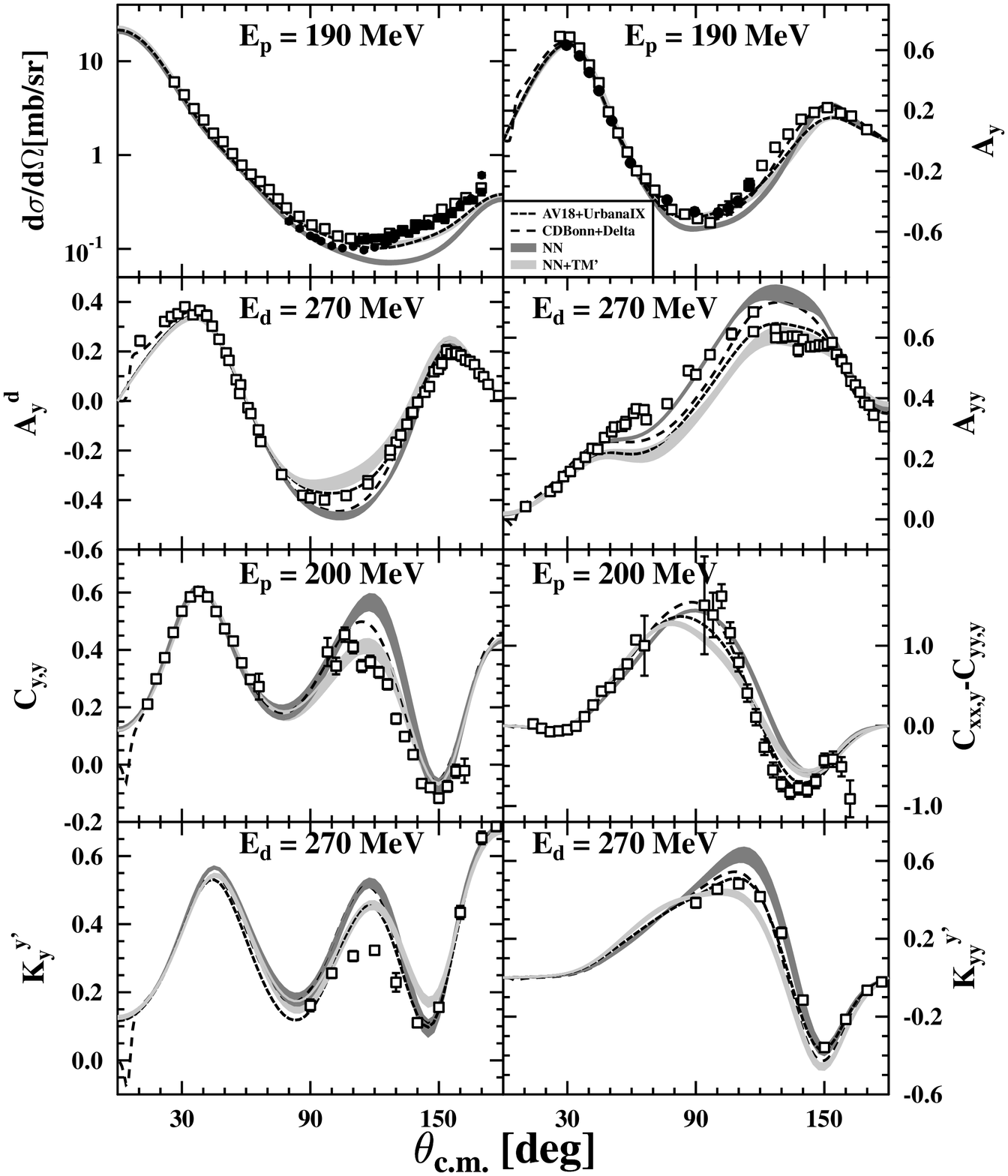} 
\vspace{-1cm}
\caption{Same as Fig. \ref{selected-elastic1} but for each observable at a higher energy. Data are from Refs.~\cite{Bieber:2000zz,Ermisch:2005kf,Igo:1972npa,Adelberger:1972pu,Sekiguchi:2004yb,vonPrzewoski:2003ig}.
}
\label{selected-elastic2} 
\end{figure}

The differential observables which are chosen for the discussions are selected at a lower and a higher energy in the energy range relevant for the discussion of this paper (between 50-250 MeV/nucleon). 
They are shown in Figs.~\ref{selected-elastic1} and \ref{selected-elastic2} for the low and high energies, respectively, and are differential cross sections and proton analyzing powers (top row), deuteron vector and tensor analyzing powers (second row), selected spin-correlation coefficients (third row) and selected spin-transfer coefficients (fourth row) for elastic scattering.

\begin{figure}[t]
\centering 
\includegraphics[width=0.8\textwidth] {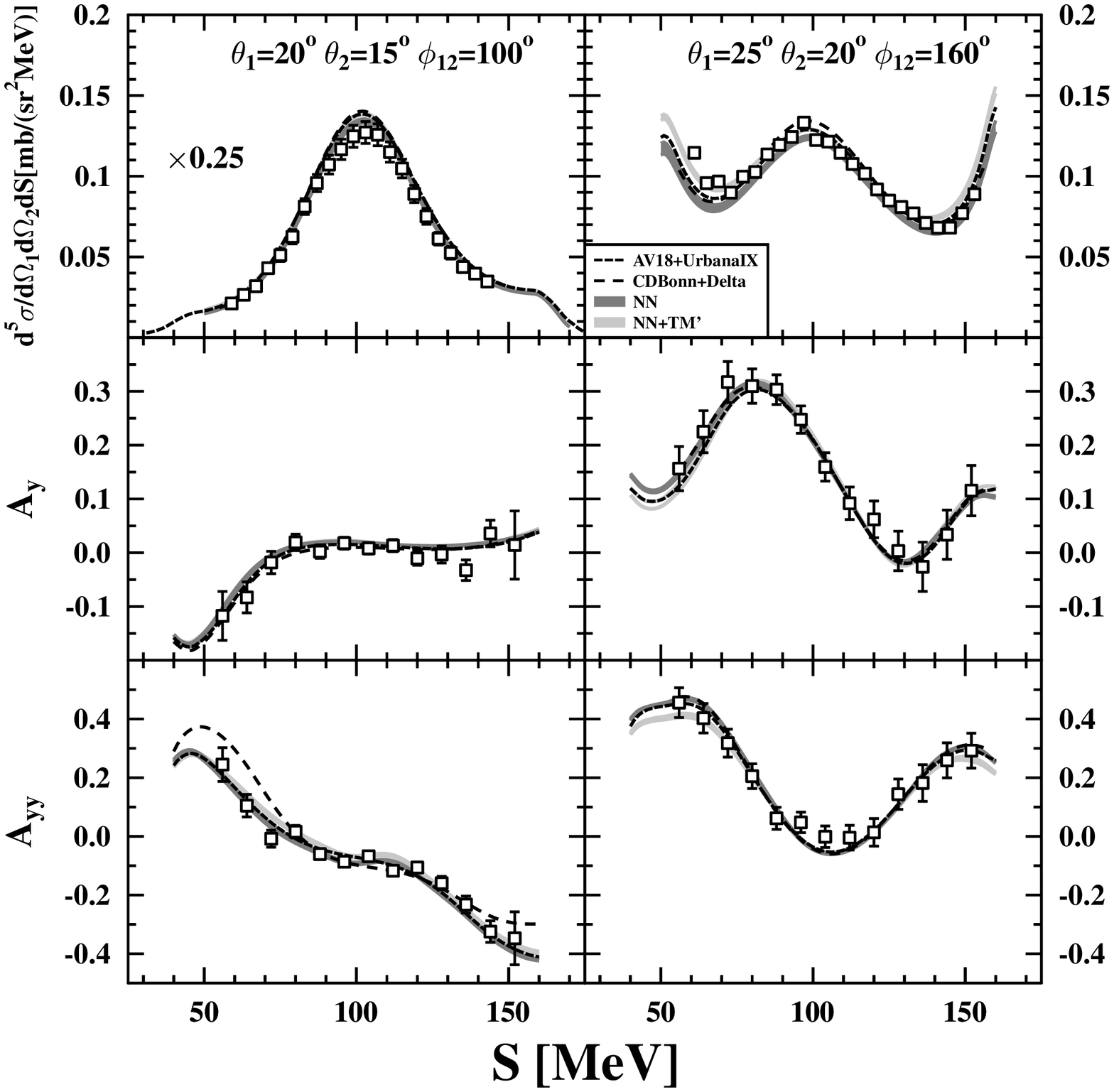} 
\vspace{-7mm}
\caption{Differential cross sections (top tow),  deuteron vector analyzing powers (second row), and deuteron tensor analyzing powers (third row) for the break-up reaction at two different kinematic configurations at an incident beam energy of 65~MeV/nucleon. For the meaning of the lines, refer to Fig. \ref{selected-elastic1}. 
}
\label{selected-breakup65} 
\end{figure}

\begin{figure}[t]
\centering 
\includegraphics[width=0.8\textwidth] {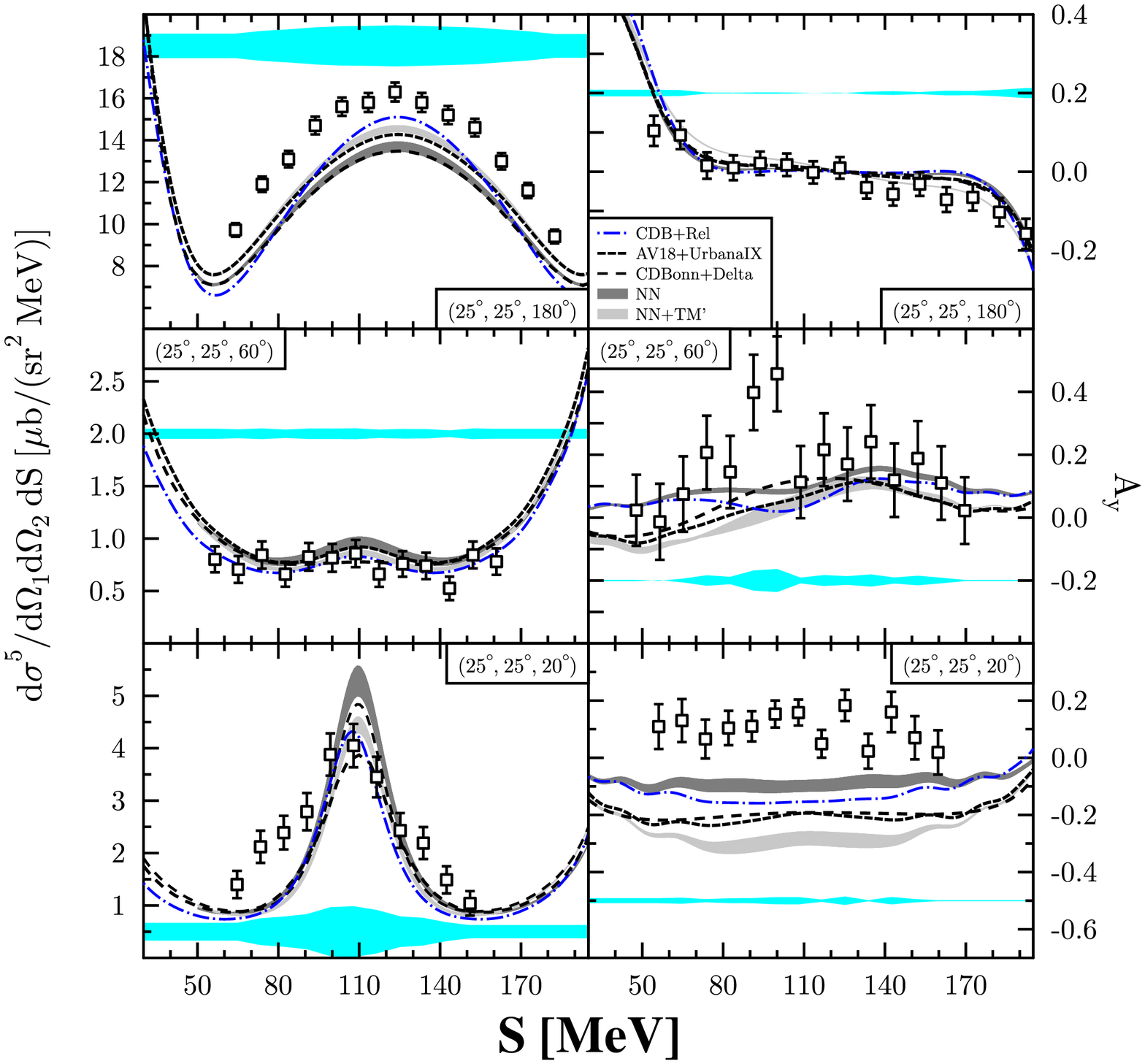} 
\vspace{-7mm}
\caption{Differential cross sections (left column) and vector analyzing powers (right column) for the break-up reaction at three different kinematic configurations. Data are taken with a polarized proton beam of 190~MeV~\cite{Mardanpour:2008phd,Mardanpour:2010zz}. The horizontal bands show the $2\sigma$ systematic uncertainty in the measurement of the observables. 
The dashed-dotted line represents the results of a calculation including relativity. 
For the meaning of all other curves, refer to Fig. \ref{selected-elastic1}.
}
\label{selected-breakup190} 
\end{figure}

In Fig.~\ref{selected-breakup65}, one can see five-fold differential cross sections, vector and tensor analyzing powers for two specific configurations at a low energy of 65~MeV/nucleon for the three-body break-up reaction. 
Other configurations but now for a high-energy beam of protons of 190~MeV are presented in Fig.~\ref{selected-breakup190}. The number of data points for spin-transfer coefficients and spin-correlation coefficients are very small \cite{Sekiguchi:2009zz,Meyer:2004prl} and therefore not presented. 

From Figs.~\ref{selected-elastic1}, \ref{selected-elastic2}, \ref{selected-breakup65}, 
and \ref{selected-breakup190} the following observations can be made. The statistical precision 
of the data set is almost everywhere 
in the phase space very high except at higher energies where 
the cross sections are generally small, and for the spin-transfer coefficients for which a secondary 
scattering is necessary and the spin-correlation coefficients which requires polarized beam and target. 
The systematic uncertainties are, on the other hand, generally the dominating sources of errors. 
For cross sections, these come from the determination of the target thickness, the beam current, 
and the acceptances and the efficiencies of the detection systems and are generally in the order of 5\% of 
the value of the cross sections. For the analyzing powers, a large fraction of uncertainties cancel out 
or remain as a second-order correction when ratios of cross sections are calculated. In this case, 
the main source for uncertainties comes from the precision at which the beam (target) polarization 
can be measured, which is usually around 2-4\% of the value of the measured spin observables. 

The second observation is that for almost all observables in elastic scattering, the calculations 
which only include 2NFs fail to a large extent to describe the data, and in particular at higher energies. 
The effect is very large in the minimum of the elastic-scattering cross section as was first pointed out by 
Wita{\l}a et al.~\cite{Witala:1998ey} and Nemoto et al.~\cite{Nemoto:1998zz}. It was exactly in this region, 
where the first searches for the effects of 3NF were conducted. It is also very intuitive that this should 
be the case, since it is in the minimum of the cross section where the 2NF effects become very small allowing 
other small effects to be relatively enhanced. However, it can also be noted that even adding the 3NF is not 
enough to account for the differences. This is also true for the analyzing powers. The discrepancies generally 
become larger at higher incident beam energies. For other observables, the message is rather mixed. 
The addition of 3NFs sometimes improves the situation but sometimes also makes the agreement with the data worse. 
The effect of the Coulomb force, which has recently been taken into account \cite{Deltuva:2005cc}, seems to be 
very small at all but the smallest angles for the elastic-scattering channel.

Many of these observables have been measured more than once at various laboratories for consistency 
checks. In most of the cases, 
the agreements are satisfactory, specially in the case of analyzing powers where the absolute normalization plays a 
minor role. In the case of the cross section, a major discrepancy has been observed between data sets taken at 
a beam energy of 135~MeV/nucleon measured at KVI \cite{Ermisch:2005kf} and RIKEN/RCNP \cite{Sekiguchi:2005vq}. 
Furthermore, data taken IUCF \cite{vonPrzewoski:2003ig} agreed with the shape of the KVI data.
Another measurement at KVI with a different setup 
has resulted in cross sections which are more compatible with the overall trend of the data at other energies and in disagreement with both earlier measurements \cite{RamazaniMoghaddamArani:2008ww}. 

\begin{figure}[t]
\centering 
\includegraphics[angle=0, width=\textwidth] {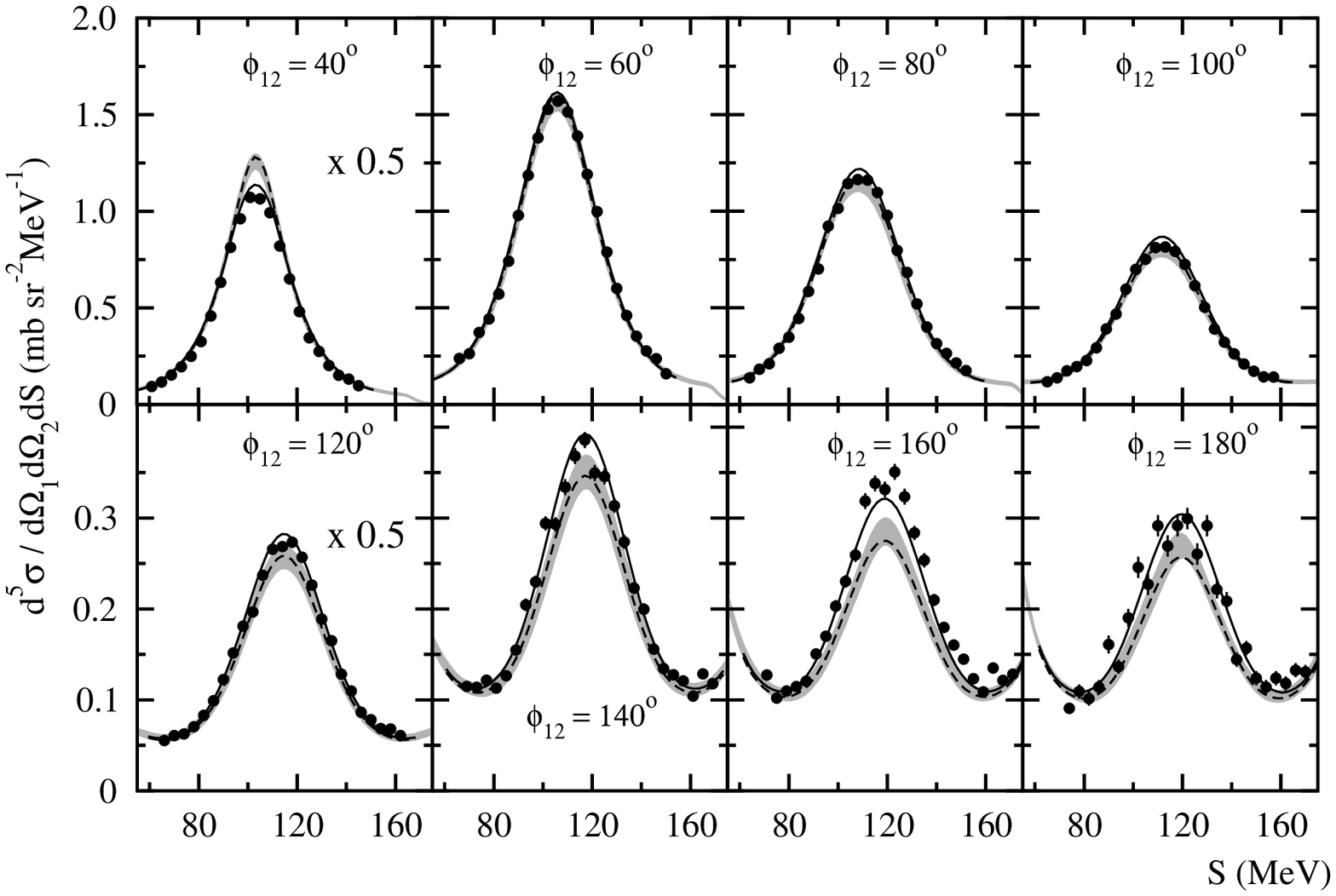} 
\caption{Results for the break-up cross sections for various coplanarity 
angles between the two outgoing protons ($\phi_1-\phi_2)$ as a function of the
kinematical variable $S$, for an incident deuteron energy of 130~MeV and the
outgoing proton polar angles of (15$^\circ$,15$^\circ$)~\cite{Kistryn:2005fi,Kistryn:2006ww}. 
The bands show the results of the chiral EFT calculations at N$^2$LO.
The curves are the 
results of the calculations of the Hanover-Lisbon group including the explicit $\Delta$ 
(dashed curve) and also including the Coulomb force (solid curve).} 
\label{Coulombeffect} 
\end{figure}

The third observation is that for the break-up reaction at 65~MeV/nucleon, the calculations with 
or without 3NF are in very close agreement with the data with a preference for the inclusion of the 3NF as 
was shown in a careful $\chi^2$ analysis \cite{Kistryn:2005fi}. With this reaction, one can also look at 
another variable, namely the relative energy of the two outgoing protons. It is intuitive and supported by  
the calculations that the Coulomb effect can be extremely large for the case where the relative energy of 
the outgoing protons is very small (see Fig.~\ref{Coulombeffect}) \cite{Kistryn:2006ww}. This has been further 
pursued in a recently-performed experiment in J{\"u}lich where the opening angle between the two protons has been further 
reduced to $5^\circ$~\cite{stephan:2011ijmpa}. At these very small angles, the Coulomb effect has even more dramatic features in the 
break-up cross sections. 

Although the 3NF effects are generally small at 65~MeV/nucleon, the 
situation changes with increasing beam energy.
The results at 190~MeV/nucleon clearly show that the 2NF is far from sufficient. Remarkably, the addition 
of a 3NF in some configurations, where the relative energy of the outgoing protons is small, makes 
the disagreements even larger (see Fig.~\ref{selected-breakup190}). Possible missing ingredients which could
contribute are relativistic effects (see section \ref{sec:theory}).
This has now been included in a model calculation by Wita{\l}a et al.~\cite{Witala:2005nw}. 
The dashed-dotted curves 
in Fig.~\ref{selected-breakup190} show the results of these calculations. Although there are improvements 
in certain parts of the phase space, the effects are clearly not sufficient to describe the bulk 
of the data.

\begin{figure}[t]
\centering 
\includegraphics[angle=0, width=\textwidth] {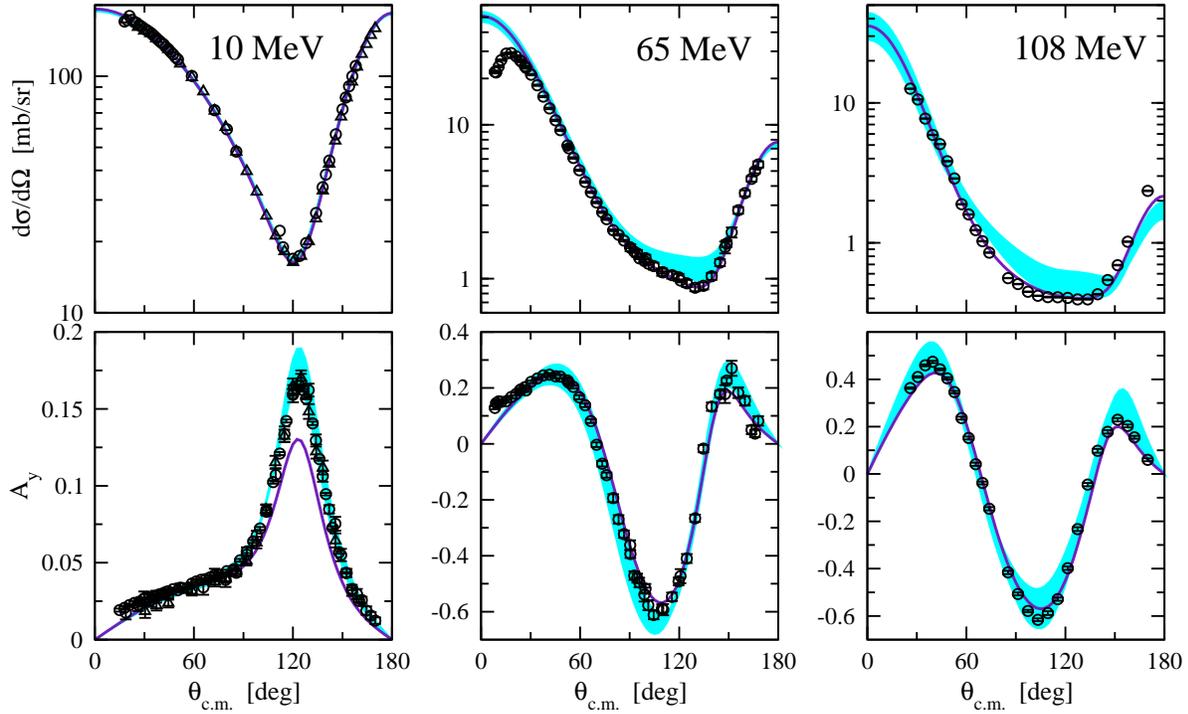} 
\caption{Results for the cross sections and analyzing powers as a function of center-of-mass scattering 
angle at three different incident energies. The bands are the results of the chiral EFT calculations at 
N$^2$LO. Also shown are the calculations based on the CD-Bonn potential combined with the Tucson-Melbourne 
3NF. At 10~MeV, triangles refer to the neutron-deuteron data from Ref.~\cite{Howell:1987fbs} while 
circles are proton-deuteron data from Ref.~\cite{Rauprich:1988fbs,Sperisen:1984npa,Sagara:1994zz} 
with the contribution due to the Coulomb interaction being subtracted. Proton-deuteron data at 65~MeV and 
108~MeV are taken from Refs.~\cite{Shimizu:1995zz,Witala:1993fbs,Ermisch:2005kf} (color online).} 
\label{chiral-energy} 
\end{figure}

Note that in all the figures in this section with the exception of Fig.~\ref{Coulombeffect}, 
only a phenomenological 3NF have been used which are ``added'' to the models which only include 
pair-wise NN interaction, and no results from the chiral EFT calculations are shown for intermediate 
energies. As mentioned in section~\ref{sec:nnforce}, much progress has been made in the past decade 
in the framework of the chiral EFT. Presently, the calculations are showing a very good agreement 
with the data at low 
energies. As one increases the energy, the theoretical uncertainties at N$^2$LO grow considerably. To illustrate this, two observables, namely cross section and 
analyzing power are shown as a function of scattering angle in Fig.~\ref{chiral-energy} for 3 different 
energies. The model calculations shown in other figures agree quite well with the results of the 
chiral EFT calculations albeit within the large error bands. It is exactly 
because of these, sometimes, large theoretical uncertainties that they are not shown everywhere. 
We further emphasize that a rather large theoretical uncertainty for $A_y$ at low energy is a 
consequence of the extreme sensitivity of this observable to the NN triplet P-waves which at N$^2$LO are 
only described within a few percent. A considerably more precise description of these partial waves at 
N$^3$LO leads to the same $A_y$-puzzle as observed for the phenomenological potential models, 
see Fig.~\ref{aypuzzle} and the discussion of Ref.~\cite{Entem:2001tj}.  

\begin{figure}[t]
\centering 
\includegraphics[angle=0, width=0.9\textwidth] {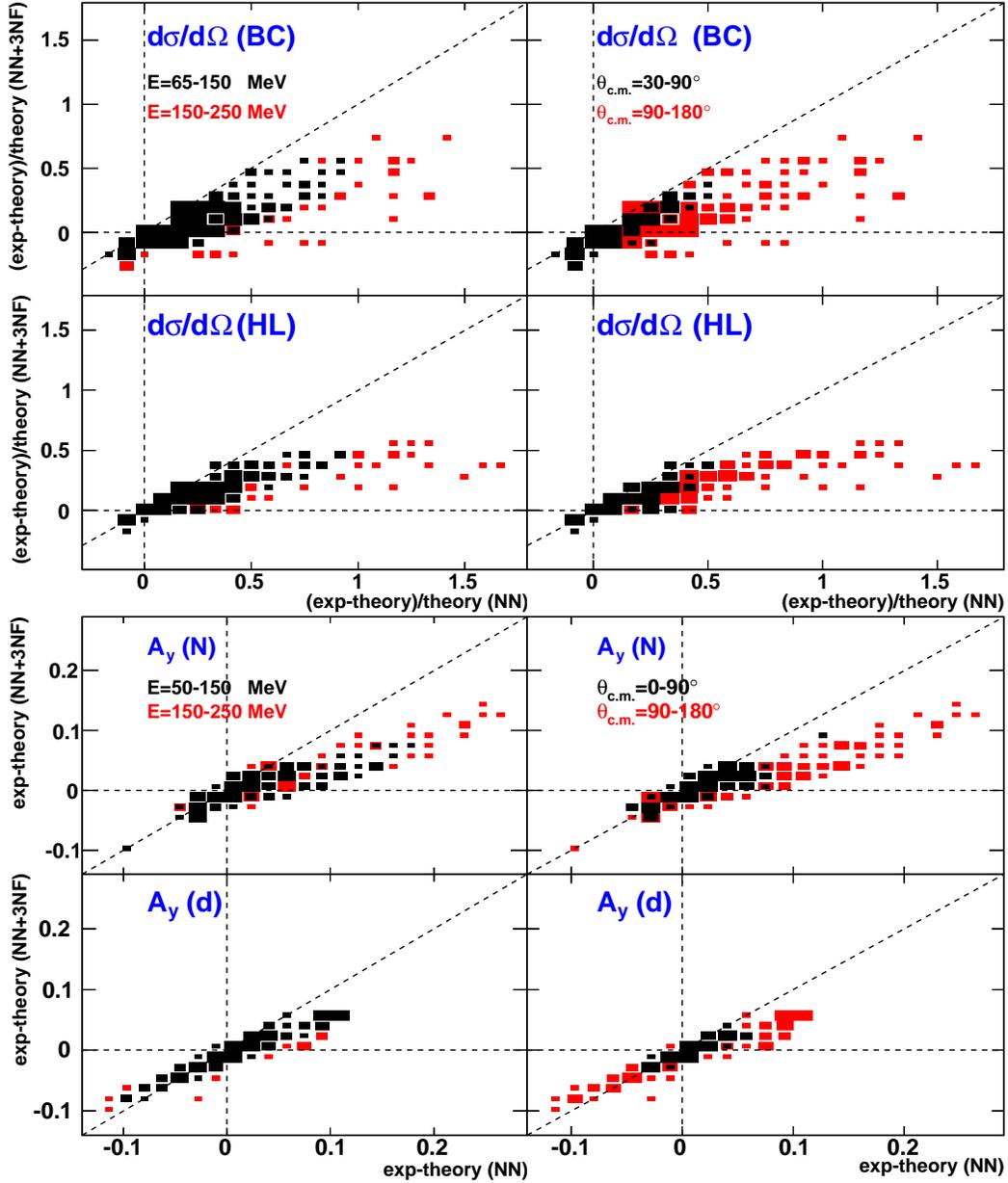} 
\caption{Results of the calculations are subtracted from all corresponding data points available in the 
literature for elastic scattering for the energy range of 50-250~MeV and center-of-mass angles
$\theta_{c.m.}$$>$30$^\circ$ and plotted as a (relative) difference between experimental data and 
calculations 
with only 2NF (x-axis) and with 3NF in addition (y-axis). The top four panels represent the relative 
differences for cross sections: on the left for two different energy ranges in two different shades 
(color online) and on the right for two different angle ranges in different shades. The label BC refers 
to a calculation from the Bochum-Cracow group based on the CD-Bonn two-nucleon potential and the TM' 3NF. 
The label HL refers to a calculation from the Hanover-Lisbon group. A similar comparison is shown in the 
bottom four panels for the proton and deuteron vector analyzing powers. In this case, only the calculations 
of the HL have been used and $\theta_{c.m.}$$>$8$^\circ$ (color online). 
}
\label{globalanalysiselastic1} 
\end{figure}

For our global analysis, all the data points have been collected along with the predictions of 
Hanover-Lisbon group. This group uses the CD-Bonn plus $\Delta$ potential based on the CD-Bonn  
refitted in order to 
accommodate the potential in a coupled-channel calculations including the $\Delta$ isobar. The reason 
for using this model is the inclusion of the Coulomb force effects which is very important in the 
comparison with data which are collected with charged-particle beams. The Coulomb effects are generally 
small but sizable at specific kinematics (small scattering angles for the elastic scattering and 
small relative energies of the outgoing protons in the case of the break-up reaction). The results of the 
predictions have, subsequently, been subtracted from the data for a certain observable. For the cross 
sections, the results are divided by the theoretical value to achieve relative deviations. For the 
polarizations observables, the absolute differences are investigated. The visualization of the 3NF effects 
shown below follows an idea of Meyer \cite{Meyer:2007tri}.
In the two-dimensional plots presented here (Figs.~\ref{globalanalysiselastic1}, 
\ref{globalanalysiselastic2}, \ref{globalanalysisbreakup}, and \ref{globalanalysisbreakup2}), these differences are shown for the 
case where the theory only includes 2NF (x-axis) and when it also includes 3NF (y-axis). All energies 
and all angles are included in the same figure. As a guide to understand the effects, three dashed lines 
have also been drawn in the figures. If the points lie on the vertical dashed line around 0, one should 
conclude that the 2NF is already sufficient in the description of the data and the inclusion of the 3NF 
only deteriorates the agreement. If the points, on the other hand, lie on the diagonal dashed line, 
it means that the effect of 3NF is on average predicted to be negligible, irregardless of the agreement 
between the theory and the data. Finally, if the points lie on the horizontal dashed line around 0, 
it means that the inclusion of the 3NF has accounted for the differences between data and models with 
only a 2NF as input. If the points lie around (0,0), the data agree with the calculations including 
only 2NF and 3NF effects are shown to be small in the calculations and by the data. 

To see trends in the data, the elastic-scattering cross sections and vector analyzing powers  
have been plotted with different shades (colors on-line) in Fig.~\ref{globalanalysiselastic1}: in the 
left panels, as a function of energy and in the right panels as a function of scattering angle in the 
center-of-mass. For the cross sections, a few observations can be made. The first one is 
that there are a number of points which lie in the lower left region of the origin. The bulk of these data 
points come from the high energy measurement at 250~MeV (obscured in 
Fig.~\ref{globalanalysiselastic1} by the black dots) \cite{Maeda:2007zza} 
and the RIKEN measurement at 135~MeV \cite{Sekiguchi:2005vq}, both at small scattering angles. 
Since Coulomb effects are very small 
at the angles chosen for this analysis and are properly taken into account in the calculations presented 
in the second row of the figure, these discrepancies with the calculations at the relatively small angles 
should be taken as a sign of normalization problems in the sense that the data lie below the calculations 
where one would expect a reasonable agreement due to the dominance of the 2NF. Note that this problem is 
worse when one uses TM' 3NF (top panels) instead of the calculation using the $\Delta$ isobar (second row). 
The relative differences between these approaches are shown explicitly in Fig.~\ref{3n-comp}. The same 
statement could also be made for the points which are on the opposite side of the origin but there, 
the signal is mixed with a genuine 3NF effect. The second observation is that the discrepancies become 
larger for higher energies and clearly become larger in the minimum of the cross section and for backward 
angles. The data are generally scattered between the diagonal line and the horizontal line. This implies 
that a 2NF is not enough to describe the data but that an additional 3NF within this model is also not 
sufficient to remedy the discrepancies. Other models of 3NFs have been examined as well and generally 
show the same pattern with the differences discussed when Fig.~\ref{3n-comp} was presented. 

For the analyzing power, $A_y$, one can again observe that the discrepancy increases as the incident 
energy increases. Here, the largest discrepancies occur at scattering angles where the cross section is 
at its minimum. For the deuteron vector analyzing power, the discrepancies are generally smaller but 
extend to both sides of zero as opposed to the proton vector analyzing power which show differences which 
are generally on the positive side.

\begin{figure}[t]
\centering 
\includegraphics[angle=0, width=0.9\textwidth] {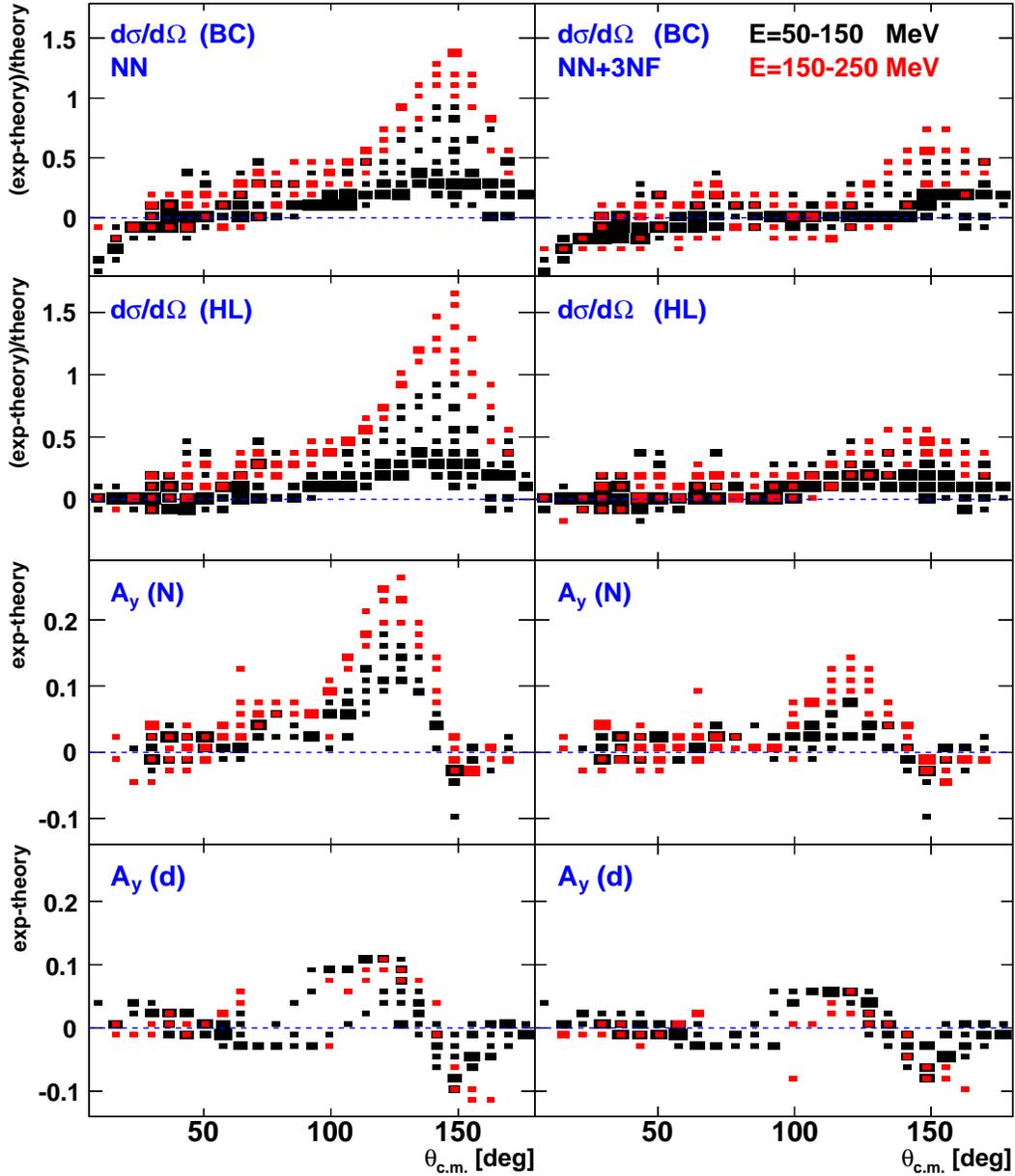} 
\caption{Cross-section (top four panels) and vector analyzing-power (third and fourth rows) differences 
for the case when only the two-nucleon forces are included in the calculations (left panels), and for 
the case where the effect of three-nucleon forces are also taken into account via the Hanover-Lisbon 
calculations (right panels) (color online). The label BC refers to calculations from the Bochum-Cracow 
group and the label HL to calculations from the Hanover-Lisbon group. 
In the top panels, the effects of the Coulomb force which becomes very strong below scattering angle 
of 25$^\circ$ are observed as they are not included in the calculations by the BC group. For the analyzing 
powers, only calculations from the HL are shown.}
\label{globalanalysiselastic3} 
\end{figure}

\begin{figure}[t]
\centering 
\includegraphics[angle=0, width=0.9\textwidth] {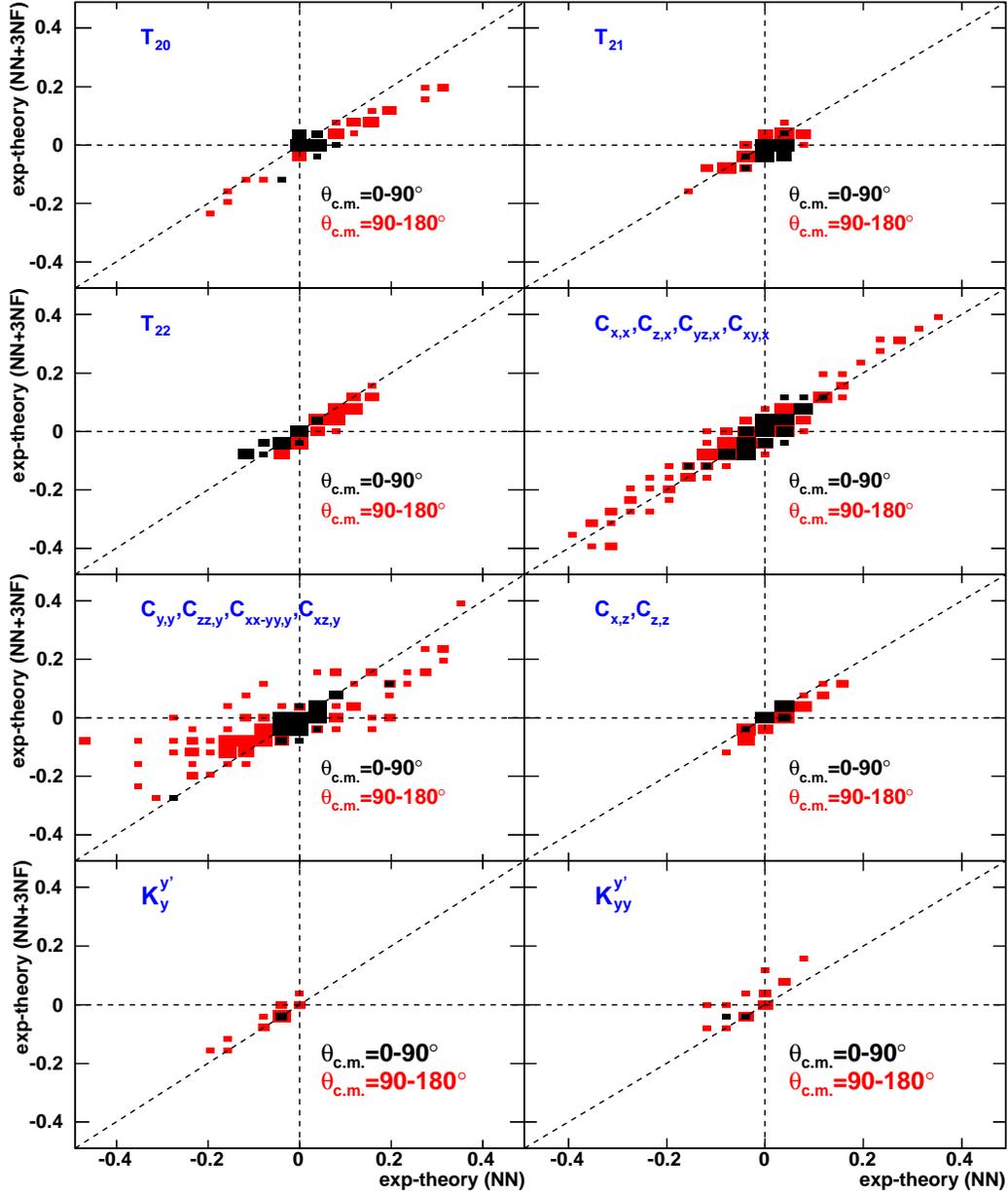} 
\caption{Results of the calculations are subtracted from all corresponding data points available in the 
literature for tensor analyzing powers, spin-transfer and spin-correlation coefficients in elastic 
scattering for the energy range of 50-250 MeV and angular range of 15-175$^\circ$ (not for all energies)  
and plotted as a difference between experimental data and calculations with only 2NF (x-axis) and with 3NF 
in addition (y-axis). The data points for all energies are mixed together and are only distinguished for 
two angular ranges with different shades (color online). 
}
\label{globalanalysiselastic2} 
\end{figure}

\begin{figure}[t]
\centering 
\includegraphics[angle=0, width=0.8\textwidth] {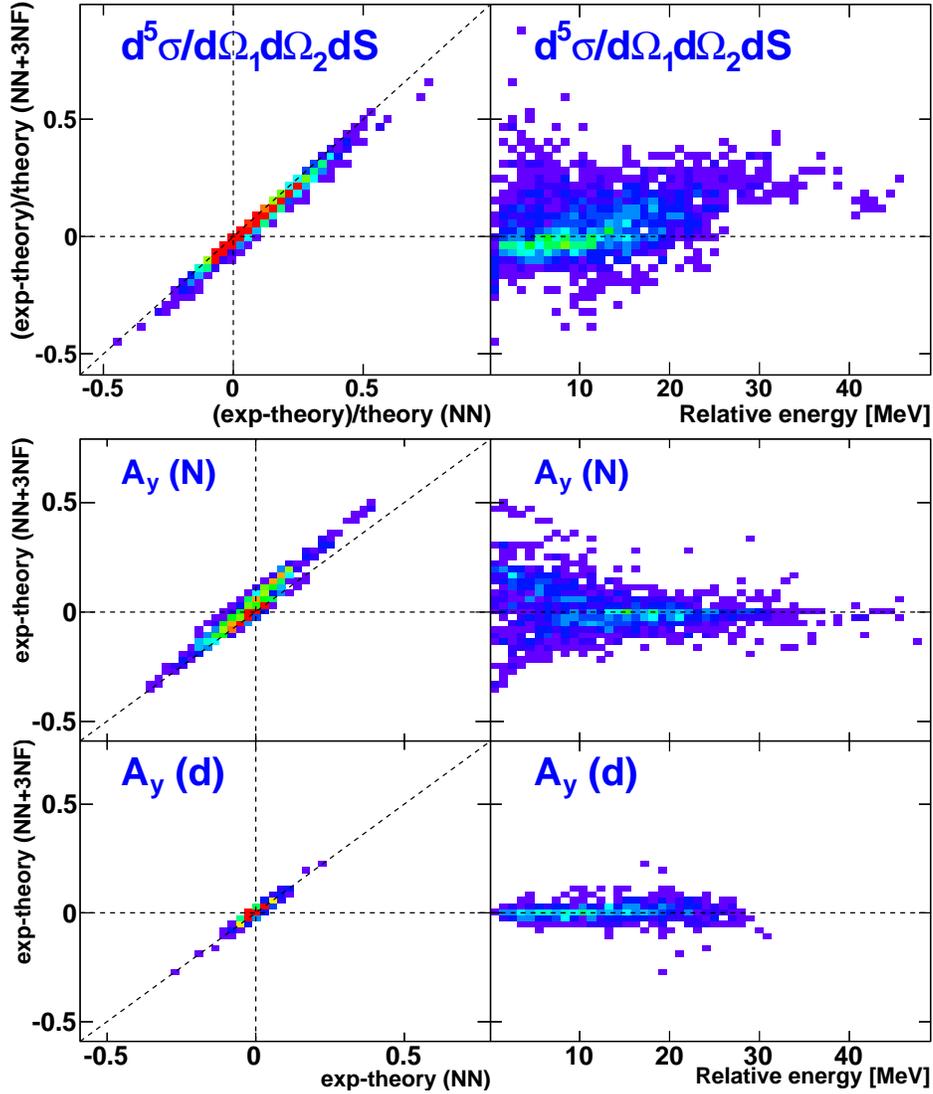} 
\caption{Results of the calculations are subtracted from all corresponding data points available in the 
literature for break-up reaction for the energy range of 65-190 MeV and various angle combinations and 
plotted, in the left panels, as a (relative) difference between experimental data and calculations 
with only 2NF (x-axis) and with 3NF in addition (y-axis). In the right panels, the differences between 
calculations including a 3NF and the data are shown as a function of the relative energy of the two 
outgoing protons. The top row shows the differences for cross sections, the second row, for the proton 
analyzing power, and the third row, for the deuteron vector analyzing power (color online). 
}
\label{globalanalysisbreakup} 
\end{figure}

\begin{figure}[t]
\centering 
\includegraphics[angle=0, width=0.8\textwidth] {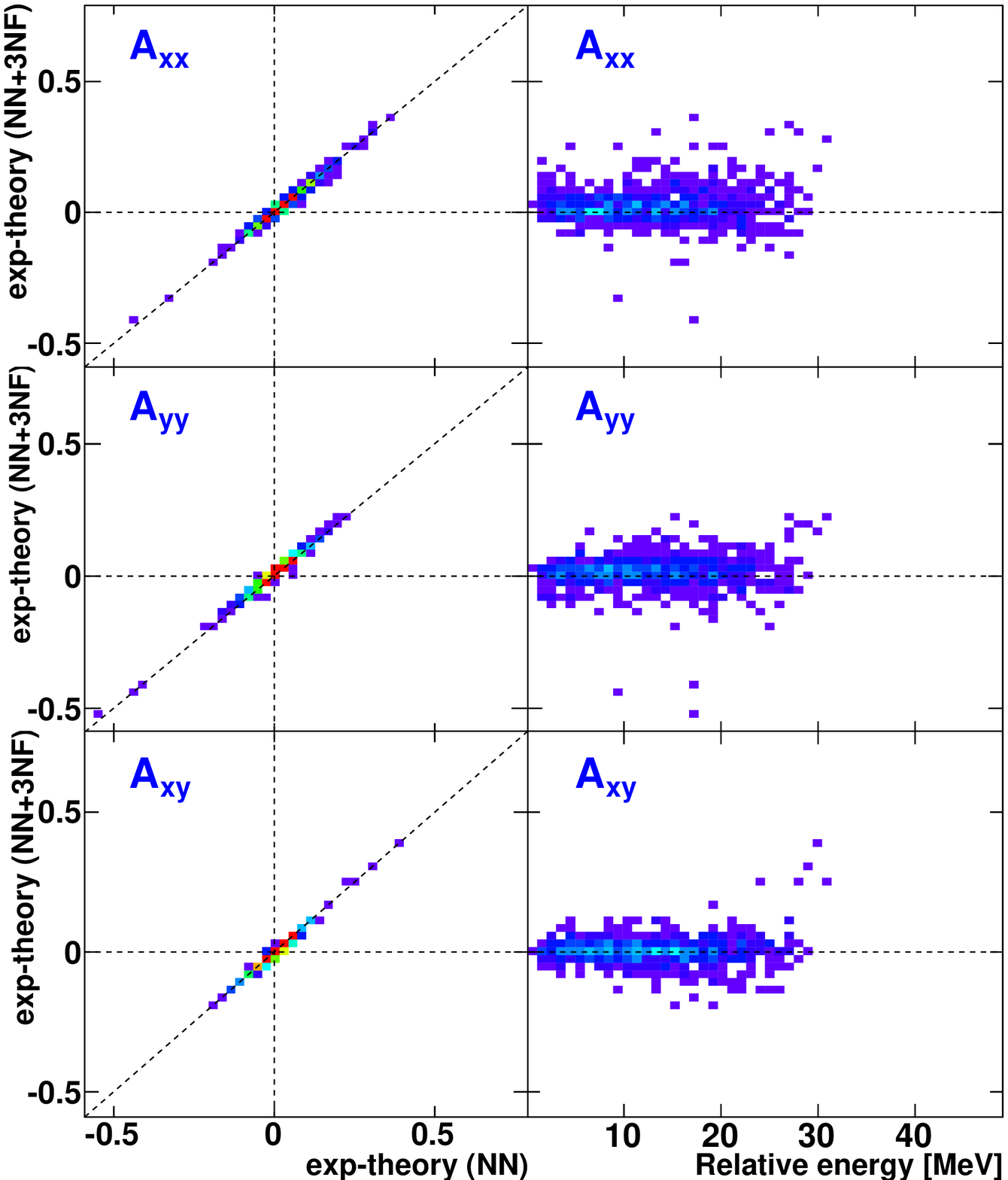} 
\caption{
Results of the calculations are subtracted from all corresponding data points available in the 
literature for tensor analyzing powers of the proton-deuteron 
break-up reaction. Presently, the database contains only a measurement taken at 65~MeV/nucleon 
for various angular combinations. See Fig.~\ref{globalanalysisbreakup} for a 
complete description of the plots (color online).}
\label{globalanalysisbreakup2} 
\end{figure}

It is interesting to see in detail how the deviations for cross sections and vector analyzing powers behave 
as a function of scattering angle. These differences are shown in Fig.~\ref{globalanalysiselastic3} for 
the case when only the two-nucleon forces are included in the calculations (left panels), and for the case 
where the effect of three-nucleon forces are also taken into account via the Hanover-Lisbon calculations 
(right panels). All three observables show a peak around the minimum of the cross section. The peak for 
the cross section vanishes only at the very backward angles whereas for the analyzing powers, 
the differences go through a zero when the cross section passes its minimum value and bend 
back towards zero at the very backward angles. The spread of the data around the peak position 
is due to the various energies that are included in the same figure. 
Generally, the higher the beam energy is, the larger the differences between data and predictions. 

All other measured observables at intermediate energies are shown in Fig.~\ref{globalanalysiselastic2}.
For the tensor analyzing powers, it is difficult to make a general statement for all observables. 
For $T_{21}$, most data points agree rather well with the calculations with and without the three-nucleon 
forces and largest discrepancies occur at backward angles. For $T_{20}$, the inclusion of three-nucleon 
forces seem to improve the agreement with the data slightly. For $T_{22}$, the discrepancies are not 
removed by the inclusion of three-nucleon forces. In fact, the contribution of the 3NF for this observable 
seems to be negligible. Also here, the largest discrepancies are observed at backward angles. The energy 
dependence of the discrepancies is very small for tensor-analyzing observables. 

A detailed study of the spin-transfer coefficients revealed similar behaviors among various groups
of coefficients when comparing data with model calculations. Those coefficients which 
showed similar trends were, therefore, combined in the global analysis as depicted in the middle
panels of Fig.~\ref{globalanalysiselastic2}. The discrepancies are 
again very large, in particular for scattering angles larger than 90$^\circ$. The different 
groups of coefficients, however, 
show different behaviors: $C_{x,z}$ and $C_{z,z}$ show good agreement with calculations and 
the effect of 3NF seem to be negligible in the calculations. For $C_{x,x}$, $C_{z,x}$, $C_{yz,x}$ and 
$C_{xy,x}$, there is a small effect due to the 3NF but the effect is in the wrong direction according
to data. This is the opposite for $C_{y,y}$, $C_{zz,y}$, $C_{xx-yy,y}$,
and $C_{xz,y}$ for which the addition of 3NF seems to bring the calculations 
much closer to the data. For these coefficients, the trends are the same for the two set of 
measurements at different energies. 

For the $K_y^{y'}$ and $K_{yy}^{y'}$, the number of data points in the literature is limited due to the 
obvious limitation in the secondary scattering of the outgoing particles. What is in the literature for 
the intermediate energies (90~MeV/nucleon and 135~MeV/nucleon) are plotted in the same way in the bottom 
row, as for the other observables in Fig.~\ref{globalanalysiselastic2}. Here, one can see that the 
discrepancies are generally small and that the effect of the 3NF in the calculations is not significant 
(as could also be observed in Figs.~\ref{selected-elastic1} and \ref{selected-elastic2}). Given the 
small number of data points, it is difficult to draw strong conclusions for these observable. For the 
points which show some deviations from the calculations, it is clear that the addition of 3NFs is not 
sufficient. 

Figure~\ref{globalanalysisbreakup} shows the results of a global analysis for the break-up reaction at 
various energies (65, 130 and 190~MeV/nucleon) for a large combination of angles. Here, the density of 
the data points is represented by the shade of grey. It is clearly seen that a large number of data points 
reside around (0,0) indicating that the 2NF is already sufficient to describe the data reasonably well 
and that the effect of the 3NF is small. Many of these points stem, however, from the measurement taken
at the lowest energy (65~MeV/nucleon). As the energy increases the length of the cluster around the 
diagonal line increases. As opposed to the elastic scattering channel, the deviations extend on both 
side of the origin: in some regions, the calculations overestimate the data and in others, they 
underestimate them. However, in almost all regions, the effect of the 3NF is not large and certainly 
not enough to remedy the differences. There is, however, a slight tendency towards repairing the 
deficiencies. The larger differences between the model predictions and the data come from 
kinematics for which the relative energy of the outgoing protons is large; see the top-right panel 
of Fig.~\ref{globalanalysisbreakup} in which the relative difference between the data and the model 
calculations including 3NF is shown. The vector analyzing powers show a different behavior. For these 
observables, there are two bands: one on the diagonal and close to (0,0) where the 3NF effects 
are small and a second band bending off from the diagonal indicating that the addition of the 3NF makes 
the agreement even worse. Further inspection of the data shows that these points belong to kinematics 
where the relative energy between the outgoing protons is small. This is illustrated in right panel 
of the figure for this observable. In the same panel, one observes several branches towards a relative 
energy of 0 MeV. The reason for the split is partly due to the fact that we have mixed data sets of two 
different beam energies (130 and 190~MeV/nucleon). Another reason is that in the analysis of the data, 
one has chosen a coarse binning for the proton angles. 
In a hypothetical figure made with data points from many energies between 65 to 200~MeV and all angles 
in a fine binning, one would then expect a cone shape with the base at 0 MeV and the summit at the 
largest possible relative energy measured in the experiment. One should therefore look at the spin 
properties of the nuclear forces at small relative energies more carefully. 

The sample of data for $A_y^d$ is very small (only at 65~MeV/nucleon) and the agreement seems to be 
reasonable, as shown in the third row of Fig.~\ref{globalanalysisbreakup}. A few data points which 
show large discrepancies occur at large relative energies. For the tensor analyzing power, the trends 
are very similar, namely that there is a good agreement between the predictions of the theoretical 
calculations including 3NFs (albeit small) and the data
as shown in Fig.~\ref{globalanalysisbreakup2}. For $A_{xy}$, a few data points show large discrepancies 
at larger relative energies. For the other two tensor analyzing powers, there seems to be no 
systematic pattern. 

\subsection{How to proceed for $A=3$ systems?}

In the study of nuclear forces, the major effort was devoted to the nucleon-nucleon system during the 
last decades of the twentieth century. Two-nucleon force models were developed and fine tuned to 
the large data sets which became available. All physical observables are presently described with 
the help of these phenomenological potentials (CD-Bonn, AV18, NijmI, NijmII and Reid93) with a remarkable value 
of 1 for the $\chi^2$/datum. With the good understanding of the long-range part of the potential, 
various attempts were made to do a model-independent analysis of all the data by bringing them together 
and performing a partial-wave analysis \cite{Stoks:1993tb,Arndt:2007qn}. The very large data set was 
brought successfully into very precise partial waves making the structure of the force very clear 
in terms of angular momenta. 


It was only the precision in the two-nucleon sector and the technical possibility of doing exact 
calculations in the framework of Faddeev equations which made it possible to undertake the efforts 
and to study the more complicated systems starting with the three-body system. These studies initially 
took place at lower energies already showing very interesting features such as the the well-known 
$A_y$ puzzle. Also, the calculations proved to be easier and more practical at lower energies. 
In the 1990s, the focus was shifted towards intermediate energies and one tried to study various 
aspects in the three-body systems. This owed itself, to a large extent, to the fact that with large 
computational capabilities, one could perform exact calculations including large angular momenta for 
various observables up to energies below the pion-production threshold. 

The interest in studying three-nucleon systems at intermediate energies required new detection systems 
and corresponding techniques; an effort which was taken up at several laboratories. The result of all 
the work in the past two decades is a database for the hadronic reactions in the three-nucleon system as shown in 
Fig.~\ref{worlddatabase}. The rich amount of available data show how well the field has matured 
and, as a result, our understanding of the nuclear forces has been significantly improved. 
The data in comparison with the state-of-the-art calculations also show (major) deficiencies in the 
models of the 3NF. A natural question that now arises is whether it is possible to also parameterize 
the 3NF as it was  done for 2NF in the past. One could argue that once 2NF and 3NF have the right 
parameters, any observable in the hadronic two- and three-nucleon sector will be described by the models 
as it should since the parameters of the model would be fitted to the data. 
The size of the database and the accuracy of the data would then determine how good our understanding 
of the underlying structure of 3NF is. Recent analysis \cite{Mardanpour:2010zz} shows that, once a 
large part of the phase space is covered, one has a tool to study specific aspects such as the 
isospin dependence of 3NFs.

Major investments for more experiments in these systems would require theoretical justifications. 
However, as was the case for the two-body system, the well-known parameters of two-nucleon forces 
were changed (albeit at a fine level) with the latest experiments from 
IUCF \cite{vonPrzewoski:1998ye} in the proton-proton system. The experimental and theoretical 
developments should, therefore, go hand in hand. Following the success in the two-body system, 
one obvious choice would be to perform a partial-wave analysis in the three-body system. The major 
challenge in performing this task is, however, the lack of an appropriate theoretical framework. 
A significant difference between the two-body and three-body systems is the very low-energy threshold of 
2.2~MeV which exists in the latter system. 
Already above this energy, an asymptotic three-body state needs to be
formulated. This problem is not solved. Therefore, a PWA above this energy
is technically impossible at the present moment.

In the absence of a solid model-independent approach for the three-nucleon system at higher energies, one would have to 
rely on the models which exist in the literature and try to refine them with the help of the data. 
These refined models could then be used in performing calculations of observables in three-body 
scattering problems but also in larger systems described in the following sections. The question of 
whether one should perform more measurements should be addressed carefully. 
Figure~\ref{worlddatabase} clearly shows that the database for the hadronic three-body reaction 
channels at intermediate energies is far from complete. The experimental situation for the elastic cross sections and analyzing powers seems to be, generally, well under control. The more exclusive spin observables, however, should be improved in the energy and 
angle range of the measurements. For the break-up channels, the database is even poorer and
many observables are missing. Due to the difficulties in the measurements of these observables, 
and with the availability of exact three-body calculations and very fast computers, 
the design of any new experiment should be guided by detailed theoretical calculations which should help single out regions of phase space where all effects such 3NF, Coulomb, relativity,
etc.~can be best studied. An example approach of this was performed by the Bochum-Cracow
theory group in Ref.~\cite{KurosZonierczuk:2002uv} in which the sensitivity to 3NF effects 
in the N$d$ break-up reaction was provided as a guide for experimental activities. This was used
as input for the design of a 4$\pi$ detection system, BINA, as shown in Fig.~\ref{BINA}.
In this context, analysis methods, such as the sampling technique, that provide a direct 
comparison between experiment and theory for the complex three-particle final state
of the N$d$ break-up reaction were applied very successfully~\cite{KurosZolnierczuk:2004xt,ThorngrenEngblom:2004ym}. Experiments that are designed to test explicitly the predictive power of
the chiral EFT in the three-nucleon continuum are recommendable. An example of this is the 
PAX experiment at COSY~\cite{Thorngren:2009loi} that will study the effects of current schemes
for 3NFs that recently were implemented at third order in the chiral EFT calculations and diagrams
appearing at fourth order. 
Aside from new measurements, there are also measurements in the
literature which require a re-measurement or re-analysis of the data due to disagreements with
each other or due to internal inconsistencies.
There are also measurements which have been performed with limited statistics or limited coverage of kinematical variables. New measurements to address these problems should also be performed. These measurements can nowadays only take place at RCNP (and to a lesser extent at RIKEN) and J{\"u}lich at intermediate energies.

\section{Three-nucleon forces and $A=4$ systems}
\label{sec:fourbody}

\subsection{Are four-nucleon forces important?}
\label{fournucleonforces}

Clearly, the 3N system is the cleanest laboratory to study 3NFs since by construction 
only NN and 3N forces can contribute. The sensitivity of 3N observables is however 
limited. We have seen that one has to go to intermediate energies and to search 
very specific configurations to identify signatures of 3NFs in 3N observables that 
are not linked to the $^3$H binding energy. Three-nucleon bound states with other spin/isospin 
configurations do not exist, so that studies of the spin-isospin structure of 3N force 
can only be performed based on the 3N scattering data in the three-body systems. 
In view of these constraints, it appears promising to study 3NF effects in 
four-nucleon (4N) systems.  It is the simplest nuclear system that supports resonance
structures besides the $^4$He bound state. It is conceivable that the resonance 
energies are similarly sensitive to 3NF contributions as the 3N binding energy. Since 
resonances of different spin and isospin exist, it should become possible to 
study the spin and isospin dependence of the 3NF in the very low energy regime. 

Obviously, to achieve this goal, it is necessary to control the size of four-nucleon force (4NF) 
contributions. Although first studies of selected 4NF contributions appeared in 
the 1980s \cite{Mcmanus:1980ep,Robilotta:1985gv}, not much is known about their 
size. The general expectation is that the 4NF contributions at least to $^4$He 
are small. Mostly, this expectation is based on the observation that the $^3$H 
and $^4$He binding energy are almost linearly correlated (Tjon line, \cite{Tjon:1975plb,Nogga:2001cz})
and that extrapolating the results of phenomenological model calculations almost hits 
the experimental 3N/4N binding energies (see footnote on page \pageref{efimovfoot}). 
Also the rather successful description of  
spectra and binding energies of light nuclei based on NN and 3N forces 
\cite{Pieper:2001ap,Navratil:2007we} is generally seen as support for the assumption that 4NFs 
are negligible. 

More quantitative is the analysis of the Hanover-Lisbon group \cite{Deltuva:2008jr}. 
Assuming that the most important 4NF contribution is due to intermediate 
$\Delta$ excitations of two-nucleons interacting with the two others, they were able 
to extract the 4NF contribution from their calculations with explicit $\Delta$ 
degrees of freedom. Both for various low-energy scattering observables 
and for the $^4$He bound state, they found the 4NF effects are not significant. 
For example, the contribution to the binding energy of $^4$He is 170 keV. 
This is approximately 0.2\% of 
the potential energy and can be considered as negligible.     

In the framework of chiral EFT, the size of 4N forces 
can be estimated based on the underlying power counting \cite{vanKolck:1994yi}. 
The first 4NF contribution appears at N$^3$LO. 
Therefore, the contribution can be expected to be less important 
than the leading 3NF. 
It is possible to check this expectation from power counting and to 
perform complete N3LO calculations in future, since
the leading 4NF has been worked out in 
Refs.~\cite{Epelbaum:2006eu,Epelbaum:2007us}. In addition to the well-known long-ranged 
pion-exchange contributions which were already discussed in the literature 
\cite{Mcmanus:1980ep,Robilotta:1985gv}, one finds that there are shorter-ranged 
contributions directly linked to the short-range part of the NN interaction. In this way, 
one obtains a consistent set of NN, 3N and 4N forces. 
Of utmost practical importance is that the leading 4NF is completely 
determined by parameters that appear in the NN interaction. No 4N datum 
is required to determine this 4NF quantitatively. In contrast to the leading 3NF,
which is driven by LECs saturated by the $\Delta$, no such LECs 
appear in the leading 4NF. In this sense, chiral EFT predicts a completely 
different 4NF than the one taken into account by the Hanover-Lisbon group. 
It is, therefore, interesting to calculate the contribution of 4NFs to nuclear systems. 
This has recently been achieved based on a perturbative calculation of the 4NF 
contribution to the $^4$He binding energy \cite{Nogga:2010epj}. It was found that for the chiral 
interactions of \cite{Epelbaum:2004fk}, the 4NF contribution is of the order of 300~keV. 
Again, this is only 0.3\% of the potential energy. It is much smaller than the typical 
contribution of the 3NF to the $^4$He binding energy of 2-4~MeV \cite{Epelbaum:2002vt}
and probably negligible  for most studies of the 4N system. Note, however, 
that for large cutoffs and for phenomenological interactions, 
the contributions were somewhat larger pointing to a possible scheme 
dependence of the importance of 4NFs. 

In summary, the preliminary results do not indicate that 4NFs give 
significant contributions for the standard low cutoff chiral interactions. It is not clear 
whether this is true for phenomenological interactions since the estimates 
based on chiral 4NF are not completely negligible. We, however, stress that in all cases 
the contribution of the 4NF is significantly less important than the contribution of the 
leading 3NFs. Based on this last insight, we can expect that calculations without 4NF 
will be sufficient to test the contribution of the leading 3NFs in 4N systems. 

\subsection{Four-nucleon systems at low energies}
\label{fournucleonlow}

The 4N systems promise to provide valuable information on nuclear dynamics. It 
is not only a simple extension of 3N dynamics to a slightly more complex system since 
it is the first system that supports a whole spectrum including the $^4$He bound state 
and several resonance structures for which data are available in different isospin channels
\cite{Tilley:1992zz}. Assuming the energies of the resonances are similarly sensitive to nuclear
interactions as the binding energies, scattering observables should significantly depend 
on 3NFs even at very low energies of a few MeV. Because the states have different spins and 
isospins, this should open the possibility to study the spin and isospin dependence of the 3NFs 
in the low-energy regime where a fast convergence of the chiral expansion can be expected. 
Obviously, the 4N system  can only  be a unique laboratory to study 3NFs if theoretical approaches are 
available that enable one to solve the 4N problem with the same accuracy and reliability as 
it is nowadays possible for the 3N problem. 

Such methods have been devised in recent years. In the following, we would like to summarize briefly the current 
status of this development. In order to be concise, we will focus on methods that have shown the reliability for 
modern nuclear interactions including all the complexity. Pioneering calculations, which unfortunately needed 
uncontrolled approximations, are summarized in Ref.~\cite{Tilley:1992zz}. We stress that such approaches 
were important milestones for the development of the methods but are not suitable for studies of subtle 
contributions of 3NFs.    

The solution of the $^4$He bound state problem has become standard by now. Many methods 
have been devised and have been benchmarked for rather sophisticated NN interactions 
(see \cite{Kamada:2001tv} and references therein). Also when 3NFs are included, there are several 
methods that have proven their reliability, e.g.~the Green's Function Monte Carlo (GFMC) approach 
\cite{Pieper:2001mp}, the no-core shell model (NCSM) \cite{Navratil:2009ut}, 
the hyperspherical harmonics expansion \cite{Viviani:2004vf}, 
solutions of Faddeev-Yakubovsky equations in momentum \cite{Nogga:2001cz,Deltuva:2008jr}
and configuration space \cite{Lazauskas:2003phd,Lazauskas:2004hq} have been used to obtain 
solutions for a wide class of NN and 3N interactions. 

Unfortunately, the information on 
3NFs which can be extracted from the $^4$He bound state is limited due to a strong 
correlation between the 3N and 4N binding energies called the Tjon line 
(which, actually represents a band) \cite{Tjon:1975plb}. 
As a consequence, although the 3NF contribution to the $^4$He binding energy 
is much larger than the 4NF contribution, the changes of the binding energy 
are rather subtle  when the $^3$H binding energy is kept constant.  Therefore, 
it is not completely clear that these changes can be  
disentangled from effects of 4NFs. Despite this, since two data points are required to fix the two unknown 
parameters of the 3NF, the $^4$He binding has been used 
to determine the $c_{D}$ and $c_{E}$ parameters of the leading chiral 3NF 
\cite{Nogga:2005hp}. It turned out that it is not always possible to find a unique set of 3NF parameters. 
These extractions were nevertheless valuable to confirm the effects of 3NFs for systems with 
$A>4$ (see next section). In the course of such investigations, it was realized that the radius 
of $^4$He is also sensitive to the $c_{D}$ and $c_{E}$ parameters and can be used 
to fix their values in a probably more reliable way \cite{Navratil:2007we}. All these investigations showed, however, 
that the $^4$He bound state is not a good laboratory to study the spin/isospin dependence of 3NFs. 
This can be traced back to the dominance of $J^{\pi}=1/2^+$ 3N partial wave states in the $^4$He 
wave function that are very similar for the $^3$H bound state.  

Much more promising is the solution of the 4N continuum problem. Although considerable progress has been 
made over the last 10 years, this problem is still not completely solved. Reliable theoretical results 
have only been obtained below the so-called three-body break-up threshold namely the energy where 
the system can break-up in two nucleons and the deuteron or in four nucleons. A first pioneering study 
below break-up threshold 
based on Yakubovsky equations was already reported in \cite{Kamada:1995aip} but 
was only continued by other groups later.


The simplest system for which data exist is the neutron-$^3$H system. Nevertheless, it shows the most 
important features of the 4N system while no Coulomb interaction needs to be taken into account. At the same 
time, the threshold for the first break-up is at rather high energies (at approximately $8.3$~MeV neutron 
laboratory energy) and data
for the energy dependence of the total cross section and some differential cross sections exist.

Obviously, the most basic observable is the zero energy cross section, which is conventionally parametrized in 
terms of the scattering lengths for spin-$0$ and spin-$1$ 
neutron-$^3$H states ($a_{0}$ and $a_{1}$). The scattering lengths 
have first been studied theoretically in detail in \cite{Viviani:1998gr,Ciesielski:1999pp}. 
Calculations based on NN interactions 
overpredict the zero energy cross section and the scattering lengths considerably. This failure could be traced 
back to the underbinding of $^3$H predicted by the NN models. Again, there is a correlation 
between the scattering lengths and the few-body binding energies \cite{Viviani:1998gr} and the experimental 
data are fairly well reproduced once the binding energies are reproduced. In this context, the calculations 
indicate that the scattering lengths are not completely consistently extracted from the two experimental results, the zero energy cross section and the so-called coherent scattering length \cite{Lazauskas:2004hq}.
Such an inconsistency calls for an independent measurements of these fundamental quantities. 

\begin{figure}[t]
\begin{center}

\includegraphics[width=0.6\textwidth]{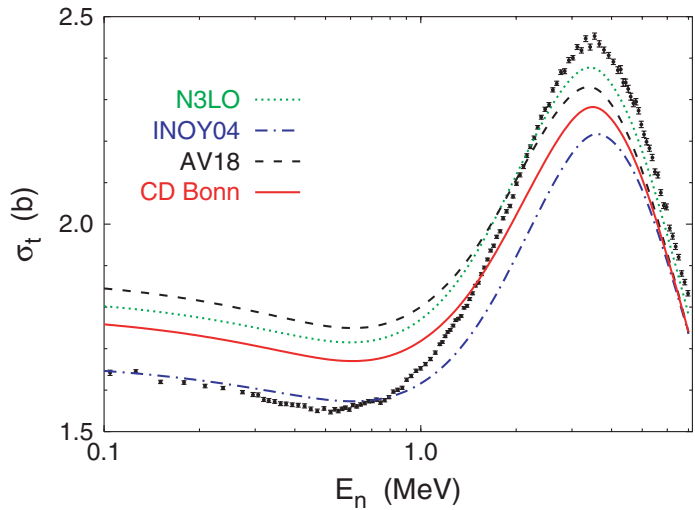}

\caption{\label{fig:4nobsxsec} 
Total neutron-$^3$H cross section $\sigma_{t}$ as a function of  
the neutron laboratory energy ($E_{n}$). The calculations are from 
Ref.~\cite{Deltuva:2006sz}. Data are from Ref.~\cite{Phillips:1980zz} (color online).
Reprinted with permission
from~\cite{Deltuva:2006sz}. Copyright (2007) by the American Physical Society. }
\end{center}
\end{figure}

The total cross section for neutron-$^3$H  scattering already has an interesting energy dependence. 
It has a broad peak at approximately 3~MeV which is generated by several P-wave resonances \cite{Tilley:1992zz}.
This energy dependence has been the focus of the early realistic calculations 
\cite{Ciesielski:1999pp,Lazauskas:2004hq}. While first calculations with simple NN interactions or with approximated 
NN interactions showed agreement with the data \cite{Ciesielski:1998sy,Fonseca:1999zz}, this was not the case 
for the complete calculations. This could finally be traced back to the simplifications in the first calculations 
 \cite{Lazauskas:2004uq} so that it is clear by now that the height of the peak is significantly underpredicted 
by realistic calculations (see Fig.~\ref{fig:4nobsxsec}). 
Studies including the 3NFs, various NN interactions or explicit $\Delta$ show that 
this failure is not correlated to the inaccurate description of the $^4$He binding energy 
and cannot be remedied by inclusion of standard 3NFs \cite{Deltuva:2008jr,Lazauskas:2004hq,Deltuva:2006sz}.
Therefore, the energy dependence of the total neutron-$^3$H cross section already constitutes a 
first puzzle in the 4N continuum calling for more systematic studies including more complex three-nucleon 
forces.  Indeed a first calculation including all topologies of the leading chiral 3NF 
(see upper part of Fig. \ref{fig:3nffeyn}) gives a much better description of the data \cite{Viviani:2008td}.

Since neutron-$^3$H experiments are difficult, the proton-$^3$He mirror system has attracted a lot 
of attention in spite of the additional complications due to the Coulomb interaction of the protons. 
The proper inclusion of Coulomb effects was first achieved in configuration space approaches, where 
the proper boundary conditions can be implemented for two cluster states 
\cite{Viviani:2001cu,Lazauskas:2005ka}. Later, Coulomb forces were also implemented in 
momentum space \cite{Deltuva:2007xv}.

\begin{figure}[t]
\begin{center}

\includegraphics[width=0.8\textwidth]{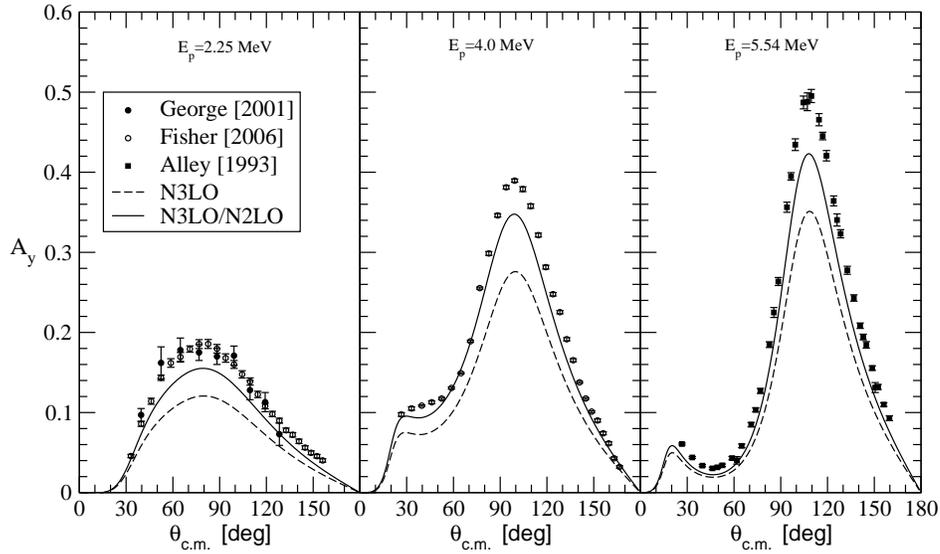}

\caption{\label{fig:4nobsay} 
                         Proton $A_{y}$ in $p$-$^3$He scattering at three different energies compared to the 
                        preliminary calculations of Ref.~\cite{Viviani:priv,Viviani:2010mf}. 
                        The solid (dashed) lines are predictions based on the chiral interaction 
                        of Ref.~\cite{Entem:2003ft} with (without) 3NFs.  Data are from 
                        Refs.~\cite{Alley:1993zza,Viviani:2001cu,Fisher:2006pm}.}
\end{center}
\end{figure}

Based on these advances, it was realized that, although the differential cross section 
of low-energy $p$-$^3$He scattering is described fairly well by phenomenological interactions, there 
is a sizable underprediction of the proton analyzing powers ($A_{y}$) 
\cite{Viviani:2001cu,Fisher:2006pm}. 
The size of this failure is much more significant than in the 3N system. 
In the 3N system, the absolute size of $A_{y}$ is only a few percent and the absolute 
deviations are of the order of a percent. In contrast, in the 4N system, the magnitude of $A_{y}$ is much 
larger so that the absolute deviations are as high as 15\%.  In view of the size of this deviation, 
it is astonishing that the predictions are not significantly dependent on NN forces 
\cite{Deltuva:2007xv} and the standard 3NFs are unable to remedy the problem 
\cite{Viviani:2001cu,Fisher:2006pm}.
However, recently a preliminary study of the Pisa group showed that the leading chiral 
3NF, properly adjusted to the $^3$He and $^4$He binding energies, significantly improves 
the description on $A_{y}$  \cite{Viviani:2010mf}, see Fig.~\ref{fig:4nobsay}. 
At the same time, the chiral 3NF does not distort the favorable description of other 
observables. This would probably be the most prominent signature of the 
topologies of chiral 3NFs observed so far. Interestingly, the same force is not able 
to improve the description of $A_{y}$ in proton-$^2$H scattering (see Fig.~\ref{aypuzzle})
leaving room for the action of higher order 3NFs. 
 
Besides the simplest elastic processes, there are also several possible reactions starting, e.g., 
with two deuterons in the initial state exemplifying the richness of the four-nucleon system for 
studies of nuclear reactions:
\begin{enumerate}
\item Elastic channel: $ d + d \longrightarrow d + d$;
\item Neutron-transfer channel:  $ d + d \longrightarrow p + t$;
\item Proton-transfer channel:   $ d + d \longrightarrow n +  ^{3}\!\rm He$;
\item Three-body final-state break-up: $ d + d \longrightarrow p + n + d$;
\item Four-body final-state break-up: $ d + d \longrightarrow p + n + p + n$. 
\end{enumerate} 

Studying the processes listed above with the same rigor becomes very involved once three- and four-body 
channels are open. In configuration space, such channels have a very complicated asymptotic behavior 
and in momentum space the singularities of the integral kernels become cumbersome. 
Below the three- and four-body thresholds, $dd$ scattering has already been studied in  
\cite{Fonseca:1999zz} where it is pointed out that also NN P-waves 
give important contributions. 
Advanced calculations of $dd$ scattering are reported in 
\cite{Deltuva:2008jr,Deltuva:2010yp}. 
This study also includes 3NFs due to $\Delta$ isobar contributions. Generally, the calculations agree 
well with experiment for elastic scattering and rearrangement to $p$ + $^3$H and $n$ + $^3$He. 
But there are some interesting discrepancies of theory and data for polarization observables.  
The contribution of the 3NFs induced by $\Delta$s is usually small. It will be interesting to check this 
for chiral 3NFs that have helped to improve the situation for elastic $n$-$^3$H scattering as 
discussed above. 

A complete overview of the different 4N reactions at low energy 
has been given by Hofmann and Hale \cite{Hofmann:2005iy}. 
The experimental data have been used for an R-matrix analysis, which is compared to 
resonating group model (RGM) calculations. 
Based on the RGM calculations, it is argued that 3NFs give important contributions to low-energy 4N reactions. 
By means of the R-matrix analysis, Hofmann and Hale identify observables that are specifically well suited 
to further constrain phase shifts and that should be measured with higher priority. 
Given the interesting sensitivity of many of the 4N observables to 3NFs discussed 
above such measurements will be important to advance our understanding of 3NFs.

\subsection{Four-nucleon systems at intermediate energies}
\label{fournucleonhigh}
In this section, the recent developments made in the four-nucleon systems at intermediate energies 
will be discussed. Contrary to the low-energy regime, where many calculations have been performed 
(see last section), there are presently no calculations done for scattering observables at these energies. 
This is partly due to the fact that the deuteron break-up threshold is very low. The main problem of the 
calculations is the treatment of break-up configurations. Already for the three-body break-up
channels, this becomes very cumbersome in configuration space calculations, which is reflected in a difficult 
singularity structure in momentum space. Therefore, Uzu, Kamada, and Koike  tried to avoid 
the singularities completely by performing the calculations for complex energies and analytically 
continuing to the real axis \cite{Uzu:2003ms}. They demonstrated that this approach is feasible and may lead 
to reliable results also for energies well above three-body break-up. It is, however, difficult numerically to ensure 
the accuracies required for such a continuation. Especially, when the energies are close to the real axis,
it is difficult to perform the 4N calculations with such a high accuracy. Therefore, so far no realistic calculations 
have been performed in this energy range.  Instead, crude approximations have been used to take 4N rescattering at high energies  into account mainly to estimate the size of possible effects 
(see e.g. \cite{Elster:1996eh,Micherdzinska:2007rd,Nogga:2006cp}). For another recent proposal 
to simplify the singularity pattern in  in the Faddeev equations which might open the way to four-nucleon 
scattering calculations the reader is referred to Ref.~\cite{Witala:2008my}. We also emphasize that 
the four-body continuum dynamics has also been studied using the Lorentz integral transform methof \cite{Bacca:2008tb}.

It is also because of the lack of theoretical predictions that no major attempts have been made in the past to 
study three-nucleon force effects in the four-body systems at intermediate energies. Nevertheless, recent 
attempts have been made to contribute to the understanding of nuclear forces in these systems. The experiments 
are very similar to those in the three-body system except that the number of channels is much higher. 
All possible channels in the interaction between two deuterons were listed in section \ref{fournucleonlow}. 

\begin{figure}[t]
\begin{center}

\includegraphics[width=\textwidth]{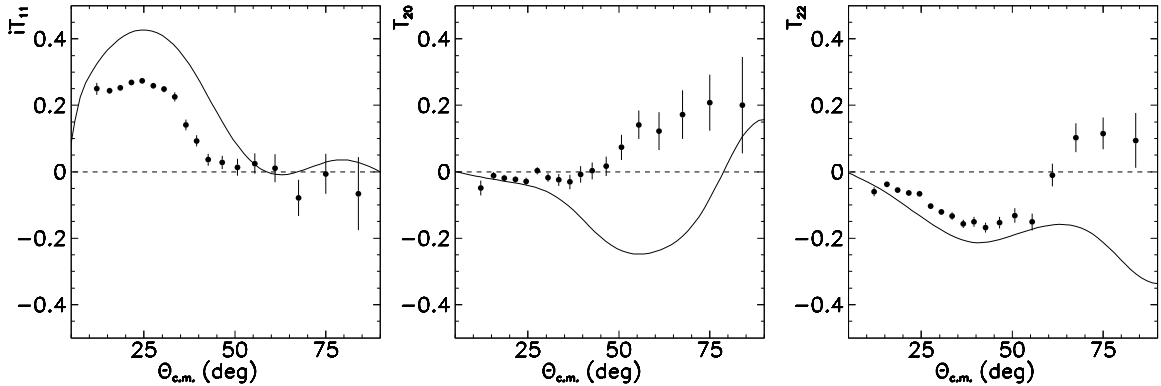}

\caption{\label{ddelastic} 
Angular distributions of various analyzing powers for deuteron-deuteron elastic-scattering at an incident beam energy of 231 MeV. Errors shown are statistical. The curve shows the result of approximated calculations
\cite{Micherdzinska:2007rd}. Reprinted with permission
from~\cite{Micherdzinska:2007rd}. Copyright (2007) by the American Physical Society. 
}

\end{center}
\end{figure}

\begin{figure}[t]
\begin{center}

\includegraphics[width=0.8\textwidth]{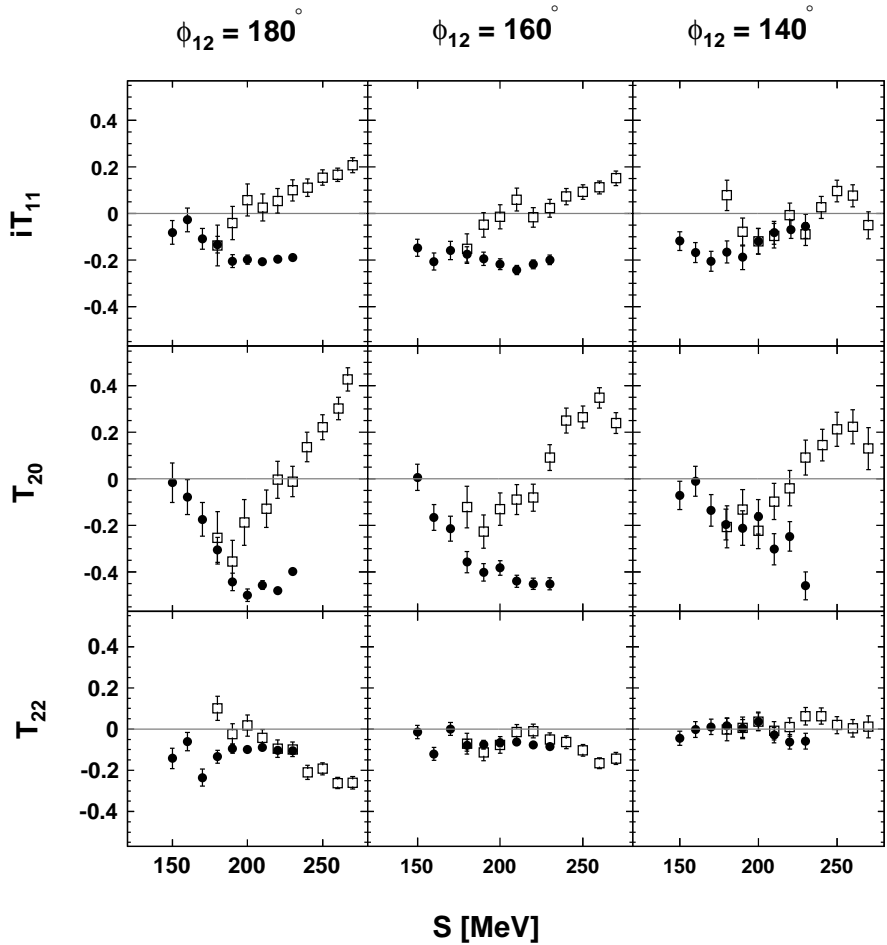}

\vspace*{-7mm}

\caption{\label{threebodybreakup} 
The vector- and tensor-analyzing powers of the reaction $ \vec d + d \longrightarrow p + n + d$ at 
$(\theta_d, \theta_p) = (15^\circ, 15^\circ)$ (open squares) and 
$(\theta_d, \theta_p) = (25^\circ, 25^\circ)$ (filled circles) as a function of $S$ for different 
azimuthal opening angles listed on top of the figure. The data were taken using a polarized
deuteron beam with an energy of 135~MeV~\cite{RamazaniMoghaddamArani:2010rb}. 
The solid lines in all panels show the zero level of the analzying powers. 
Only statistical uncertainties are indicated. Reprinted with permission 
from~\cite{RamazaniMoghaddamArani:2010rb}. Copyright (2010) by the American Physical Society. }

\end{center}
\end{figure}

The channels with the two-body final state are best studied using a magnetic spectrometer, as a 
good measure of the scattering angle of one of the outgoing particles is enough to determine the kinematics, 
provided that the particles are well identified. For the study of the 
other channels, one has to resort to detection systems such as that shown in Fig.~\ref{BINA} which span a large 
part of the available phase space of the reaction. Results of a recent measurement at 
IUCF \cite{Micherdzinska:2007rd} of elastic deuteron-deuteron scattering is shown in 
Fig.~\ref{ddelastic} together with calculations that are the results of very approximate  
calculations. More recently, two high-precision measurements were performed at 
KVI aiming to measure all reaction channels mentioned in the previous section. In Fig.~\ref{threebodybreakup}, 
the results of vector and tensor analyzing powers for the three-body break-up channel 
(reaction ($iv$) in the previous section) are shown for a few selected kinematics \cite{RamazaniMoghaddamArani:2010rb}. 
From the figure, it is clear that there is a rich set of data with high precision at this energy which can be 
used to test the models as they become available at intermediate energies. 

\section{How do 3NF manifest themselves in $A>4$ systems?}
\label{sec:manybody}

Whereas the few-body systems are ideal laboratories to study nuclear 
forces since calculations can be done with high accurcy, the application 
of the forces to more complex systems is of even more physical interest. 
Although much progress has been made in recent years, it is still not possible to 
reliably predict nuclear levels and binding energies for medium mass and heavy 
nuclei based on NN and 3N interactions. 
Most approaches are based on effective two-nucleon interactions, which are fitted 
to a range of nuclear data. Such interactions often fail to predict observables, especially when 
one explores exotic systems \cite{Bender:2003jk}. Assuming the nucleons are the relevant 
degrees of freedom to describe nuclei, it should be possible to predict such observables 
based on the underlying nuclear interactions. It is well known that for such studies not only 
2N forces, but also 3N forces are quantitatively relevant. In this section, we review approaches to calculate observables in light nuclei based on 2N and 3N interactions 
and summarize the quality of the predictions for these systems.  

We start with a section on the quality, advantages and disadvantages of several 
methods and, in the second section, focus on physical issues. We also note that we will 
not distinguish 3NFs due to truncations of the model space and due to other 
left-out degrees of freedom. We stress that these contributions cannot be disentangled
as becomes clear when we discuss 3NFs in conjuction with the renormalization 
group. In this context,  standard 3NFs are adjusted to absorb the model space (cutoff) dependence 
of observables. 
The results are in many ways interesting since they show the significance of 3NFs and at the same 
time they point to deficiencies of today's interactions. 
 
\subsection{Computational methods}

Reaching solid conclusions on the structure and significance of various parts of the nuclear Hamiltonian 
requires computational methods that allow one to reliably predict binding energies, 
excitation spectra and other properties of light nuclei. 
Several of such schemes have been developed over years, which provide such solutions for different 
types of interactions. Many methods imply constraints on the forces or systems to be studied. But they have been benchmarked for observables where several methods can be applied, e.g. for the $A=4$ binding energy in 
\cite{Kamada:2001tv}. Therefore, the numerical results are very reliable in their respective range of applicability. 

A very broad overview of nuclear binding energies has been obtained using the GFMC method \cite{Pieper:2001mp}.  As can be seen in Fig.~\ref{BEs} in the
introduction, the binding energies of the ground state and the low lying excited states have 
been calculated for many nuclei with $A \le 12$. The calculation starts from a variational trial 
wave function. Using an imaginary time propagation, the ground state is projected out of this 
trial state. For this, a path integral method is applied. Thereby, integrals are calculated with a 
Monte Carlo method, but sums over quantum numbers are kept explicitly. 
Therefore, the calculations grow exponentially with $A$ so that the 
applicability is limited to systems within the $p$-shell. 
In this way, it is possible to calculate expectation values for many operators. 
Due to the complexity 
of the wave functions, it is however not possible to generate them. 
Usually, the calculation allows one to study the properties of  the lowest level for a given set of quantum numbers. 
It is however also possible, but computationally much more difficult, to 
obtain excited states for a given quantum number \cite{Pieper:2004qw}.  

Results are usually obtained for the class of Argonne interactions since they are given in a suitable local 
operator form that is required in the implementation of the method. For this class of interactions,
however, even exotic, very weakly bound systems like $^8$He can be studied. 
Also, the GFMC approach has been generalized to scattering of nucleons on light nuclei 
\cite{Nollett:2006su} enabling the investigation of unbound systems, i.e. $^5$He.  
These features make GFMC especially interesting to study the charge and isospin 
dependence of the nuclear force based on the Argonne-Urbana-Illinois set of interactions.  

The no-core shell model (NCSM) is more flexible with respect to the form of the nuclear 
interaction. Here, the Schr\"odinger equation for the nucleus is solved in a 
harmonic oscillator (HO) basis. Thereby, no inert core is assumed. 
A direct solution for standard interactions would, in 
general, require a much too large number of HO basis states and is not possible. 
Therefore, the problem is first solved for a cluster of two or three nucleons. 
This solution can be used to define an effective interaction, which is then used in the nuclear structure 
calculation \cite{Navratil:2009ut}. Alternatively, a direct solution is possible for very soft interactions 
obtained in renormalization-group-based approaches to the nuclear interaction. This will be briefly 
discussed below.  At this point, it is possible to calculate and apply the effective interaction up
to three-body cluster level \cite{Navratil:2002zz} taking into account also the 3NFs discussed 
in the previous sections  \cite{Navratil:2003ef}. The NCSM gives very reliable predictions for the 
excitation energies of states that are dominated by the lowest HO states. This implies that so-called 
intruder states such as  e.g.~the $^{12}$C Hoyle state are not well described \cite{Navratil:2000ww}. 

In the standard formulation of NCSM, the calculations are feasible for $p$-shell nuclei. For more complex systems, 
the number of basis states increases dramatically so that converged calculations cannot be obtained 
anymore. But recently, the approach has been extended to medium mass nuclei  \cite{Roth:2007sv} 
based on a restricted NCSM basis. Here not all basis states are taken into account. Instead the importance 
of each state is a priori estimated perturbatively. In this way, a major fraction of the states can be dismissed 
before the Schr\"odinger equation is solved. 
This so-called importance truncated NCSM has been criticized for its numerical accuracy
(see \cite{Dean:2007vi,Roth:2008at}) but promises an interesting possibility for 
ab-initio calculations of much heavier systems or intruder states. 

Another recently developed extension is the combination of the NCSM with the 
resonating group method (RGM) 
\cite{Quaglioni:2008sm,Quaglioni:2009mn,Navratil:2010jn,Navratil:2011ay}. 
So far this combination has been implemented for nucleon-nucleus 
and deuteron-nucleus scattering  but at this point omitting 3NFs.
The scattering problem is solved 
within a basis of cluster states and relative coordinates of the clusters. 
The cluster states are obtained from NCSM calculations. 
It is then possible to implement the antisymmetrizer for states of different clusters 
(or the spectator nucleon with respect to the nuclear cluster). 
This enables to calculate a realistic norm kernel for an RGM calculation. 
The effective RGM interaction can, therefore, be systematically 
calculated from the underlying nuclear interaction.  This approach works nicely for 
very low energy scattering, where only a few excitations of the clusters are relevant. 

For interactions based on EFT, it recently became feasible 
to solve the problem not based on a nucleonic Hamiltonian but directly on the 
Lagrangian of the effective field theory \cite{Epelbaum:2009pd}. 
The solution is performed for a discretized space-time lattice. 
The strength of this approach is its wide range of applicability. Starting 
from NN systems \cite{Borasoy:2007vy}, 
which enable to determine the low energy constants for the particular 
regularization implied by the lattice, even light nuclei 
\cite{Epelbaum:2009pd,Epelbaum:2010xtepj} and neutron matter \cite{Borasoy:2007vk} 
have been investigated.  
The lattice calculations for nuclear structure result in binding 
energies accompanied with reliable error bounds. The uncertainty estimates do not 
only take  into account the numerical errors due to statistical fluctuations of the applied 
Monte Carlo scheme and lattice artifacts 
but also the uncertainties of the nuclear Lagrangian by considering 
several orders of the chiral expansion.  In this approach,
as is also true for GFMC, nuclear wave functions cannot be computed.
Instead, expectation values of operators, e.g. correlation functions, 
can be computed. Recently, even excited states for $^{12}$C have been extracted including 
the weakly bound Hoyle state that is relevant for nucleosynthesis \cite{Epelbaum:2011drf}. 
Since NN and 3NFs are automatically included when the pertinent terms of the 
Lagrangian are taken into account, this calculation is in fact the first calculation of this 
state based only on a few-nucleon input. 

For low-momentum interactions, many-body perturbation theory (MBPT)  is another very 
flexible approach to the nuclear many-body problem. It has been applied to finite nuclei 
\cite{Roth:2005ah,Coraggio:2010jz}  as well as to nuclear matter 
\cite{Bogner:2005sn,Tolos:2007bh,Hebeler:2010xb}. 
For a review, we refer to Ref.~\cite{Bogner:2009kx}. 
In this scheme, the solution is based on the Hartree-Fock approximation to the observables, which 
is particularly simple for closed-shell nuclei and for nuclear matter. 
For low momentum interactions, this first-order approximation is already sufficiently close to the solution,
so that a perturbative expansion for the higher-order contributions becomes meaningful. 
It has been shown that second-order contributions are significant.  Higher-order contributions 
have only partly been taken into account, but the results indicated that they 
are negligible.  So far only the nuclear matter calculations involve 3NFs. 
These can be exactly taken into account for the leading, Hartree-Fock,
term. For higher orders in the MBPT expansion, state-of-the-art calculations require approximations 
of the 3NF. Since these higher-order contributions are generally small, it is expected that the 
approximations only lead to insignificant errors of the calculation.  

Also the coupled-cluster approach \cite{Kummel:1978zz}, 
which is especially suitable for medium mass closed-shell nuclei, 
has been applied to Hamiltonians that involve 3NFs  \cite{Mihaila:1999vn,Coon:1977zfp,Hagen:2007ew}. 
Early calculations were performed approximating the 3NF by an effective NN force 
\cite{Mihaila:1999vn,Coon:1977zfp}. 
Based on such calculations, it has already been argued  that the residual 3NFs are indeed 
small \cite{Coon:1986cx}.
So far, the calculations involving the complete 3NFs \cite{Hagen:2007ew} have only been performed for $^4$He since 
the computational demands are very high. It turned out, for this system and a specific 
choice of NN interaction, that the proper 3NF contributions are tiny. The bulk of the 
3NF effects could be absorbed in effective NN forces, see also Ref.~\cite{Coon:1977zfp} for a related 
earlier study.  At this point, for more complex 
systems, the importance of 3NFs is often estimated based on variations 
of the NN force (e.g. via a cutoff of a low-momentum interaction). In some case, 
this indicates that a complete treatment of 3NFs might be required in such systems 
\cite{Hagen:2010gd}. 

This concludes our survey on methods that have been applied to the nuclear many-body 
problem based on NN and 3N interactions. We would like to mention briefly that other 
techniques have been developed that promise the capability to employ NN and 3N forces 
to systems with $A>4$, i.e. the Fermion Molecular Dynamics (FMD) \cite{Feldmeier:2000cn} 
and  the effective interaction hyperspherical harmonics approach (EIHH) \cite{Barnea:2001ak}.
FMD has been used with modified NN interactions for a wide variety of nuclear systems 
including the Hoyle state \cite{Chernykh:2007zz} and has proven to provide a flexible ansatz 
for the nuclear wave function that can accommodate shell-model-like states and cluster-like 
states simultaneously. Unfortunately, 3NFs have not been incorporated yet. 
EIHH  has already been  used  for $A=4$ including 3NFs \cite{Barnea:2004ac,Barnea:2010prc},
but, for $A>4$ ,  it has only been applied using simplified NN interactions \cite{Bacca:2004dr}.

\subsection{Signatures of 3NFs for $A>4$}

Based only on the phenomenological NN interactions, the binding energies of light nuclei 
are underpredicted \cite{Pieper:2001mp,Navratil:1998uf} as can be expected from 
the results for $A \le 4$ systems. This underbinding is indeed much reduced, once 
3NFs are taken into account that are adjusted to reproduce the binding energies 
of light systems \cite{Pieper:2001mp,Navratil:2003ef}. The comparably small deviations 
from the experimental binding energies observed for the combined interactions is 
nevertheless a first interesting hint that $p$-shell nuclei are much more sensitive to details 
of the 3NFs than the $^4$He nucleus.  

First of all, one observes that the binding energies predicted 
based on AV18 (or a simplified version AV8' \cite{Pudliner:1997ck}) 
in combination with the Urbana-IX or TM 3NFs are smaller than obtained from experiment. 
The deviations are small compared to the overall binding energy and on the 
percent level, when compared to the potential energy. Nevertheless the deviations 
can be significant in many cases since the ordering of levels is affected. A prominent 
example is the ordering of the $J^\pi = 3^+$ and $J^\pi = 1^+$ levels in $^{10}$B. 
Experimentally, the splitting of both states is 720~keV where $J^\pi = 3^+$ is the ground 
state of $^{10}$B. Clearly, the level splittings of $p$-shell nuclei are tiny compared to 
binding energies/potential energies. Therefore, it is not surprising that AV18 
alone and in combination with Urbana-IX leads to a $J^\pi = 1^+$ ground state. 
This remains true if other phenomenological NN interactions are used \cite{Navratil:1998uf}. 
Interestingly, the addition of the TM 3NFs at least results in the correct ordering 
of levels although the splitting seems to be smaller than experiment \cite{Navratil:2003ef}.
These results make two facts obvious: the binding energies and splittings of levels of $p$-shell 
nuclei are extremely sensitive to details of the nuclear Hamiltonian and rather small 
deviations can lead to qualitatively different results since the ordering of levels 
can be affected with the obvious impact on possible transitions between the levels.
At the same time, one can expect an especially large sensitivity  
to 3NFs that one only finds in very specific kinematical conditions in 3N and 4N scattering, 
e.g. in cross section minima. The systematic studies discussed 
in the previous chapters are, therefore, not only academic exercises but are required to determine
nuclear Hamiltonians that are accurate enough for systematic predictions of nuclear structure. 

The GFMC studies revealed another interesting failure of the combination of AV18 and Urbana-XI. 
The deviations of predictions from the experimental results increased with an increasing number 
of neutrons \cite{Pieper:2001ap}. This clearly points to a deficiency in the isospin dependence 
of this combination of nuclear interactions. 
This observation triggered the development of new 3NFs models. The GFMC collaboration  
identified so-called ring diagrams, 3$\pi$-exchanges with intermediate $\Delta$ excitations, 
as a possible source of additional isospin $T=3/2$ contributions of the 3NFs. They added the expected 
new spin/isospin structures to their model which culminated  
the series of Illinois interactions (IL1 to IL5) \cite{Pieper:2001ap}. Indeed, their 
study showed that the new interaction terms further improved the description of the spectra 
of $p$-shell nuclei. The new terms resolved the underbinding problem for neutron-rich systems 
and the deviations in the level orderings at the same time \cite{Pieper:2002ne}. 
These improvements could also be identified for $n$-$^4$He scattering  \cite{Nollett:2006su}.
Note that these original calculations suffered from a bug in the three-nucleon force code 
\cite{Pieper:priv} which is resolved in the meantime, and lead to the development of a new parameter set (IL7) of the Illinois 3NF. Figure~\ref{BEs} summarizes the corrected results 
which qualitatively agree with the old ones. 

A close look  reveals that the 3NF contributions due to triples of neutrons 
are very small for $p$-shell nuclei probably because the additional neutrons are usually well 
separated from each other and from the core of the neutron-rich nucleus \cite{Wiringa:priv}. 
This is very different for neutron matter and it turned out that the Illinois model, that 
describes $p$-shell  nuclei very well, is not suitable for neutron matter \cite{Gandolfi:2010za}. 
An improved version is under development \cite{Pieper:priv}.

Although the results for TM' are in some aspects different to the ones of Urbana-IX and/or 
Illinois, the  calculations always showed that the overall agreement with experiment is clearly 
improved by the addition of 3NFs. Looking more into details, the results still show 
deviations from data similarly to calculations based on Urbana-IX. This situation can already 
be expected from the fact that also the $pd$ and $nd$ scattering data are not satisfactorily described. 

Since today's combinations of NN interactions and 3NFs do anyway not describe nuclei 
perfectly and  because of the technical complexity to apply 3NFs in nuclear 
structure calculations, there have been attempts to get around 3NFs by modifying 
NN interactions. Most prominently the INOY (inside non-local outside Yukawa) model 
\cite{Doleschall:2000rp} and the JISP (J-matrix inverse scattering potentials)
\cite{Shirokov:2007plb} are phase-shift equivalent realistic NN interaction that 
are engineered to improve the description of binding energies. The INOY model 
reproduces the $^3$H binding energy exactly.  But calculations for 
$p$-shell nuclei showed discrepancies to charge radii, spectra and binding energies again 
\cite{Navratil:2004cns}.
One is lead to conclude that also this model requires the addition of 3NFs 
although $s$-shell systems might be described without them. 
Interestingly, a modified version of INOY, where P-wave NN phase shifts 
have been altered so that the $nd$ $A_{y}$ is properly described, leads to 
a good description of some spin-orbit (LS) splittings in $p$-shell nuclei. This makes a relation 
of both discrepancies likely. As the name suggest, the long-range 
part of the interaction nevertheless is driven by the $1\pi$-exchange so that it 
is conceivable that INOY results can be improved by adding 3NFs based on 
$2\pi$ exchange to improve this situation. JISP, in contrast, is not only fitted 
to NN data. Using a scheme of unitary transformations, JISP could be fitted to 
$p$-shell spectra while keeping the description of the NN data. Therefore, both, 
$p$-shell spectra and NN data are described well. Despite this success, it remains 
to be seen to what extent predictions are possible in more complex systems. 
Since the unitary transformations also change the long-range part of the interaction, 
it is also not clear how JISP could be improved by 3NFs, if discrepancies to data are found. 

The huge number of possible operator structures, estimated to be at 
least 120, makes a purely phenomenological 
approach to the 3NFs impossible. Clearly, it would be helpful to improve the 3NFs 
according to some theoretical guidance and chiral
EFT promises such a guidance. As discussed before, the leading 3NF in chiral 
EFT is given by the usual $2\pi$-exchange  and additional $1\pi$ and contact 
interactions (see Fig.~\ref{fig:3nffeyn}). Therefore, using the NCSM, the impact of these 
additional structures on observables involving $p$-shell nuclei has been studied. In a first step,
$^7$Li was investigated \cite{Nogga:2005hp}. In this first study, the spectrum of $^7$Li was 
compared based on chiral interactions. The parameters of the chiral 3NFs were chosen to be 
consistent with the NN interaction used and the two parameters related to the $1\pi$ and contact 
terms used to fix the binding energies of $^3$H and $^4$He exactly at their experimental values. 
As expected, the results for $^7$Li differed for both choices of these parameters showing 
that the additional shorter-range contributions are significant for $p$-shell nuclei. 

It was, however, not clear which choice of parameter set resulted in a better description. 
More research on the impact of the shorter-range contributions revealed that 
the fitting procedure of \cite{Nogga:2005hp} missed that also the density of $^4$He
considerably depended on the parameters of the 3NF. Given that the binding energy 
of $^4$He might get contributions from several subleading few-nucleon 
interactions including the 4NF, it became clear that fitting the parameters 
to the density (or matter radius) of $^4$He and the $^3$H binding energy is a more 
reliable approach. This was first done in  \cite{Navratil:2007we}. This work studied  a wide range 
of observables for several $p$-shell nuclei. Using the density of $^4$He, a preferred 
set of strength constants for the leading 3NF was found. Then the combinations of the 
strength were modified keeping the $^3$H binding energy constant and allowing for 
slight variations  of the predictions for the density. 
In this way, it was found that most excitation 
energies and transition matrix elements were only mildly dependent on the choice 
of strength constants. 

A few of the observables turned out to be very sensitive to the 
contribution of the short range 3NFs: e.g.~the quadrupole moment of $^6$Li, the ratio of the 
electric quadrupole strength $B(E2)$ for transitions of the ground state of $^{10}$B to the 
first and second excited $J^\pi=1^{+}$ states, and the magnetic dipole strength $B(M1)$ 
for transitions from the ground state of $^{12}$C to the  $J^\pi=1^{+}$ isospin 0 state.  
Within the accuracy of the calculations, it was possible to find a choice for the 
strength of the short-range contributions which consistently describes all these observables 
and the density of $^4$He. The spectra of the considered $p$-shell nuclei
generally improved by adding the chiral 3NF in this form. Still, a few discrepancies  
hint to missing contributions or inaccurately determined parameters of the 
interactions. Again, this is in line with the situation in few-nucleon systems, where 
a few observables are still not well described. Since many of the discrepancies are 
related to LS splittings, it is conceivable that such small deviations might 
be related to the $A_{y}$ problem. It will be interesting to study the effect of subleading 
3NFs on such observables and a possible correlation with $A_{y}$. 

The nuclear lattice simulations carried out so far out so far are in line with the results 
that the properties of nuclei are qualitatively described at N$^2$LO.
Here, the convergence 
of the ground and excited states with respect to the order of the chiral expansion 
was studied in more detail. Looking at predictions at NLO and N$^2$LO,  
it became clear that a quantitative description requires 
N$^3$LO contributions which have not been completely included yet in any 
of the approaches \cite{Epelbaum:2011drf}. 

A large fraction of studies involving 3NFs is based on renormalization group (RG)
approaches to the nuclear force. The $V_{low-k}$  \cite{Bogner:2003wn,Epelbaum:1998hg,Fujii:2004dd}, 
the similarity renormalization group (SRG) \cite{Bogner:2006pc} and the
unitary correlation operator method (UCOM) \cite{Roth:2004ua} belong to this 
category. By different means one of the phenomenological NN interactions is softened such that 
the long range part of the interactions is unaltered. 
The procedures guarantee that NN phase shifts do not change for momenta 
below a cutoff momentum. 
Such a softening is a pre-requisite to solve the nuclear many-body problem 
based on NN and 3N interactions. For $V_{low-k}$ and SRG it was explicitly  
shown that the softened NN interaction are universal meaning that they do not depend 
on the NN interaction from which one starts the procedure. The softening implies 
a dependence of the results on the RG parameter (or cutoff) which was 
shown to be of the size expected for 3NFs and should be mainly absorbed by 3NFs 
\cite{Nogga:2004ab}. The similarity of the 
interactions to chiral interactions in some cutoff range makes it conceivable that RG interactions 
may be consistently augmented by chiral few-nucleon forces. 
The application of such combinations to nuclear structure revealed that, for a limited cutoff 
range, the NN interaction and the 3N interaction can be perturbatively  added to 
a Hartree-Fock solution \cite{Bogner:2005sn,Hebeler:2010xb} (for a review see \cite{Bogner:2009bt}). 
The results show that the 3NF
contribution is of the size expected. But this implies that it is necessary to obtain a realistic 
saturation point for nuclear matter. The cutoff dependence of the results is very much reduced 
by the 3NFs indicating that 4N and higher-order interactions are still less important. 
The calculations for nuclear matter also showed a large sensitivity of the results 
on the strength parameter of the 2$\pi$-exchange part of the chiral 
3NF \cite{Tolos:2007bh,Hebeler:2010xb}. 
Since the uncertainties of these parameters are rather large, a possible determination 
from few-nucleon data, as started in \cite{Kievsky:2010zr}, is certainly well motivated. 
Such determinations should, however, carefully address the theoretical uncertainty resulting 
from the truncation of the chiral expansion for the nuclear forces.

Similar calculations for UCOM for finite nuclei were performed without 3NF and, for 
an optimized correlation parameter, showed a rather good description of binding energies
\cite{Roth:2005ah}. 
But it turned also out that radii are not well reproduced and again the need for 3NFs 
was confirmed.  Very recently short-ranged 3NFs could be included in similar 
calculations \cite{Gunther:2010cs} improving the agreement with data.

The RG approaches have the important advantage that a cutoff parameter 
can be varied in certain limits similarly to the chiral EFT approach. The cutoff 
dependence of results is then a lower bound of the accuracy achieved in calculations 
and gives a first estimate of omitted contributions, e.g. higher-order few-nucleon 
interactions. It can also be used to identify correlations between observables
as the extensions of the Tjon line to more complex systems \cite{Bogner:2007rx}.  
For SRG, this has even been extended to generate some 
parts of the missing 3NFs \cite{Jurgenson:2009qs}. Such developments enable 
one to give better estimates for, e.g., the 4NF contribution (which was shown 
to be small in this case) but they also allow to determine the 3NF for 
arbitrary  cutoffs starting from a single fit to data \cite{Jurgenson:2009qs}. 

A further interesting application of 3NFs is reported in \cite{Otsuka:2009cs}. Within 
the standard shell model, the oxygen isotopes are studied in more detail. It is shown 
that calculations based on phenomenological NN interactions fail to reproduce the 
position of the drip-line correctly since they results in too low single energies for the 
neutron $d_{3/2}$ orbitals. 3NFs correct this failure leading to the correct 
position of the drip-line at $^{24}$O. 

This short overview concentrated on the results for binding energies. 
Indeed, these are especially sensitive to 3NFs, but we stress 
that many nuclear structure calculations showed that 
3NFs also have impact on transition matrix elements. In some 
cases, transition matrix elements are strongly dependent on a suitable 
interplay of wave function components. This has been exemplified e.g. in  
\cite{Hayes:2003ni,Maris:2011as,Holt:2009uk} for neutrino scattering and the lifetime of 
$^{14}$C, see also Ref.~\cite{Holt:2007ih} for a related recent work. 

In summary, the  3NFs have been shown to be important ingredients for nuclear 
structure calculations.  Although, for the $p$-shell, the predictions are already 
in fair agreement with the data, there are still discrepancies. It is hoped that 
such discrepancies can be resolved by higher-order EFT interactions.
It will be interesting to observe how nuclear structure 
and few-body results are affected by the next generation of 3N and higher-order 
interactions. 
 At the same time, the development of RG methods to make the nuclear many-body 
 problem tractable highlighted the importance of 3NFs to remove the scale dependences
 of these approaches when only NN interactions are used.

\section{Summary, conclusions and outlook}
\label{sec:conclusions}

In this review we have summarized recent theoretical and experimental achievements 
towards understanding the role of the three-nucleon force in few-nucleon systems
and its application in many-nucleon systems.  
Resolving the structure of the three-nucleon force 
requires accurate experimental data and the 
ability to accurately solve the few-body Schr\"odinger equation for a given 
nuclear Hamilton operator. Considerable progress has been made in the past 
decade on both fronts. 

Clearly, the most natural place to test the 3NF
is the three-nucleon continuum. Neutron-deuteron scattering calculations can nowadays 
be routinely performed with and without 3NFs both for the elastic and the break-up channels. 
Since most of the data are proton-deuteron data, one needs to take into account the 
long-range electromagnetic interaction in the Faddeev equations which is  
especially important 
in various proton-deuteron break-up configurations. 
Recently, impressive progress towards resolving this long-standing 
challenge was made albeit the final word is perhaps still to come. 
The experimental techniques have improved drastically
in the course of time and were partly driven by the precision requirements 
in the field of few-nucleon systems. 
The experimental situation in N$d$ scattering at intermediate energies is 
summarized in Fig.~\ref{worlddatabase}. It clearly demonstrates that 
the existing database is far from being complete.
Most of the data corresponds to differential cross sections and nucleon and deuteron 
analyzing powers. A restricted number of more complicated spin observables 
such as the spin transfer coefficients have also been measured. 
The global impact of the  
phenomenological 3NF models on the available data is visualized in 
Figs.~\ref{globalanalysiselastic1}, \ref{globalanalysiselastic3} and
\ref{globalanalysiselastic2} for elastic scattering and 
Figs.~\ref{globalanalysisbreakup} and \ref{globalanalysisbreakup2} 
for the break-up observables. The experimental data are confronted with the 
theory calculations by the Hanover-Lisbon group based on the 
coupled-channel version of the CD-Bonn potential which allows one to identify 
the effects of the 3NF due to intermediate $\Delta$ excitations.  
This particular choice can serve as a representative example 
of three-nucleon calculations based on phenomenological high-precision 
potentials accompanied by 3NF models, see Fig.~\ref{globalanalysiselastic1}. 
The elastic cross section and the nucleon vector 
analyzing power show a clear signature of the 3NF with the deviations 
from the data being significantly reduced once the 3NF is included. On the other hand, 
sizable deviations, typically increasing with energy, still persist after 
inclusion of the 3NF especially at backward angles  indicating 
deficiencies in the 3NF models. It is difficult to make a definite 
global conclusion for tensor analyzing powers and spin-transfer/correlation 
coefficients where the situation is more controversial (which is, in part, due to 
the less precise and much smaller amount of data available). The analysis of the 
break-up data shows that the effects of the 3NF are predicted to be, in most cases, 
fairly small and insufficient to remove the discrepancies with the theory. 
This is especially true for the vector analyzing power indicating the possible 
deficiencies in the spin structure of the current 3NF models. 
  
Given the recent progress towards rigorous solution of the four-body problem 
in the continuum which is, however, presently limited to the low-energy region,  
four-nucleon systems are expected to be a promising testing ground for three- and 
four-nucleon forces in the near future. It should also be emphasized that 
the four-nucleon systems feature certain properties, such as the 
low-lying resonances, which are absent in 3N scattering and are expected 
to increase the sensitivity to the details of the nuclear Hamiltonian. 
One well-established puzzle in the 4N continuum is related to the total 
cross section for neutron-triton  scattering showing a broad peak 
around $E_{\rm lab} \sim 3$ MeV due to several P-wave resonances, which appears to be  
underpredicted by the existing two- and three-nucleon force models. 
The $A_y$ puzzle also persists in the 4N system although a recent calculation 
by the Pisa group indicates a significant improvement due to inclusion 
of the chiral 3NF at N$^2$LO. Data at intermediate energy are rather scarce for this
system. The lack of data for this system is primarily due to the lack of exact calculations. 
Only recently, precision data for various reaction channels and covering 
a large part of the phase space have become available. 
The four-nucleon continuum will clearly be 
an important frontier area of research in the next years. 

Spectra of light nuclei provide another interesting testing ground 
for three-nucleon forces. Calculations performed within the GFMC and NCSM 
approaches typically show a clear improvement  
once the 3NFs (chiral 3NF at N$^2$LO, Urbana IX, Illinois 3NF models and TM') are included. 
Sizable deviations, however, still persist pointing to the deficiencies
in the spin- and isospin structure of these 3NF models. 

The results of the present analysis demonstrate clearly that the momentum-spin-isospin 
structure of the 3NF is not properly described by the existing 
phenomenological models. 
What is needed is a \emph{systematically improvable theoretical framework}
which allows to derive \emph{consistent} two- and many-nucleon 
interactions and currents. Such a framework is provided by chiral effective field 
theory. Two- and many-nucleon forces in this approach 
are derived from the underlying effective Lagrangian and do 
not suffer from conceptual problems typically arising 
in the context of phenomenological models (such as off-shell effects). 
In the past decade, NN potentials at N$^3$LO in the 
chiral expansion have been developed and demonstrated to allow
for an accurate description of the two-nucleon data comparable to the 
one of the high-precision potentials. The three-nucleon force
has so far only been explored at order N$^2$LO where it first 
starts contributing to the nuclear Hamiltonian. Its short-range part 
depends on the two new LECs $c_D$ and $c_E$ which cannot be determined in the 
2N system. The $c_D$-term is governed by the NN$\to$NN$\pi$ transition 
which figures importantly in strong, electromagnetic and weak few-nucleon 
reactions.  The exciting possibility to bridge these very different processes 
is already being explored by several groups, but more quantitative studies 
should be carried out in the future. The description of the low-energy 
neutron-deuteron scattering observables at N$^2$LO is, in general, 
comparable to the one obtained from the high-precision potential models showing 
similar puzzles such as the cross section in the space-star break-up
configuration. While the results at intermediate energies are generally 
in a good agreement with the data, the theoretical uncertainty becomes 
rather large reflecting similar pattern in the 2N sector at this order
in the chiral expansion. Promising results were obtained in calculations 
based on the N$^3$LO NN potentials accompanied with the N$^2$LO 3NF. In particular, 
the inclusion of the 3NF was shown to improve description of the 
spectra of light nuclei and to reduce the $A_y$ puzzle in the 4N system.  
Another interesting ab-initio approach to few- and many-nucleon systems 
combines effective field theory with numerical lattice simulations. Recently, 
this method has been successfully applied at N$^2$LO to compute the ground state 
energies of $^4$He, $^8$Be and $^{12}$C and the Hoyle state which presents 
a major challenge for nuclear theory. 

The first corrections to the 3NF at N$^3$LO are 
becoming available and currently being implemented in 3N scattering calculations. 
It is worth mentioning that 
they do not involve any additional 
free parameters. While definite 
conclusions about the importance of subleading corrections to the 3NF can 
only be made after performing explicit  calculations, the preliminary 
estimations of the longest-range contributions indicate that the effects in 
nucleon-deuteron scattering 
might be fairly small \cite{Ishikawa:2007zz}. 
If so, it would be necessary to extend the calculations 
to an even higher order in the chiral expansion to resolve the puzzles 
in the 3N continuum. A more promising approach to improve the convergence 
of the chiral expansion is to explicitly include the 
spin-$3/2$ degrees of freedom corresponding to the $\Delta$(1232) isobar
which is well known to play  an important
role in nuclear dynamics  due to its low excitation energy and strong coupling
to the $\pi N$ system. Such \emph{explicit} treatment of
the $\Delta$ in the EFT leads to more natural values of the LECs and 
allows one to resum a certain class of important
contributions. The improved convergence of the resulting approach in the 
NN sector was recently confirmed in peripheral nucleon-nucleon scattering 
at N$^2$LO. The 
implications for 3N scattering are yet to be worked out. While the complete
expressions for the 3NF at N$^2$LO are exactly the same in both approaches 
(althouth certain contributions in the $\Delta$-full theory are shifted
from N$^2$LO to NLO), one expects sizable long-range N$^3$LO contributions 
to the 3NF and 4NF due to intermediate $\Delta$-excitations, 
see also \cite{Machleidt:2009bz} for a related discussion.   
The calculations by the Argonne group using the phenomenological Illinois 
3NF models motivated by the ring diagrams with an intermediate $\Delta$-excitation 
seem to indicate the importance of such contributions.  It is worth mentioning that 3NFs  
due to two and three intermediate $\Delta$-excitations neglected 
by the Argonne and Hanover-Lisbon groups also appear at N$^3$LO, and the resulting potentials 
turn out to be of a similar size.    
Last but not least, the four-nucleon force also starts to contribute at N$^3$LO.   
The parameter-free expressions have recently been worked out in the $\Delta$-less 
formulation and shown to yield a small contribution to the $\alpha$-particle 
binding energy of the order of a few hundred keV. It remains to be seen whether 
this conclusion will still hold after explicit inclusion of the $\Delta$-isobar
although the results in the 4N system obtained by the Hanover-Lisbon group 
seem to indicate a small contribution. 

The question to be raised is whether one should do more measurements and if so, what should 
they be? As shown in Fig.~\ref{worlddatabase}, cross sections and (proton and deuteron) 
analyzing powers have been measured at several energies and for a large range of scattering
angles, and aside from some normalization problems, the database is in good shape for these
two observables. The situation is, however, different for more exclusive spin observables such 
as spin-transfer and spin-correlation coefficients which have been measured only for a couple 
of energies and in a limited angular coverage. Different combinations of spins in these 
observables clearly show different behavior making them suitable for these studies.  

In the break-up sector, there is certainly room for improvement for all observables. However, due to
the difficulties in performing the experiments, the measurements should be guided by theoretical 
input as to where the effects would be largest. Also, for this reaction channel, more exclusive spin
observables should be measured in order to fully understand the underlying dynamics. It should,
however be mentioned that observables like spin-transfer coefficients are the most difficult to 
measure for a large part of the phase space as one would require a polarimeter in several 
regions of the phase space. First attempts have been made to measure this observable for a
selected kinematics \cite{Sekiguchi:2009zz}. The only laboratories which are, in principle, 
capable of carrying out this type of measurements at intermediate energies are RIKEN, RCNP 
and J{\"u}lich.  

The situation for four-nucleon systems is similar to that of the three-nucleon system of about 20 years ago. Measurements exist only to a limited extent and primarily at energies below the 
break-up threshold.
Note that the large efforts undertaken in the three-nucleon sector in the
last two decades were due to the availability of exact calculations justifying large investments for 
these measurements. In that respect, the theoretical efforts in the four-nucleon sector are clearly
lagging behind and should be vigorously pursued to prove the feasibility of doing exact 
calculations at intermediate energies before major efforts are put into the actual measurements of 
observables. 

In short, a lot has been learned about nuclear forces and few-nucleon systems 
in the past two decades making this field of science almost an exact field 
where the conclusions are drawn based on precise, quantitative arugments. Accurate calculations
can be performed routinely in two- and three-body systems using various potential models. The
results of these calculations can be examined with the precise data which are available for these
systems. All observables can be investigated and remarkable agreements have been reported in 
a large part of the phase space for all possible reactions. However, there are also clear 
discrepancies which are much larger than any experimental or theoretical uncertainties pointing to
the fact that the nuclear Hamiltonian still needs refinement. 
It has been shown that the disagreements in cross sections and
analyzing power for the elastic channel increase with incident energy. More 
exclusive spin observables show discrepancies to various degrees depending on the combination
of the spin components. In selected break-up kinematics, S-wave proton-proton pairs were observed 
at small relative energies and sizable discrepancies were observed with theory predictions
for the vector analyzing power, $A_y$. Strikingly, data taken at similar elastic kinematics 
in which the deuteron was detected, were described significantly better by the same theory.  
All these observations call for a systematic and quantitative theoretical approach for
the nuclear forces and few-nucleon dynamics. To establish such an approach in the complete
energy range up to the pion-production threshold still poses a challenge 
for a theory. The ongoing efforts in the chiral effective field theory should
be further pursued to extend 3N calculations which already exist to higher 
energies and to reduce the theoretical uncertainty. Based on the experience from the 2N system, 
this will require to go to at least N$^3$LO in the chiral expansion of the 3NF and, possibly, 
to include the $\Delta$ isobar as an explicit degree of freedom. 
Work along these lines is in progress, and will certainly stimulate dedicated 
experiments in N$d$ scattering to explore and test the spin structure of the emerging, novel 3NFs. 
In parallel, efforts should be increased towards accurate calculations of
four-nucleon scattering observables as they
promise to be a good testing ground to study nuclear forces.
Once these systems are under control, 
steps can be taken to understand the full dynamics of nuclear systems with more nucleons 
involved. On the computational front, various groups have developed techniques to calculate 
complicated many-body systems including 3NFs. Some of these new developments rely on RG 
evolved nuclear interactions, where  
the 3NF is an essential ingredient to remove the scale dependence of results. Independent of
this issue, the input from few-nucleon systems will clearly be crucial to eventually understand 
the more complex systems. Nuclear structure and reaction studies will undoubtedly 
benefit from the present developments.  

\vspace{1cm}
{\bf ACKNOWLEDGEMENTS} 

Writing a review article entails discussing various aspects with many people. 
The input of so many colleagues has been essential in bringing this article 
to a successful end. Here, we mention the names of all colleagues who have given 
very valuable comments on various parts of the article alphabetically and would like 
to express our gratitude toward them. They are: S.~Coon, A.~Deltuva, A.C.~Fonseca, 
R.J.~Furnstahl, W.~Gl{\"o}ckle, J.~Golak, F.~Gross, H.-W.~Hammer, M.N.~Harakeh, 
H.~Kamada, A.~Kievsky, St.~Kistryn, U.~van~Kolck,  D.~Lee, R.~Machleidt, J.L.~Matthews, 
U.-G.~Mei{\ss}ner, H.O.~Meyer, P.~Navratil, G.~Orlandini, T.~Papenbrock, S.~Pieper, K.~Sagara, 
H.~Sakai, P.U.~Sauer, H.~Paetz~gen.~Schieck, A.~Schwenk, K.~Sekiguchi, E.~Stephan, E.~Stephenson, 
P.~Th\"orngren~Engblom, R.G.E.~Timmermans,
W.~Tornow, M.~Viviani, and H.~Wita{\l}a. Further, we would like to thank all 
our collaborators for having helped us in all aspects of our work which led to this review article. 
Part of the numerical calculations were performed at the JSC, J\"ulich, Germany.
N.~Kalantar-Nayestanaki and J.G.~Messchendorp are financially supported by the University of Groningen 
(RuG) and the Helmholtzzentrum f\"ur Schwerionenforschung GmbH (GSI), Darmstadt.  
E.~Epelbaum would like to acknowledge the financial support of the Helmholtz Association 
(contract number VH-NG-222) and the European Research Council (ERC-2010-StG 259218 NuclearEFT).  

\vspace{1cm}
{\bf References}
\vspace{2mm}

\bibliographystyle{lit-database/h-physrev5}
\bibliography{lit-database/lit}    

\end{document}